\documentclass{jfm}
\usepackage[dvipsnames]{xcolor}
\usepackage{amsmath}
\usepackage{mathtools}
\usepackage{amssymb}
\usepackage{bm}
\usepackage{subfig}
\usepackage{graphicx}
\usepackage{tikz}

\begin{document}

\newtheorem{lemma}{Lemma}
\newtheorem{corollary}{Corollary}
\newcommand{\rom}[1]
    {\MakeUppercase{\romannumeral #1}}
\shorttitle{Vortex line topology} 
\shortauthor{B. Sharma, R. Das and S. S. Girimaji} 

\title{Local vortex line topology and geometry in turbulence}

\author
 {
 Bajrang Sharma\aff{1}
  \corresp{\email{bajrangsharma@tamu.edu}},
  Rishita Das\aff{1}
  \and 
  Sharath S. Girimaji\aff{1, 2}
  }

\affiliation
{
\aff{1}
Department of Aerospace Engineering, Texas A\&M University, College Station, TX 77843, USA
\aff{2}
Department of Ocean Engineering, Texas A\&M University, College Station, TX 77843, USA
}

\maketitle

\begin{abstract}
The local streamline topology classification method of \cite{chong1990general} is adapted and extended to describe the geometry of infinitesimal vortex lines. Direct numerical simulation (DNS) data of forced isotropic turbulence reveals that joint probability density function (PDF) of the second ($q_\omega$) and third ($r_\omega$) normalized invariants of the vorticity gradient tensor asymptotes to a self-similar bell form beyond $Re_\lambda > 200$. The same PDF shape is also seen at the late stages of breakdown of Taylor-Green vortex suggesting the universality of the bell-shaped pdf form in turbulent flows. Additionally, vortex reconnection from different initial configurations is examined. The local topology and geometry of the reconnection bridge is shown to be identical in all cases with elliptic vortex lines on one side and hyperbolic filaments on the other. Overall, topological characterization of vorticity field provides a useful analytical basis for examining vorticity dynamics in turbulence and other fluid flows.
\end{abstract}

\section{Introduction}

The origins of the fields of vortex dynamics and topology are intricately intertwined \cite{moffatt2008vortex}.  \cite{helmholtz1858integrale} developed equations describing vorticity field evolution in idealized fluid flow and proposed the notion of vortex lines. Helmholtz’s seminal work in German was translated into English with some enhancements by \citep{tait1867translation}. The concept of vortex line served as an inspiration to \citep{kelvin1867vortex,kelvin1869vortex} who hypothesized the `Vortex Theory of Atoms’. According to this theory, matter is constituted of interconnected and knotted vortex filaments. The theory of atoms motivated a series of papers by \cite{tait1877knots,tait1884knots,tait1885knots} to characterize and classify knotted filaments.  Although the vortex theory of atoms has long been disavowed, Tait’s investigation of knots served as the foundation of the discipline of topology \citep{epple1998topology}. In more recent times, \cite{moffatt1969degree} conducted extensive investigations of knottedness of vortex lines and developed the relation between knot topology and energy spectrum \citep{moffatt1990energy}.

Vortex dynamics plays a crucial role in many fluid flow phenomena \citep{saffman1992vortex}. Indeed, \cite{kuchemann1965report} suggests that vortices are the `sinews and muscles’ of fluid motion.  The structure and evolution of vortex lines and sheets provide valuable insight into aerodynamic lift, wake dynamics and chaotic character of fluid flows. Vortices also play a central role in hurricanes, tornadoes, and astrophysical flows. Large-scale coherent vortices provide structure and drive many complex turbulent flows.  At smaller scales, vortex-stretching provides the central energy cascade mechanism in turbulence. Scalar mixing is also critically dependent on vortices for large scale stirring (entrainment) and diffusive enhancement at small-scale. Therefore, understanding vortex dynamics and characterization of underlying vortex-line topology at all scales of motion is important for many fields of engineering and nature. While large scale features of vortices such as knots, links and coherent structures have been reasonably studied, local topology and geometry of vortex lines have not received much attention.

The objective of this work is to develop a mathematical framework to characterize and classify the local topology and geometry of  vortex line elements. Infinitesimal vortex line elements are building blocks of finite-sized vortex lines and their study will lead to a deeper understanding of vorticity field. Further, infinitesimal vortex line topology can also provide insight into many important phenomena such as vortex stretching, vortex reconnection, and helicity. 

Vortex reconnection process is crucial in turbulent cascade \citep{yao2020physical}, noise generation in jets \citep{zaman1980vortex} and fine scale mixing in turbulence \citep{hussain1986coherent,hussain2011mechanics}.  The problem of vortex reconnection in symmetric configurations such as interaction of anti-parallel  vortex tubes \citep{melander1988cut} and collision of vortex rings \citep{kida1991collision} have been examined extensively in literature. Vortex inter-linkage at oblique angles are also commonly observed in propeller tip vortex interactions \citep{johnston1990propeller} and flow over pitching wings \citep{freymuth1989visualizing}. Moreover, in 3D flows vortex reconnection between orthogonally offset tubes is more likely than the symmetric parallel configuration. Thus, investigation of the vortex line structure at different alignments warrants further attention.

Much like infinitesimal material-element \citep{batchelor1952effect,orszag1970comments,monin2013statistical,girimaji1990material} and local streamline topology \citep{chong1990general,martin1998dynamics,elsinga2010evolution,das2020characterization}, the investigation of infinitesimal vortex-line elements can yield deeper understanding of various turbulence processes. 
Toward the stated objective, we undertake various tasks as follows:
\begin{enumerate}
    \item Adaptation and extension of the streamline local topology classification framework \citep{chong1990general} to describe the structure of infinitesimal vortex line elements. The adaptation requires performing the critical point analysis in a rotating reference frame. Additionally, we demonstrate that vorticity being a pseudovector does not affect this analysis. 
    \item Classification of the infinitesimal vortex line geometry (which is distinct from topology) is performed following the approach of \cite{das2020characterization}.
    \item Investigation of the universal features of probability density function (PDF) of the vorticity-gradient invariants and vortex-line topology distribution in turbulence. Two types of flows are considered: (a) statistically-stationary forced isotropic turbulence at various Reynolds numbers, and (b) breakdown of Taylor-Green vortex.
    \item Characterization of the local vortex line topology during different stages of vortex reconnection process initiated from different configurations. The different initial configurations considered are (a) anti-parallel \citep{melander1988cut} and (b) orthogonal \citep{boratav1992reconnection} vortex tubes.
\end{enumerate}

\section{Vorticity gradient tensor and local vortex line geometry}\label{sec:theory}

\cite{perry1987description} and \cite{chong1990general} characterized topological properties of local streamlines in terms of the velocity gradient tensor ($\bm{A} \equiv \nabla \vec{u}$, where ${\vec{u}}$ is the field velocity) using critical point analysis.
Our goal in this section is to derive a similar framework relating the vorticity gradient tensor ($\boldsymbol{\Phi}$) to the local vortex line geometry.

The vorticity gradient tensor $\boldsymbol{\Phi}$ is defined as: 
\begin{equation}
    \label{eq:wgt_def}
    \Phi_{ij}\equiv \frac{\partial \omega_i}{\partial x_j} \;\;\; \text{where} \;\; {\vec{\omega}} = \nabla \times {\vec{u}}
\end{equation}
where $\vec{\omega}$ is the vorticity vector. It is well known that vorticity is a \textit{pseudovector}, i.e., it has an additional sign change under an improper rotation such as reflection. Similarly, it can be easily demonstrated that the vorticity gradient tensor is also a \textit{pseudotensor}. 


Analogous to streamlines, a vortex line is defined as a curve that is instantaneously tangential to the vorticity vector ($\Vec{\omega}$) at any point in the flow. Mathematically, the local tangent vector $d\Vec{X}$ at any point on a vortex line $\Vec{X}$ is related to the vorticity vector as follows
\begin{equation}
d\Vec{X}\times\Vec{\omega}=0 \quad\mbox{where} \quad d\Vec{X}=dX_1\hat{i}+dX_2\hat{j}+dX_3\hat{k}
\label{eq:vline0}
\end{equation}
which implies
\begin{equation}
     \omega_3dX_2-\omega_2dX_3=0\mbox{,}\quad\quad\omega_2dX_1 - \omega_1dX_2=0\mbox{,}\quad\quad\omega_3dX_1-\omega_1dX_3=0
     \label{eq:vline0b}
\end{equation}
Equation \eqref{eq:vline0b} can be expressed as the following set of differential equations dependent on an arbitrary parameter $s$:
\begin{equation}
     \cfrac{dX_2/ds}{dX_3/ds}=\cfrac{\omega_2}{\omega_3}\mbox{;}\quad\quad\cfrac{dX_1/ds}{dX_2/ds}=\cfrac{\omega_1}{\omega_2}\mbox{;}\quad\quad\cfrac{dX_3/ds}{dX_1/ds}=\cfrac{\omega_3}{\omega_1}
    \label{eq:vline}
\end{equation}
Equivalently, (\ref{eq:vline}) can be written as
\begin{equation}
    \frac{d\Vec{X}}{ds}=\Vec{\omega}
    \label{eq:vline_diff}
\end{equation}
Solution trajectories obtained by integrating \eqref{eq:vline_diff} for a frozen vorticity field represent instantaneous vortex lines.

We intend to examine the local vortex line structure in the immediate neighborhood of some reference point ($\vec{x}_0$) by applying the critical point analysis.
Toward this end we first introduce the relative vorticity vector $\vec{\tilde{\omega}}(\vec{x};\vec{x}_0)$ at any point in the field surrounding $\vec{x}_0$: 
\begin{equation}
    \label{eq:def_vortrel}
    \vec{\tilde{\omega}}(\vec{x};\vec{x}_0)=\vec{\omega}(\vec{x})-\vec{\omega}(\vec{x}_0)
\end{equation}
 We define ``relative vortex lines'' as curves wherein, the relative vorticity vector is tangent to any point ($\vec{x}$) in the curve. Similar to vortex lines, relative vortex lines can be obtained by integrating the following differential equation for a frozen vorticity field. 
\begin{equation}
    \frac{d\Vec{x}}{ds}=\Vec{\tilde{\omega}}(\vec{x};\vec{x_0})
    \label{eq:rvline_diff}
\end{equation}
It can be shown that relative vortex lines are vortex lines as observed from a frame rotating with half the reference angular velocity $\vec{\omega}(\vec{x}_0)$.
We present the formal analysis of the relation between vortex lines and relative vortex lines in Appendix \ref{app:A}. It is important to note that vorticity equation is invariant to uniform reference-frame rotation unlike Navier-Stokes equation. 
Thus, the local vortex line topology in a uniformly rotating coordinate frame is similar to that of an inertial frame, whereas, the streamline topology in the two frames might not be similar.

The relative vorticity field in the immediate neighborhood of a reference point ($x_0$) can be approximated by the first order term of a Taylor series expansion about the reference point. Therefore, from equation (\ref{eq:rvline_diff}), we have
\begin{equation}
    \frac{dx_i}{ds} \approx \frac{\partial \tilde{\omega}_i}{\partial x_j}x_j
    \label{eq:ts1}
\end{equation}
 At the reference point $x_0$, the right hand side of \eqref{eq:ts1} is zero, i.e. $x_0$ is a critical point. 
 The gradient of relative vorticity in terms of the vorticity gradient tensor $\boldsymbol{\Phi}$ is given by:
\begin{equation}
    \frac{\partial \tilde{\omega}_i}{\partial x_j}=\frac{\partial }{\partial x_j}(\omega_{i}(\vec{x})-\omega_{i}(\vec{x}_0))=\Phi_{ij}
    \label{eq:relVGT}
\end{equation}
From equations \eqref{eq:ts1} and \eqref{eq:relVGT}, the equations for relative vortex lines can be written as
\begin{equation}
    \label{eq:vline_final}
    \frac{dx_i}{ds}=\Phi_{ij}x_j
\end{equation}
The form of equation \eqref{eq:vline_final} is identical to the local streamline ($\vec{x}'$) equation given by \cite{chong1990general} :
\begin{equation}
    \frac{dx'_i}{dt}= A_{ij}x'_j
    \label{eq:streamline}
\end{equation}

\textit{Topological classification of streamlines:} \cite{chong1990general} used equation \eqref{eq:streamline} to classify topology of local streamlines based on phase-space analysis given by \cite{kaplan1958ordinary}.
Thus, analogous to the characterization of local streamline topology in terms of velocity gradient tensor, the topology of local vortex lines can be characterized on the basis of vorticity gradient tensor. Chong \textit{et al.}'s work demonstrates that the invariants of velocity gradient tensor are sufficient to classify the local streamline topology. Specifically, in incompressible flows (wherein $\partial u_i /\partial x_i = 0$), the second ($Q$) and third ($R$) invariants of $\bm{A}$ exclusively classify the topology of local streamlines.

\subsection{Topological classification of vortex lines}

In a similar fashion, \eqref{eq:vline_final} can be used to classify the local vortex line topology in terms of the invariants of vorticity gradient tensor $\bm{\Phi}$, 
\begin{equation}
    \label{eq:invar}
    P_\omega=-\Phi_{ii}=0, \quad Q_\omega=-\frac{1}{2}\Phi_{ij}\Phi_{ji}, \quad R_\omega=-\frac{1}{3}\Phi_{ij}\Phi_{jk}\Phi_{ki} 
\end{equation}
Vorticity, curl of a vector, is divergence free by construction. Thus, the vorticity gradient tensor is trace free much like the velocity gradient tensor in incompressible flows. This key result allows us to draw analogues between the analysis of the invariants of velocity gradient tensor in incompressible flows and those of the vorticity gradient tensor. Therefore, $Q_\omega$ and $R_\omega$ completely classify the local vortex line topology. 

Unlike $R$, $R_\omega$ is not invariant under frame reflection, as $\Phi_{ij}$ is a pseudotensor. However, since frame reflection is not employed in the methodology of \cite{chong1990general} or \cite{kaplan1958ordinary}, we consider $R_\omega$ to be invariant for the purposes of topological classification. 
The governing equations for evolution of vorticity gradient tensor and its invariants are derived in Appendix \ref{app:B}. The equations demonstrate that the evolution of $Q_\omega$ and $R_\omega$ depends on $\Phi_{ij}$, velocity gradient tensor ($A_{ij}$) and their derivatives. However, unlike the velocity gradient evolution, vorticity gradient evolution equation does not have an explicit dependence on the pressure field.

\begin{figure}
    \centering
    \includegraphics[width=0.8\textwidth]{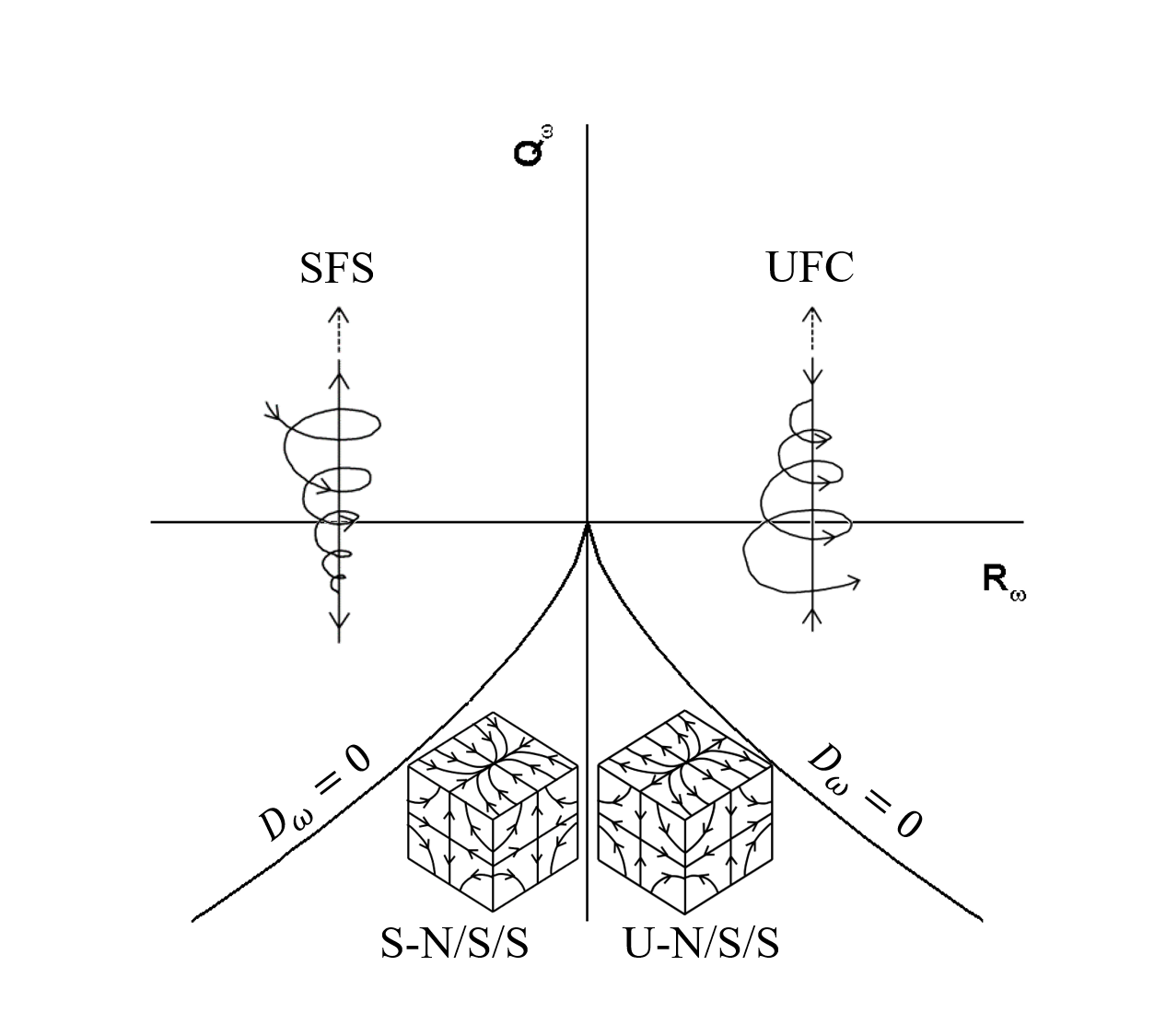}
    \caption{Canonical vortex line shapes in the invariant space of $\Phi_{ij}$}
    \label{fig:QR_chong}
\end{figure}

Vortex lines are classified into four distinct topologies based on the local values of $Q_\omega$ and $R_\omega$ and the canonical shape for each topology is displayed in figure \ref{fig:QR_chong}. Discriminant $D_\omega$ plays a key role: 
\begin{equation}
    \label{eq:dis}
    D_\omega=Q_\omega^3+\frac{27}{4}R_\omega^2
\end{equation}
Below the discriminant line ($D_\omega<0$), all three eigenvalues are real and two of the eigenvector planes contain saddle points while one contains a stable/unstable node, resulting in saddle-node combinations \citep{perry1987description}.  
In the saddle-node combination region, stability of node is determined by $R_\omega$.  
For $R_\omega<0$, the vortex lines converge in the nodal plane and this topology is referred to as stable-node/saddle/saddle (S-N/S/S).
Similarly, $R_\omega>0$ region represents diverging vortex lines in the nodal plane and this vortex line topology is therefore called unstable-node/saddle/saddle (U-N/S/S). 
Above the discriminant line ($D_\omega>0$), $\boldsymbol{\Phi}$ has two complex conjugate and one real eigenvalues. This region represents vortex lines that spiral around the only real eigenvector, forming a stable/unstable focus. When $R_\omega<0$, vortex lines spiral towards the center and out of the focal plane and the vortex line topology is termed as stable focus stretching (SFS). Similarly in the region $R_\omega>0$, vortex lines spiral away from the center into the focal plane and the topology is termed as unstable focus compression (UFC). The vortex line topologies above the discriminant lines, i.e. SFS and UFC, are spiraling in nature and are hereby referred to as focal topologies. On the other hand, vortex line topologies below the discriminant lines, i.e. UN/S/S and SN/S/S, do not spiral about a focus and are therefore termed as non-focal topologies.

It has been shown in a recent study \citep{das2020characterization}, that the topological description of streamlines in the invariant plane of the velocity gradient tensor ($Q$-$R$) does not uniquely specify the streamline shape, i.e. each point in the $Q$-$R$ plane can represent multiple streamline shapes of the same topology that are not geometrically similar. 
Similarly, the topological framework for vortex lines described above does not specify the vortex-line shapes uniquely. Additionally, the tensor components $\Phi_{ij}$ can be arbitrarily large and its invariants can increase unboundedly. Thus, it is expedient 
to construct a compact invariant space uniquely characterizing the vortex line shape. 

\subsection{Normalized vorticity gradient tensor}

Along the lines of \cite{das2019reynolds}, we normalize $\boldsymbol{\Phi}$ by it's Frobenius norm to compute the normalized vorticity gradient tensor ($\boldsymbol{\chi}$). 
\begin{equation}
    \label{eq:def_nwgt}
    \chi_{ij}\equiv\frac{\Phi_{ij}}{\|\Phi\|}\quad\mbox{where}\quad \|\Phi\|=\sqrt{\Phi_{mn}\Phi_{mn}}
\end{equation}
The normalized vorticity gradient tensor ($\boldsymbol{\chi}$), like $\boldsymbol{\Phi}$, is trace-free. Additionally, each component of the tensor is bounded.
Following \cite{das2019reynolds,das2020characterization} these bounds can be determined. For the sake of brevity only the key results are presented here.

The bounds of the diagonal elements of $\bm{\chi}$ are a consequence of its trace-free nature combined with the constraint imposed by normalization.
\begin{equation}
    \label{eq:diag_bound}
    -\sqrt{\frac{2}{3}}\leq\chi_{ij}\leq\sqrt{\frac{2}{3}}\quad\forall\thinspace\thinspace i=j
\end{equation}
Off-diagonal components of the tensor constrained simply by normalization are bounded as follows:
\begin{equation}
    \label{eq:odiag_bound}
    -1\leq \chi_{ij} \leq 1\quad\forall\thinspace\thinspace i\neq j
\end{equation}
The tensor $\boldsymbol{\chi}$ has three invariants denoted by $p_\omega$, $q_\omega$ and $r_\omega$.
\begin{equation}
    \label{eq:norm_invar}
    p_\omega=-\chi_{ii}=0, \quad q_\omega=-\frac{1}{2}\chi_{ij}\chi_{ji}=\frac{Q_\omega}{\|\Phi\|^2}, \quad r_\omega=-\frac{1}{3}\chi_{ij}\chi_{jk}\chi_{ki}=\frac{R_\omega}{\|\Phi\|^3} 
\end{equation}
The invariants - $q_\omega$ and $r_\omega$ - are also bounded. To obtain the bounds, first the tensor $\boldsymbol{\chi}$ is decomposed into a symmetric ($\boldsymbol{\chi^s}$) and a skew-symmetric ($\boldsymbol{\chi^w}$) tensor. The resulting tensors are then expressed in principal frame of $\boldsymbol{\chi^s}$. The trace free constraint of $\boldsymbol{\chi^s}$ and the normalization restrictions are used to establish the bounds of $q_\omega$ and $r_\omega$.
The second invariant of $\boldsymbol{\chi}$ is bounded as \citep{das2020characterization}:
\begin{equation}
    \label{eq:bound_invarq}
    \begin{split}
        -\frac{1}{2}\leq q_\omega \leq \frac{1}{2}
    \end{split}
\end{equation}
For a given value of $q_\omega$, $r_\omega$ is bounded by:
\begin{equation}
    \label{eq:bound_invarr}
    -\frac{1+q_\omega}{3}\sqrt{\frac{1-2q_\omega}{3}}\leq r_\omega \leq \frac{1+q_\omega}{3}\sqrt{\frac{1-2q_\omega}{3}}
\end{equation}
Both the minimum and maximum values of $r_\omega$ occur at $q_\omega=0$ leading to the following absolute bounds for $r_\omega$
\begin{equation}
    \label{eq:bound_invarr_abs}
    -\frac{\sqrt{3}}{9}\leq r_\omega \leq \frac{\sqrt{3}}{9}
\end{equation}
The bounds of $q_\omega$ (\ref{eq:bound_invarq}) and $r_\omega$ (\ref{eq:bound_invarr}) represent the boundaries of the realizable $q_\omega$-$r_\omega$ plane. It is important to note here that unlike the normalized velocity gradient invariants, these bounds of $\boldsymbol{\chi}$-invariants are valid in compressible flows as well (as $\bm{\chi}$ continues to be divergence free).

\subsection{Vortex line shape in the normalized invariant space}

\begin{figure}
    \centering
    {\includegraphics[width=0.7\textwidth]{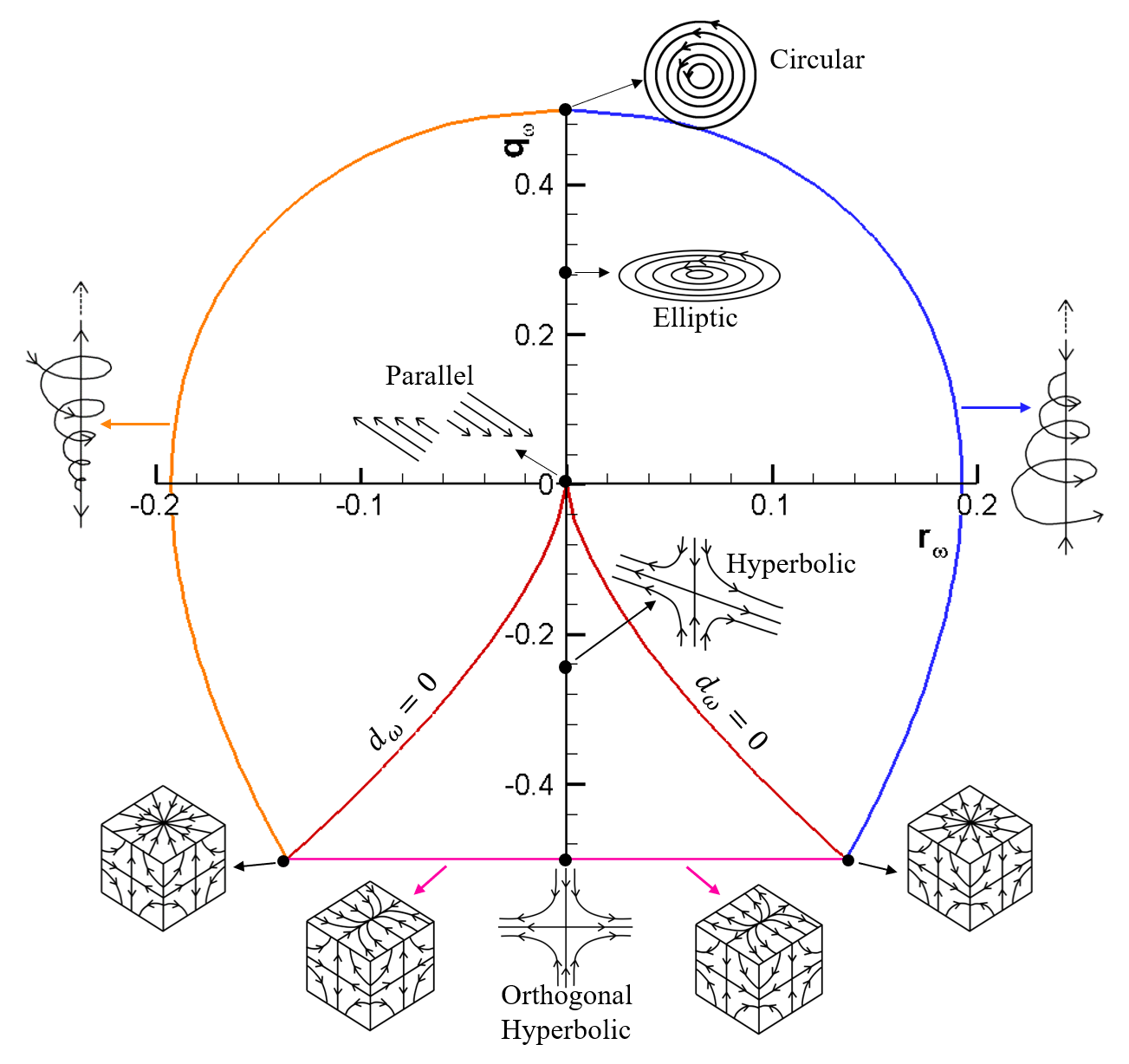}}
    \caption{Schematic of vortex line shapes represented by different points in the $q_\omega-r_\omega$ plane.}
    \label{fig:qrshapes}
\end{figure}

The invariants of $\bm{\chi}$ uniquely characterize the shape of the local vortex lines and $\|\Phi\|$ specifies the scale-factor \citep{das2020characterization}.
All the vortex line shape features discussed in this section can be obtained by performing phase space analysis \citep{kaplan1958ordinary} of the following system of ordinary differential equations, obtained from (\ref{eq:vline_final}) and (\ref{eq:def_nwgt}):
\begin{equation}
    \frac{dx_i}{ds'}=\chi_{ij}x_j  \quad \text{where} \quad s'=\|\Phi\|s 
    \label{eq:nvline}
\end{equation}

We now detail the local vortex line shape features, as represented by the different regions in the $q_\omega$-$r_\omega$ space, in figure \ref{fig:qrshapes}.
\begin{enumerate}
    \item \textit{Two-dimensional vortex lines:} Along the $r_\omega=0$ line, $\bm{\chi}$ has a zero eigenvalue ($\lambda_1=0$) and this results in planar vortex line shapes. 
    For points lying on the negative $q_\omega$ axis, all the eigenvalues are real, leading to open hyperbolic vortex lines. Moving down the line as $q_\omega$ becomes more negative, the oblique eigenvectors of the two non-zero eigenvalues approach orthogonality. At the bottom-most point $(q_\omega=-0.5,r_\omega=0)$ $\boldsymbol{\chi}$ is symmetric and has orthogonal eigenvectors, leading to converging and diverging vortex lines perpendicular to each other. At the origin $(q_\omega=0,r_\omega=0)$ all eigenvalues of $\bm{\chi}$ are zero and it represents straight vortex lines. On the positive $q_\omega$ axis, $\boldsymbol{\chi}$ has two purely imaginary eigenvalues resulting in closed vortex lines that are planar elliptic in shape. At the topmost point ($q_\omega=0.5, r_\omega=0$), $\boldsymbol{\chi}$ is skew symmetric and the corresponding vortex lines are perfectly circular in shape.  
    
    \item \textit{Three-dimensional vortex lines (non-degenerate topologies):} 
    The interior of the $q_\omega$-$r_\omega$ plane represent all possible three-dimensional vortex line shapes that can be classified into four distinct topologies.
    The four regions of the $q_\omega$-$r_\omega$ plane demarcated by the $r_\omega=0$ and the discriminant $d_\omega=q_\omega^3+({27}/{4})r_\omega^2 = 0$ lines, represent these four topologies -- SFS, UFC, U-N/S/S and S-N/S/S, similar to the $Q_\omega$-$R_\omega$ plane (figure \ref{fig:QR_chong}).
    Note that inside each of the topology regions, the actual vortex line shapes differ from the canonical form given in figure \ref{fig:QR_chong} and vary depending upon the ($q_\omega$,$r_\omega$) value. For example, inside the
    UFC region of the $q_\omega$-$r_\omega$ plane (above discriminant line \& $r_\omega>0$), the axis of spiraling of the vortex line is in general oblique with respect to the direction of compression. 
    
    \item \textit{Three dimensional vortex lines (degenerate cases):} Specific shapes emerge at the boundaries of the $q_\omega$-$r_\omega$ plane. Such degenerate three-dimensional vortex line shapes are discussed below.
    \begin{enumerate}
        \item \textit{Left and right curved boundaries:} On the right boundary (blue line in the figure), vortex lines spiral out while being compressed along the axis of real eigenvector, perpendicular to the focal plane. The vortex line shape at this boundary is the same as the canonical shape for UFC topology given in the $Q_\omega$-$R_\omega$ plane. Similarly, on the left boundary (orange line in the figure) vortex lines spiral in while being stretched along the axis of real eigenvector resembling the canonical shape for SFS vortex line topology. 

        \item \textit{Bottom boundary:} Along the $q_\omega=-0.5$ line, the tensor $\chi_{ij}$ is symmetric. The vortex line shapes here resemble the canonical S-N/S/S or U-N/S/S shapes depending on the sign of $r_\omega$. Vortex lines corresponding to left half of the bottom boundary ($q_\omega=-0.5$, $r_\omega<0$) undergo compression along two orthogonal directions and an expansion in the third direction forming a tubular structure. Similarly, vortex lines corresponding to right half of the bottom boundary ($q_\omega=-0.5$, $r_\omega>0$) expand in two orthogonal directions and are compressed in the third direction resulting in disc like shapes. 
        
        \item \textit{Intersection of (a) and (b):} At the corners of the plane where the discriminant lines intersect with the boundary, i.e. at $q=-0.5, r=\pm 1/(3\sqrt{6})$, $\bm{\chi}$ has two equal eigenvalues resulting in a star node. The corresponding vortex line shapes are termed as ``axisymmetric vortex compression" at the left corner and ``axisymmetric vortex expansion" at the right corner.
    \end{enumerate}
    
\end{enumerate}
\section{Numerical simulation details}\label{sec:NS_details}

Vortex line geometry can provide novel insight into various flow processes. In this work, we focus on the characteristic features of local geometry in turbulent flows and flows exhibiting vortex-line reconnection.
We use direct numerical simulation (DNS) data to examine the local vortex line geometry for different flows:
\begin{enumerate}
    \item Forced homogeneous isotropic turbulence
    \item Breakdown of Taylor-Green vortex flow
    \item Vortex reconnection of anti-parallel vortices
    \item Vortex reconnection in orthogonally interacting tubes
\end{enumerate}
As mentioned in the introduction, these flows involve important vortical processes.

\subsection{Forced homogeneous isotropic turbulence}

DNS datasets of incompressible forced homogeneous isotropic turbulence from Turbulence and Advanced Computation lab at Texas A\&M University are employed. 
The simulations are performed in a periodic box of dimensions $2\pi\times2\pi\times2\pi$, with random forcing applied at large scales to maintain statistical stationarity. 
The datasets have been well validated and previously used to study intermittency \citep{donzis2008dissipation,donzis2010short}, anomalous scaling \citep{yakhot2017emergence,yakhot2018anomalous} and velocity gradient dynamics \citep{das2019reynolds}. The datasets used here span a Taylor Reynolds number range of $Re_\lambda \in (1,588)$. The Taylor Reynolds number is based on the Taylor microscale ($\lambda$)  and is given by
\begin{equation}
    \label{eq:tre}
    Re_\lambda=\frac{u'\lambda}{\nu}\quad;\quad\lambda=\sqrt{\frac{15\nu u'^2}{\epsilon}}
\end{equation}
where $u'$ is the rms velocity, $\nu$ is the kinematic velocity and $\epsilon$ is the mean dissipation rate. The details of all the datasets used, including the numerical resolution based on the maximum wavenumber resolved $\kappa_{max}$ and the Kolmogorov length scale $\eta$ are given in Table 1.
\begin{table}
 \begin{center}
  \begin{tabular}{lcccc}
    $Re_\lambda$  & Grid resolution   &   $\kappa_{max}\eta$ & Source  \\[3pt]
       $1$   & $256^3$ & $105.6$ & \cite{yakhot2017emergence}\\
       $6$   & $256^3$ & $34.8$ & \cite{yakhot2017emergence}\\
       $9$   & $256^3$ & $26.6$ & \cite{yakhot2017emergence}\\
       $14$   & $256^3$ & $19.87$ & \cite{yakhot2017emergence}\\
       $18$   & $256^3$ & $15.59$ & \cite{yakhot2017emergence}\\
       $25$   & $256^3$ & $11.51$ & \cite{yakhot2017emergence}\\
       $86$   & $256^3$ & $2.83$ & \cite{yakhot2017emergence}\\
       $225$   & $512^3$ & $1.34$ & \cite{donzis2008dissipation}\\
       $385$   & $1024^3$ & $1.41$ & \cite{donzis2008dissipation}\\
       $588$   & $2048^3$ & $1.39$ & \cite{donzis2008dissipation}\\
  \end{tabular}
  \caption{Details of forced isotropic turbulence data sets}{\label{tab:res}}
 \end{center}
\end{table}

\subsection{Taylor-Green vortex flow}\label{sec:TGV}

Direct simulations of the time evolution of incompressible Taylor-Green vortex flow are performed in a periodic box of dimension $2\pi$, starting from the initial field given by:   
\begin{equation}
    \label{eq:tg}
    \begin{split}
    u=U_0\sin{x}\cos{y}\cos{z} \\
    v=-U_0\cos{x}\sin{y}\cos{z} \\
    w=0
    \end{split}
\end{equation}
The pressure field is initialized as follows:
\begin{equation}
    \label{eq:tg_pressure}
    p=p_0+\frac{\rho_0U_0^2}{16}\left[\left(\cos{\frac{2x}{L}}+\cos{\frac{2y}{L}}\right)\left(\cos{\frac{2z}{L}}+2\right)\right]
\end{equation}
where $\rho_0=1$, $L=1$. 
Following \cite{chapelier2012inviscid} and \cite{bull2015simulation}, the Reynolds number ($Re$) is chosen to be
\begin{equation}
    Re=\frac{U_0L}{\nu}=1600
\end{equation}

The simulations are performed using a finite volume solver based on Gas kinetic methods (GKM) given by \cite{xu1998gas}. Instead of solving the Navier-Stokes equation, GKM solves the modelled Boltzmann equation for the single particle distribution function $f$. The solver employs a first order Bhatnagar-Gross-Krook  (BGK) model for the collision terms in the Boltzmann equation. Subsequently, the distribution function $f$ is then used to compute the fluxes for the conservative variables. The solver has been well validated for a variety of compressible flows: wall bounded flows \citep{xie2014instability,mittal2020nonlinear}, decaying and homogeneous shear turbulence \citep{kumar2013weno,kumar2014stabilizing} and mixing layers with Kelvin-Helmholtz instability \citep{karimi2016suppression,karimi2017influence}. Although, GKM is well suited for non-equilibrium and rarefied effects, it is equally applicable in the context of incompressible continuum regime. {We provide validation for the applicability of the solver in incompressible flows in the following sub-section.} 

\subsubsection{Numerical Validation}\label{sec:num_val}
 \begin{figure}
    \centering
    \subfloat[]{\includegraphics[width=0.45\textwidth, height=0.225\textheight, keepaspectratio]{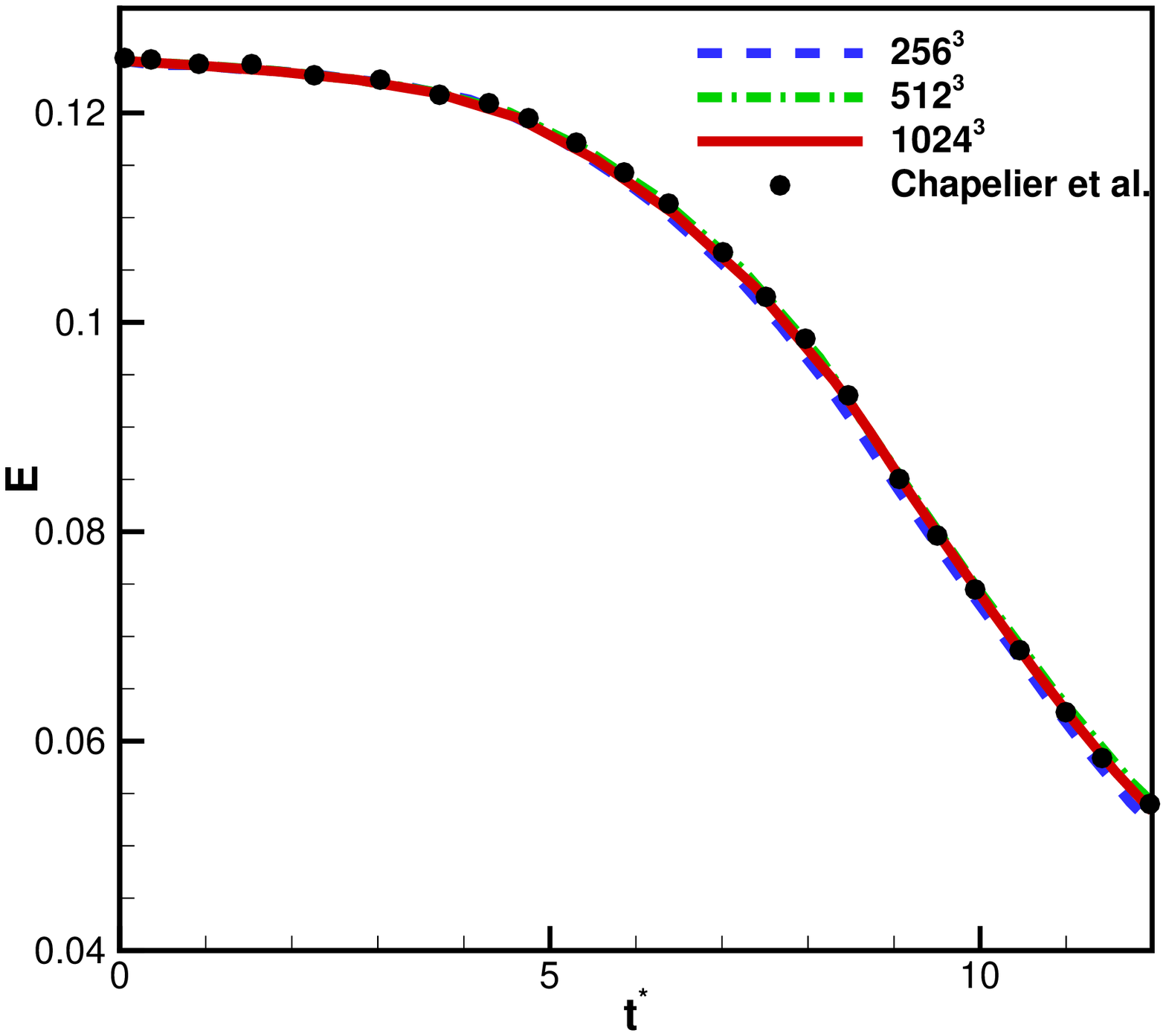}}
    \subfloat[]{\includegraphics[width=0.45\textwidth, height=0.225\textheight, keepaspectratio]{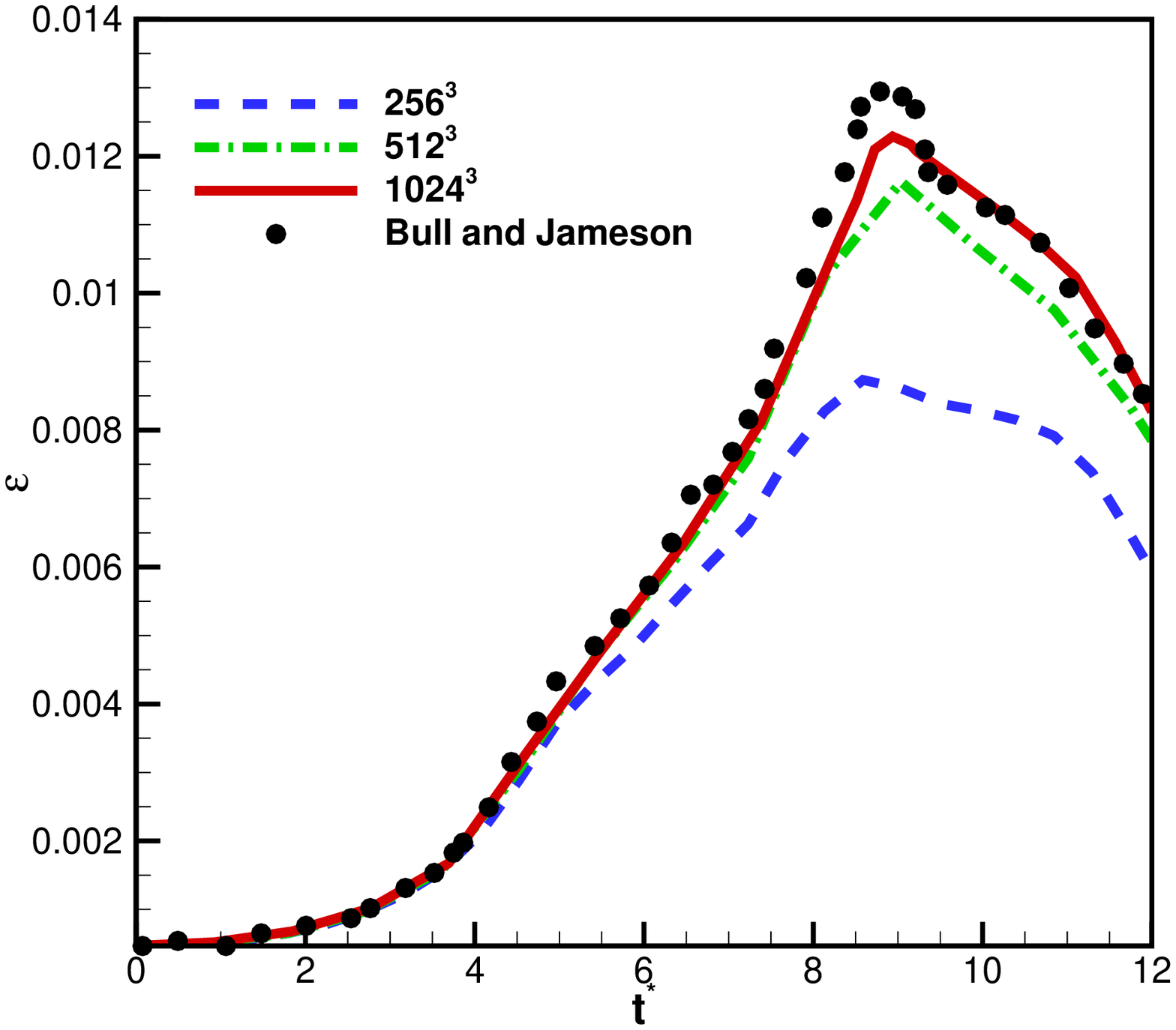}}
    \caption{Time evolution of (a) normalized kinetic energy ($E/U_0^2$) and (b) normalized mean dissipation rate ($\epsilon/(U_0^3/L)$) for a Taylor Green vortex}
    \label{fig:val_TG}
\end{figure}

We simulate the Taylor-Green vortex flow on three sets of grid with $256^3$, $512^3$ and $1024^3$ points. The evolution of turbulent kinetic energy with normalized time, $t^*$, is shown in figure \ref{fig:val_TG}(a). The turbulent kinetic energy ($E$) is normalized by $U_0^2$ and $t^*$ is defined as:
\begin{equation}
    t^*=\frac{tU_0}{L}
\end{equation}
The results for kinetic energy decay agree very well with the results of \cite{chapelier2012inviscid} for all three grids. Additionally, figure \ref{fig:val_TG}(b) plots the evolution of volume averaged dissipation rate $\epsilon=2\nu\langle S_{ij}S_{ij}\rangle$ normalized by $(U_0^3/L)$, where $S_{ij}$ is the strain-rate tensor. The current results are compared against those obtained from a high-order flux reconstruction based method by \cite{bull2015simulation}. The initial growth of dissipation (upto $t^*=5$) on the $256^3$ grid agrees well with the reference solution, however, there is significant undershoot in the peak value. 
As the grid resolution is improved ($512^3$ and $1024^3$ grids), a much better agreement with the reference solution is observed. 
\begin{figure}
    \centering
    \includegraphics[width=0.5\textwidth, keepaspectratio]{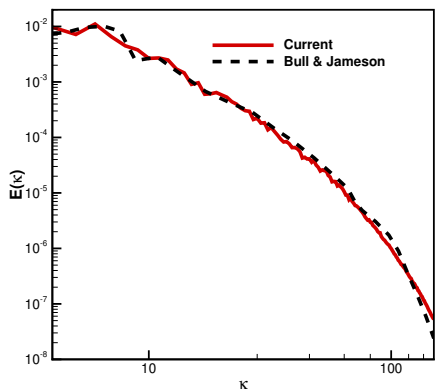}
    \caption{Kinetic Energy spectrum just after peak dissipation at $t^*=9$}
    \label{fig:spectrum}
\end{figure}

Figure \ref{fig:spectrum} shows the kinetic energy spectrum just after the dissipation peaks at $t^*=9$. Turbulence at this stage is well developed upto the smallest dissipative scales. The spectrum as obtained from \cite{bull2015simulation} is also plotted here for comparison and we observe good agreement between the two data sets. Overall, the results for kinetic energy spectrum and dissipation rate evolution on the $1024^3$ grid agree very well with the benchmark data from literature.   
The flow field from the $1024^3$ grid are used for further analysis in this paper.

\subsection{Vortex reconnection of anti-parallel vortices}

\begin{figure}
    \centering
    \includegraphics[width=0.5\textwidth]{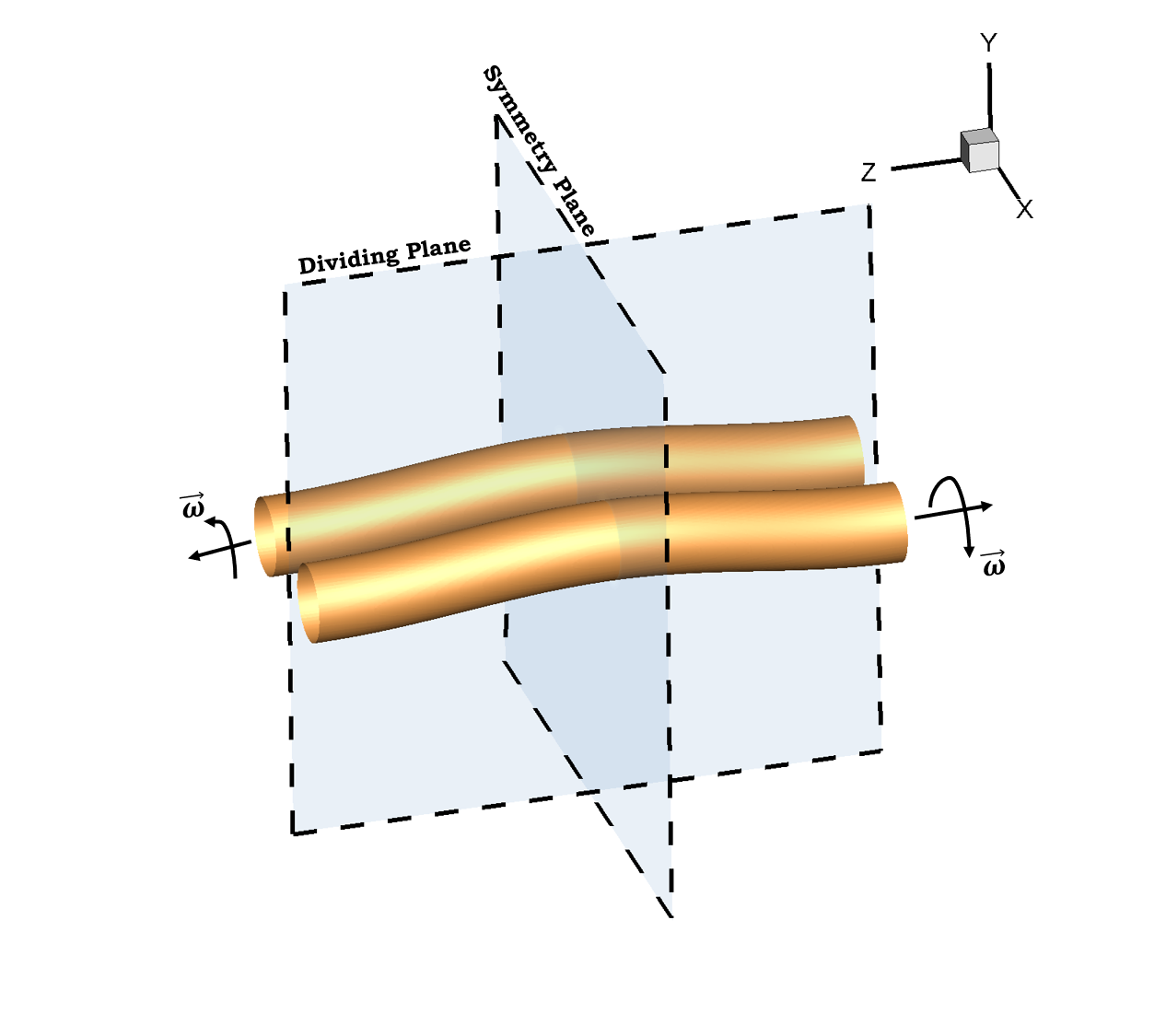}
    \caption{Schematic of initial configuration for interaction of anti-parallel vortices}
    \label{fig:schematic_parallel}
\end{figure}

We simulate the interaction of two perturbed {anti-parallel} vortex tubes (figure \ref{fig:schematic_parallel}) in a periodic box of dimension $2\pi$ using initial conditions as outlined in \cite{melander1988cut}. The core of the vortex tubes are specified by the following parametric curve:
\begin{equation}
\label{eq:vcore}
\begin{split}
     x=x_c+p\cos{\alpha}\cos{t}\\
    y=y_c+p\sin{\alpha}\cos{t}\\
    z=t\\    
\end{split}
\end{equation}
Here, $(x_c,y_c)$ is the centroid of the unperturbed tube and is specified as $(\pm0.81,0)$, $\alpha=\pi/3$ is the inclination angle and $p=0.2$ is the perturbation amplitude. To ensure vorticity is zero outside the tubes, a compact Gaussian function \citep{melander1988cut} is used for vorticity distribution within the tube's cross section of radius $r_c=0.666$.
\begin{equation}
\label{eq:compact}
    \omega(r) = \left\{
        \begin{array}{ll}
            \omega_0(1-f(r/r_c)) & \quad r \leq r_c \\
            0 & \quad r > r_c
        \end{array}
    \right.
\end{equation}
where, $f(\eta)=\exp{(-K\eta^{-1}\exp{(1/\eta-1)})}$, $K=1/2\exp(2)\log(2)$ and $\omega_0=20$. Vorticity at every point in the cross section is tangential to the parametric curve describing the vortex core \eqref{eq:vcore}. This is done to ensure that circulation ($\Gamma$) is {conserved along the vortex tube.} The vorticity and velocity fields are related by the following equation,
    \begin{equation}
        \label{eq:vel_invert}
        \nabla^2\vec{v}=-\nabla\times\vec{\omega}
    \end{equation}
This Poisson equation \eqref{eq:vel_invert} is solved to generate a solenoidal velocity field to initialize the present simulations. The ensuing vorticity field is divergence free and approximately compactly supported in the tubes. The Reynolds number based on the circulation $\Gamma$ is set to: 
\begin{equation}
    Re=\frac{\Gamma}{\nu} = 3000
\end{equation}
The solver outlined in section \ref{sec:TGV} is used for simulating the flow on a uniform grid with $256$ points in each direction. This resolution was found to be reasonable for the present problem.

\subsection{Vortex reconnection in orthogonally interacting tubes}

\begin{figure}
    \centering
    \includegraphics[width=0.5\textwidth]{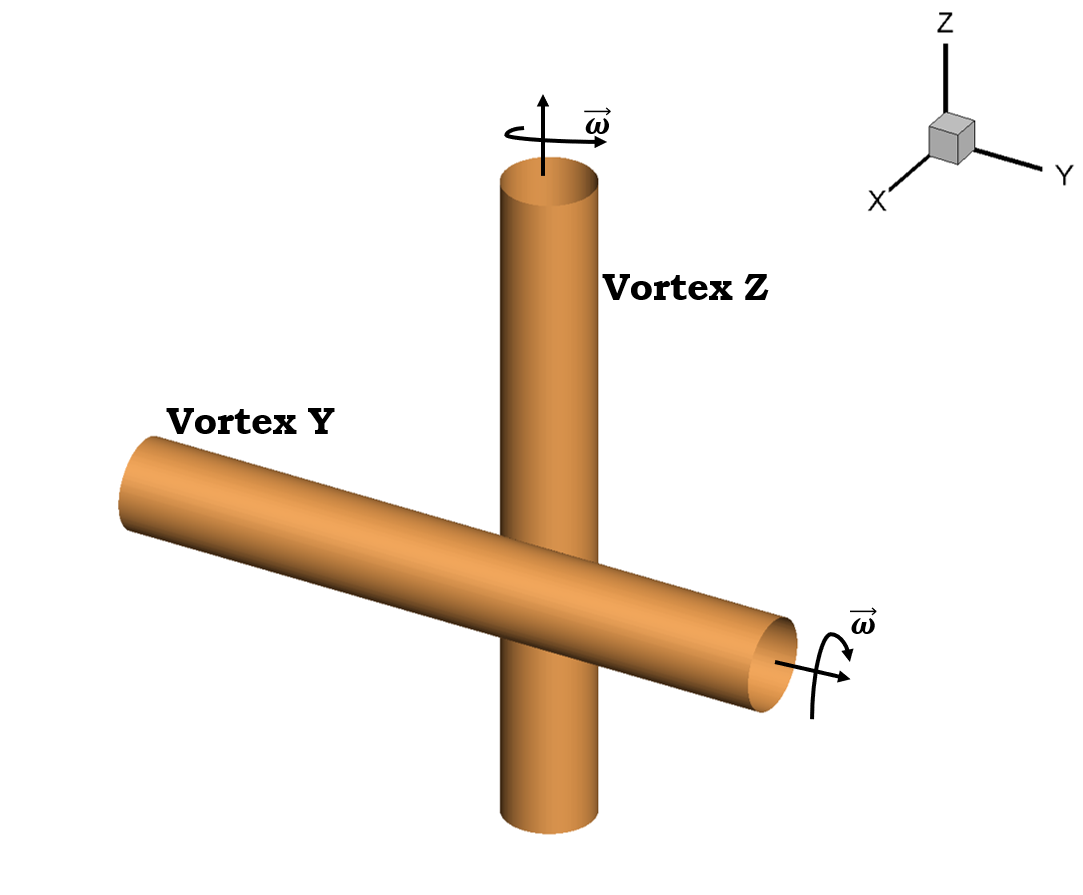}
    \caption{Initial configuration for interaction of orthogonally offset tubes}
    \label{fig:schematic_orthogonal}
\end{figure}

We also simulate interaction of two orthogonally offset vortex tubes in a periodic box of dimension $2\pi$ \citep{boratav1992reconnection}. The initial configuration is shown in figure \ref{fig:schematic_orthogonal}. We specify vorticity along the axes of the tubes, specifically vorticity in ``Vortex Y" is along $-\hat{y}$ axis and vorticity in ``Vortex Z" is along $+\hat{z}$ axis. The compact Gaussian function, described previously in \eqref{eq:compact}, distributes vorticity in the tube's cross-section. This ensures vorticity is non-zero only inside the tube of radius $r_c=0.666$. Both tubes have an initial circulation of $\Gamma=7.665$, and the Reynolds number based on circulation is set to $Re=1400$. As discussed previously, the velocity field is initialized by solving the Poisson equation \eqref{eq:vel_invert}.

\section{Local vortex line shapes in turbulent flows}

The probability distribution of local vortex line shapes in turbulent flow fields generated from (i) randomly initialized isotropic field with large-scale forcing and, (ii) Taylor-Green vortex field without any external forcing, are investigated in detail in this section. The vortex-line shapes are analyzed in the framework of normalized vorticity gradient tensor invariants ($q_\omega$,$r_\omega$). 
A comparison is drawn between the probability distributions of vortex line shapes in the two different turbulent flows. 

\subsection{Forced isotropic turbulence}

\begin{figure}
\centering
    \subfloat[]{\includegraphics[width=0.32\textwidth, keepaspectratio]{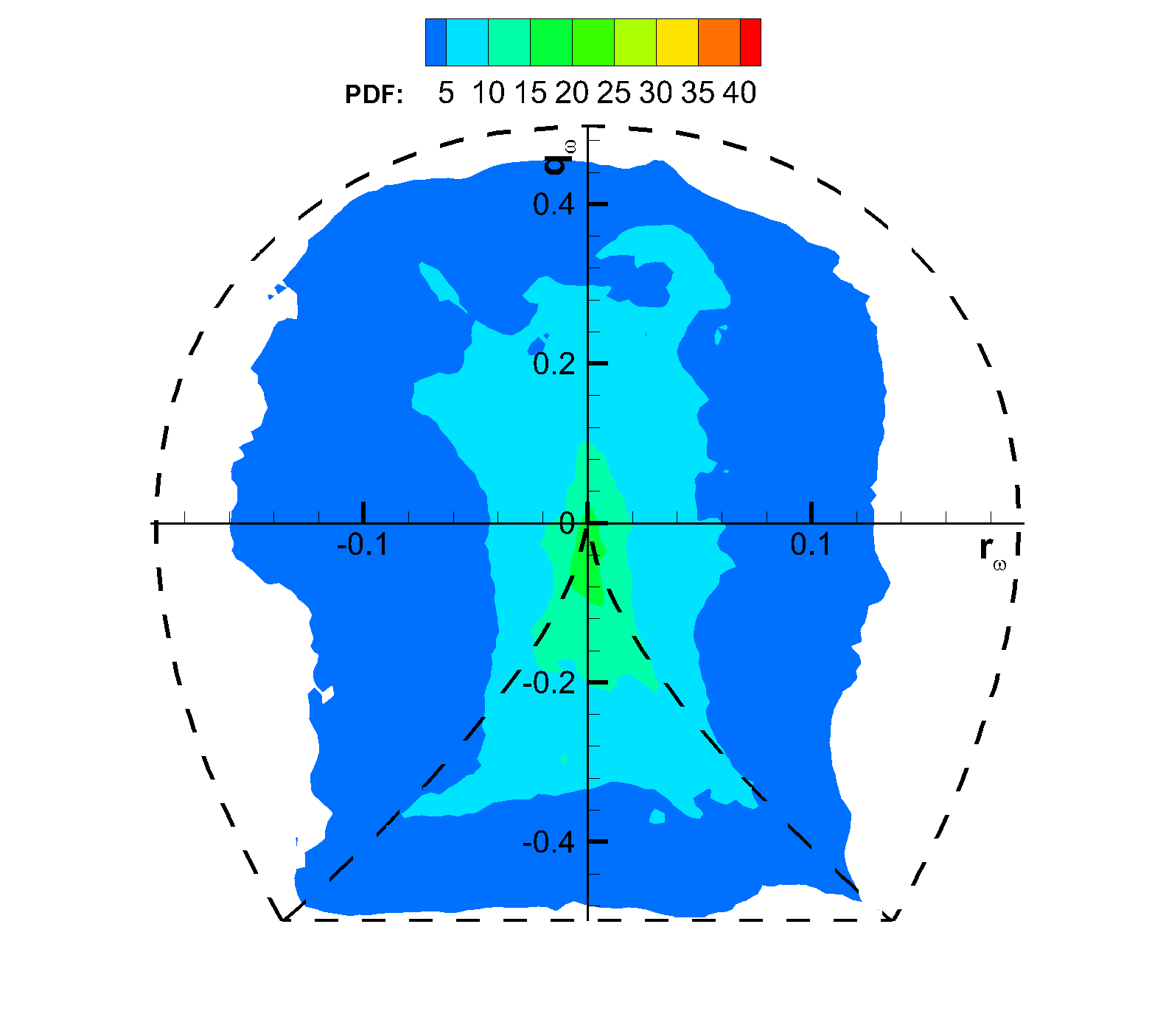}} 
    \subfloat[]{\includegraphics[width=0.32\textwidth, keepaspectratio]{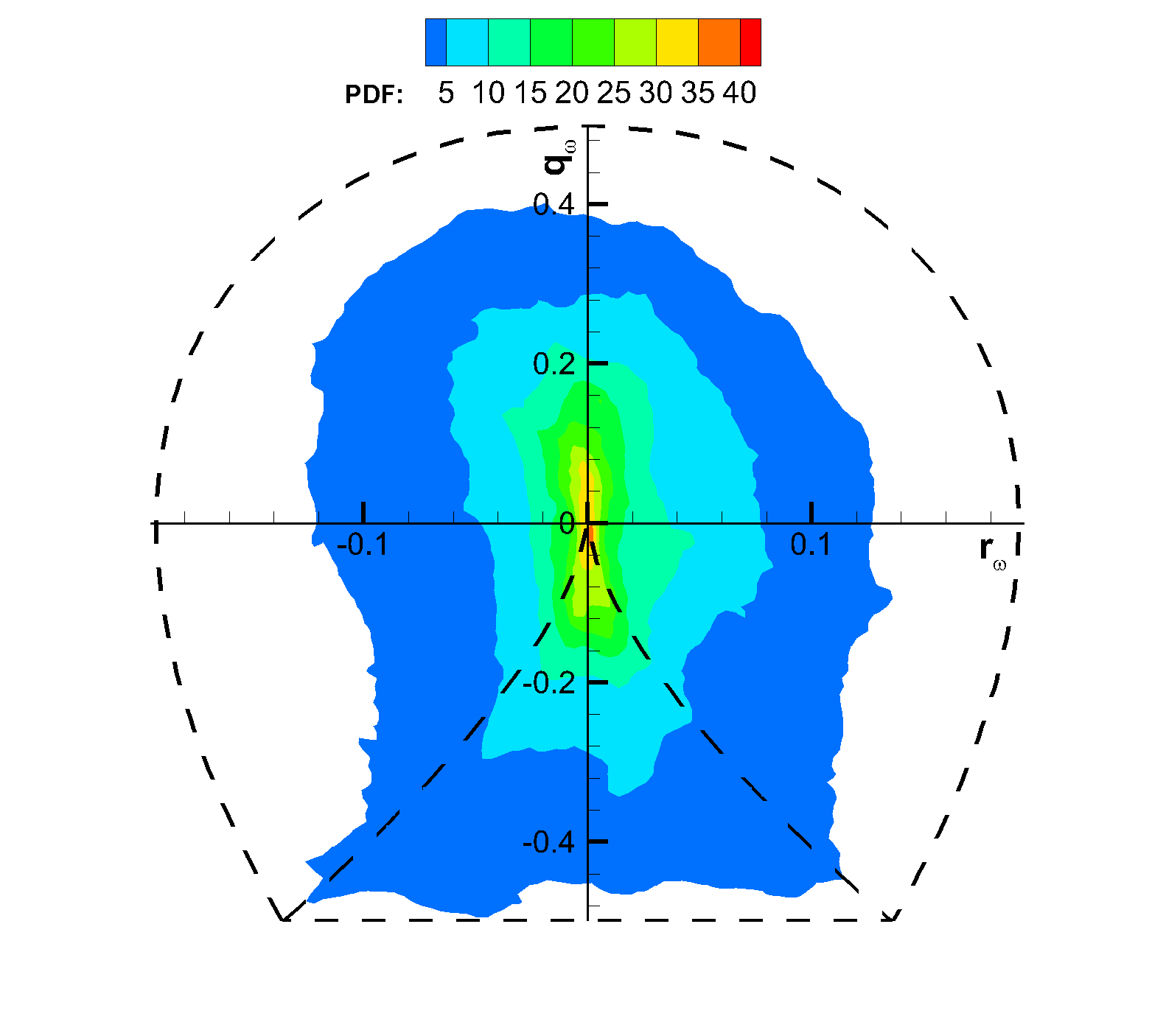}}
    \subfloat[]{\includegraphics[width=0.32\textwidth, keepaspectratio]{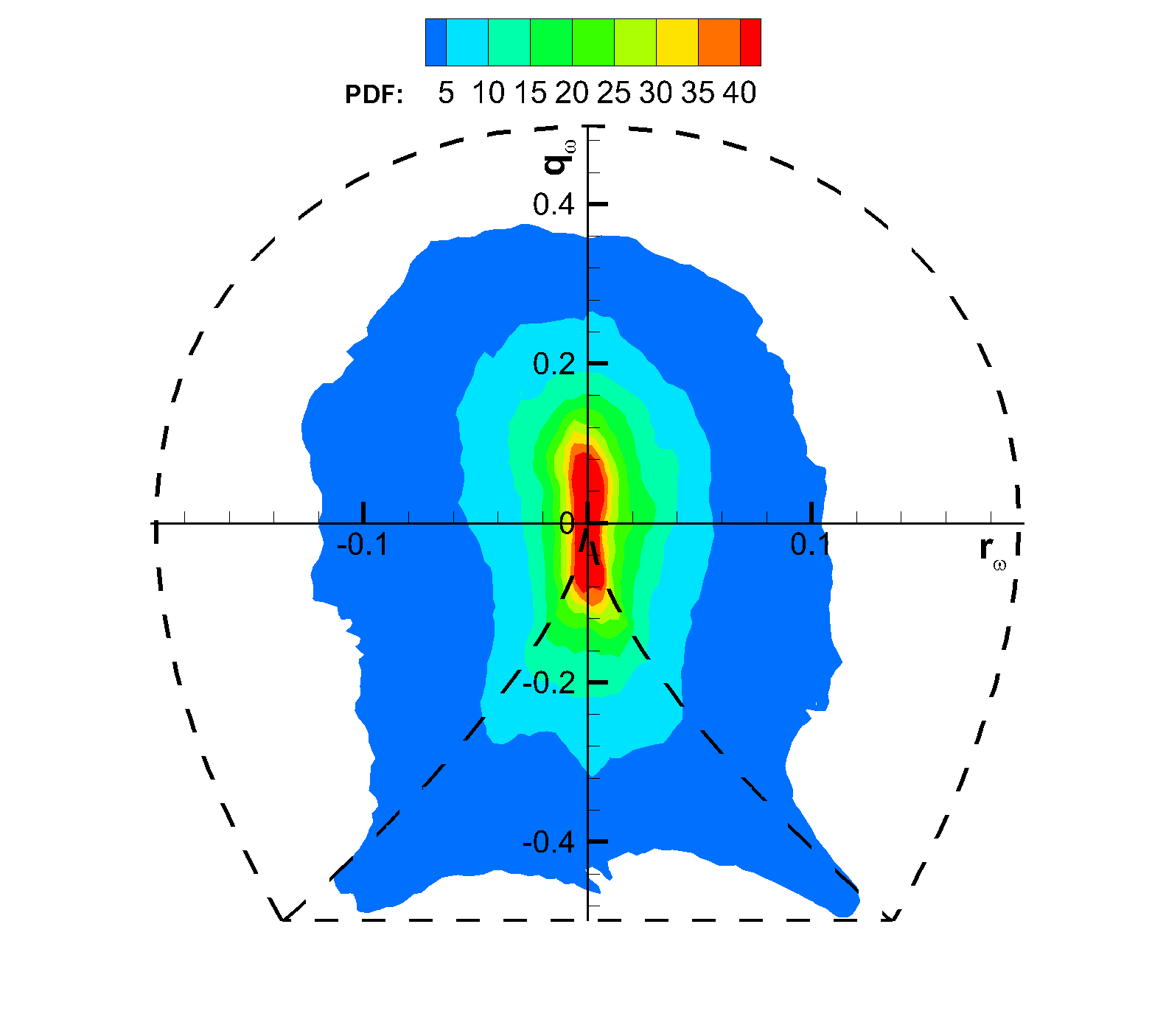}} \\
    \subfloat[]{\includegraphics[width=0.32\textwidth, keepaspectratio]{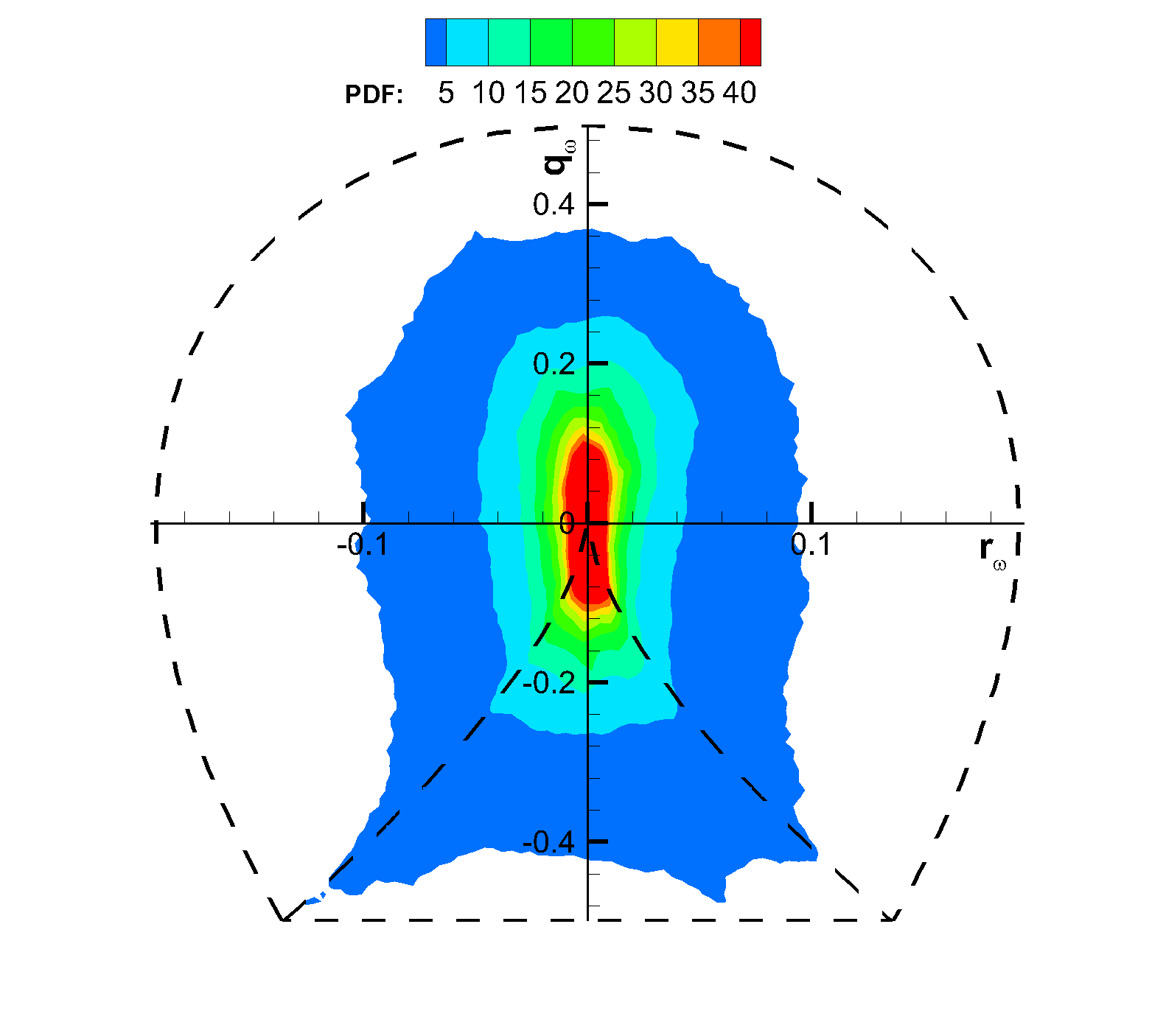}}
    \subfloat[]{\includegraphics[width=0.32\textwidth, keepaspectratio]{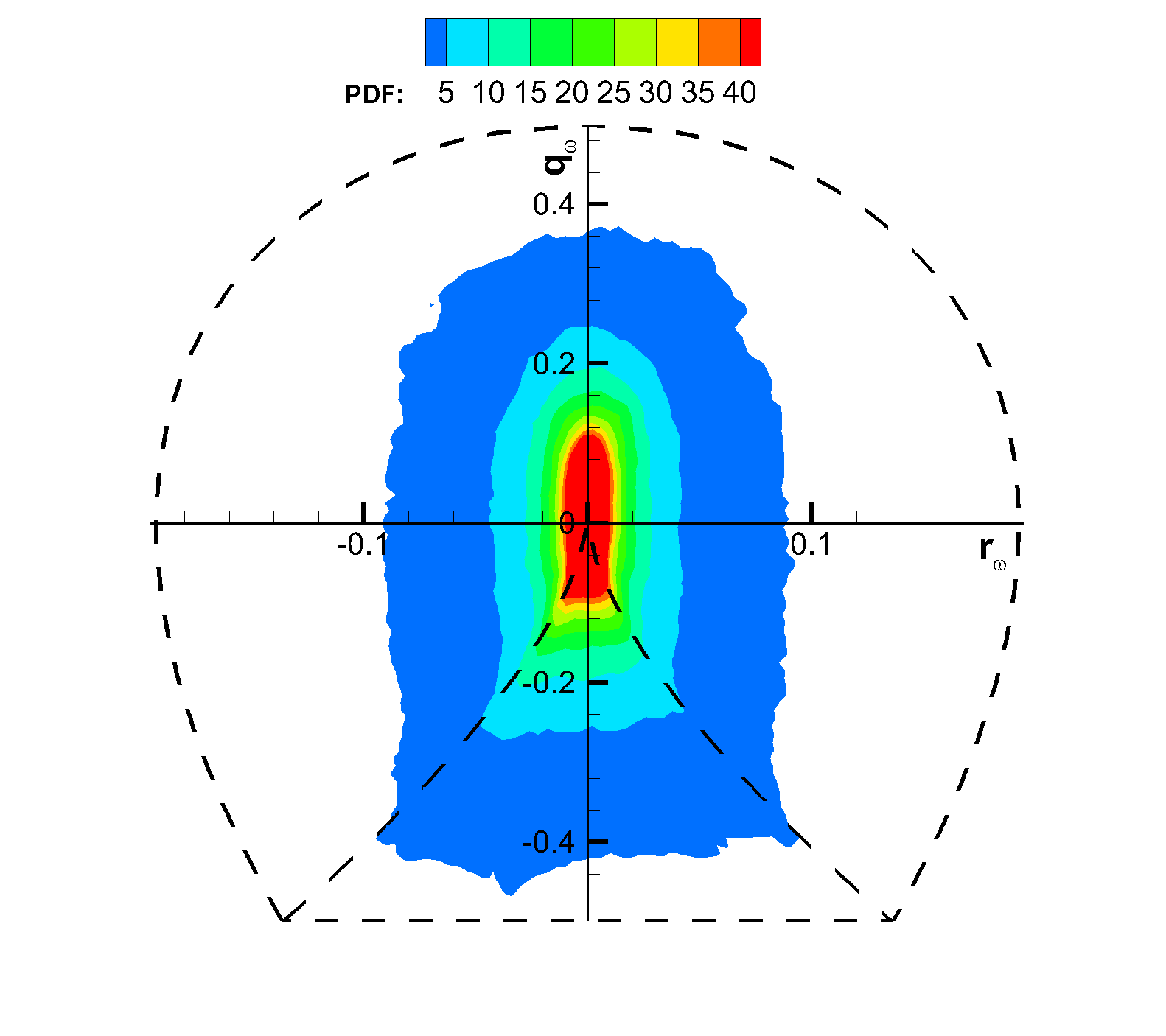}}
    \subfloat[]{\includegraphics[width=0.32\textwidth, keepaspectratio]{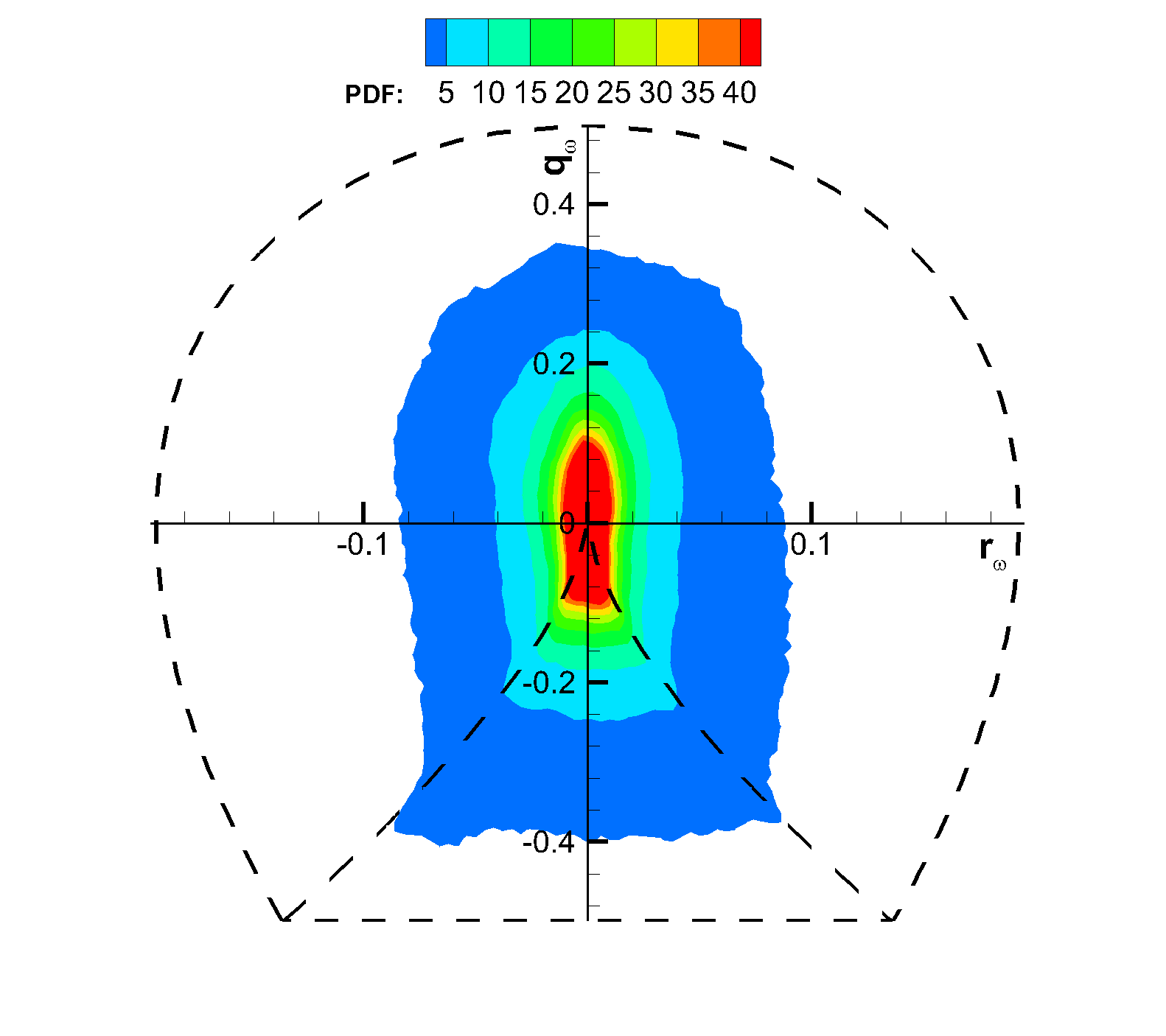}} \\
    \subfloat[]{\includegraphics[width=0.32\textwidth, keepaspectratio]{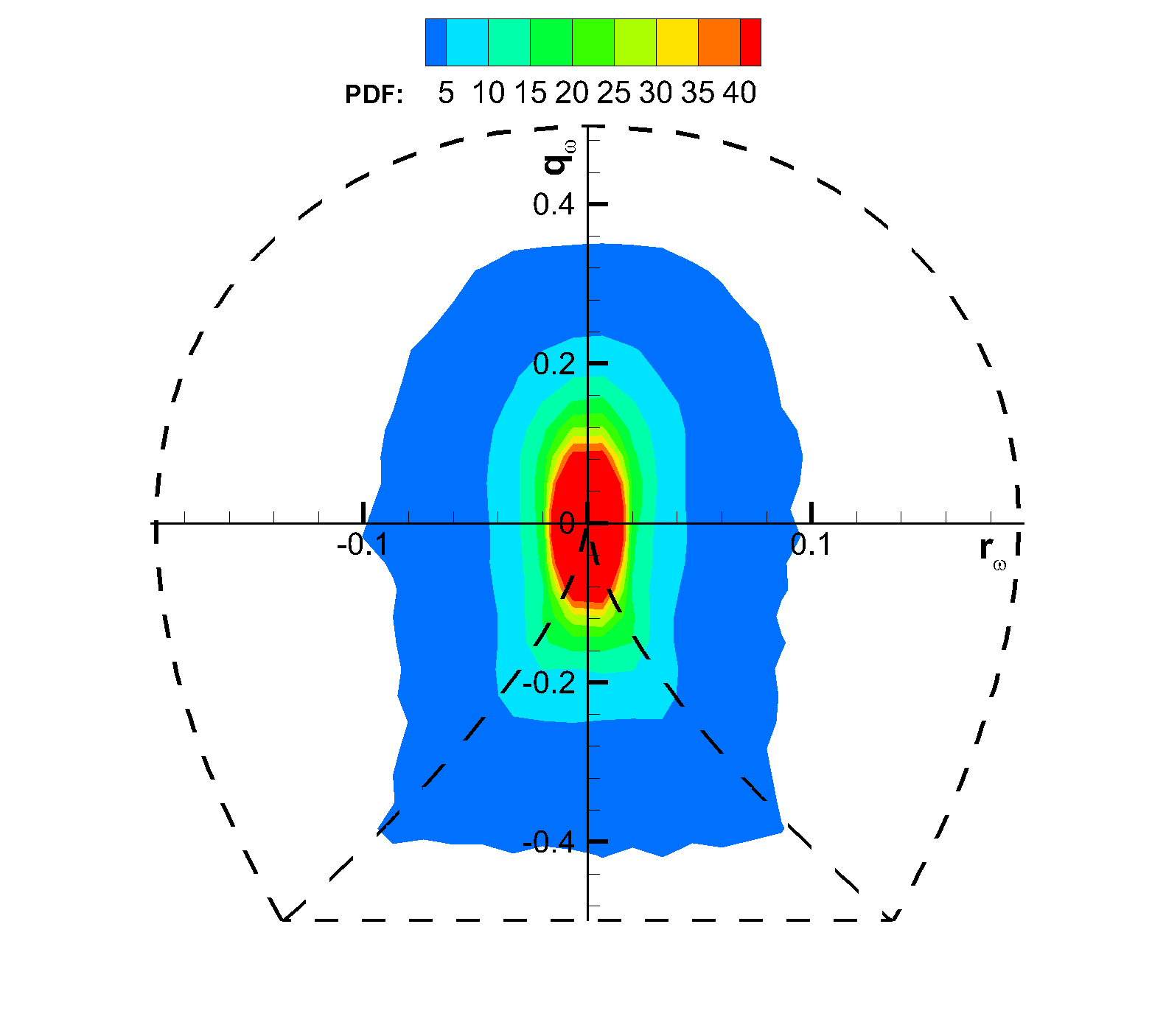}}
    \subfloat[]{\includegraphics[width=0.32\textwidth, keepaspectratio]{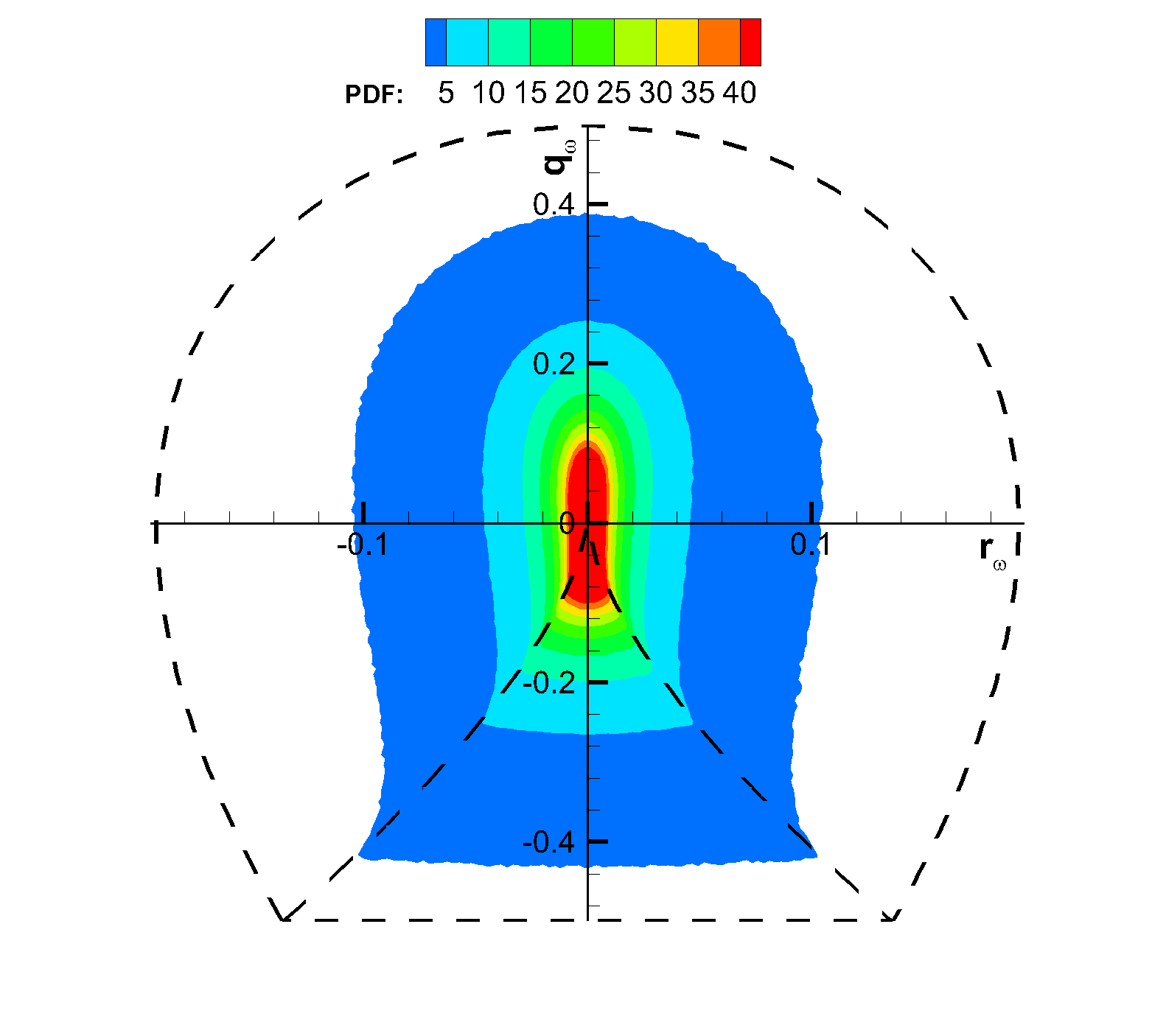}}
    \subfloat[]{\includegraphics[width=0.32\textwidth, keepaspectratio]{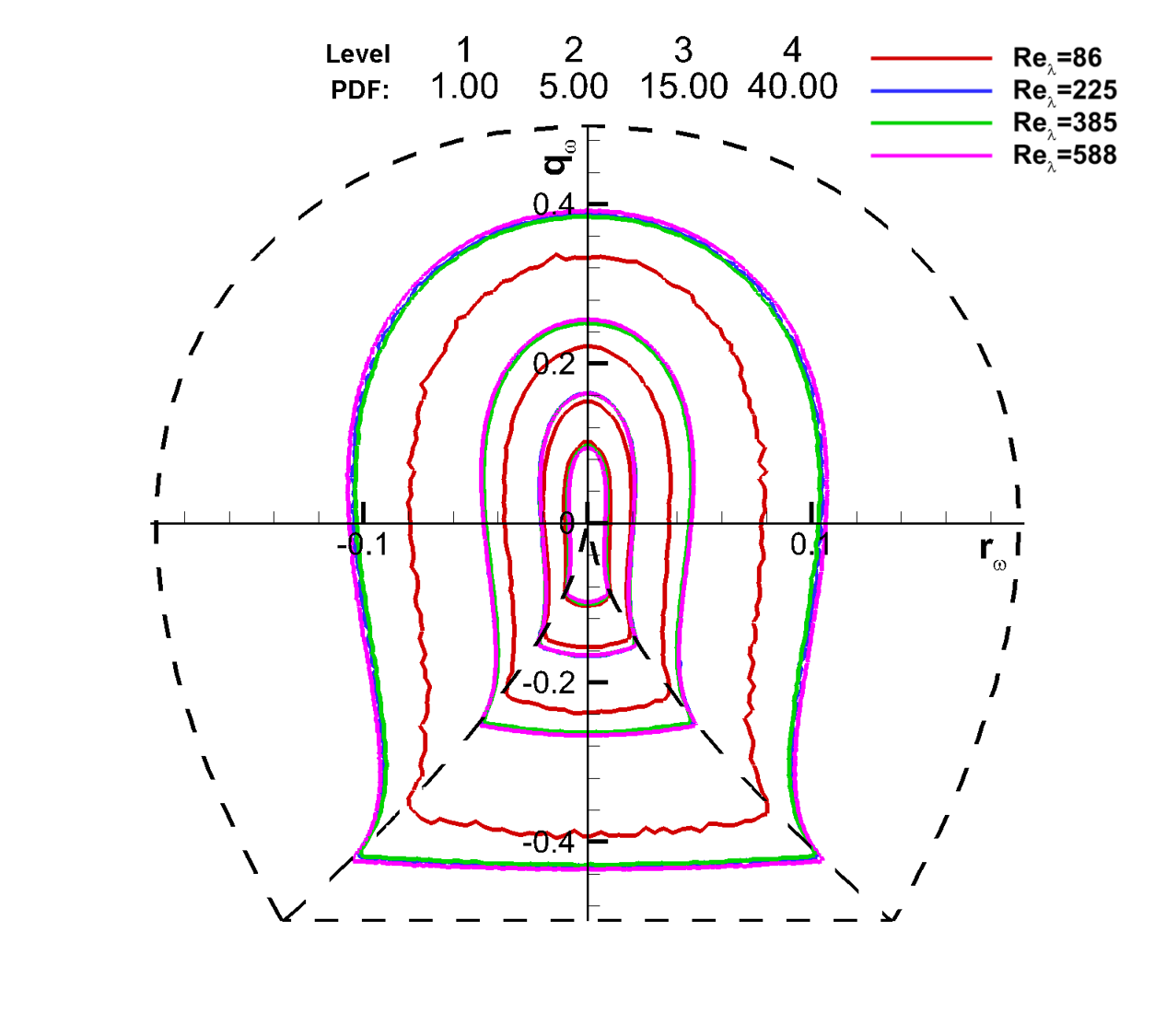}} 
    \caption{$q_\omega-r_\omega$ joint PDF filled contour plots for $Re_\lambda$ = (a) $1$, (b) $6$, (c) $9$, (d) $14$, (e) $18$,  (f) $25$, (g) $86$ and (h) $225$ (i) line contour plots for $Re_\lambda=86-588$ }
    \label{fig:jpdf}
\end{figure}


The joint probability density functions (PDFs) of $q_\omega$-$r_\omega$ in forced isotropic turbulent flows of different $Re_\lambda$ are plotted in figure \ref{fig:jpdf}. 
The dashed lines mark the realizable region of the $q_\omega$-$r_\omega$ plane. 
In all the $Re_\lambda$ cases, the joint PDF is fairly symmetric in $r_\omega$ and the symmetry is more pronounced at higher Reynolds numbers.
This symmetry indicates that vortex lines in a turbulent flow field are equally likely to be stable (converging towards a center or a node) as unstable (diverging from a center or a node). 
It is well-known that the joint PDF of velocity gradient tensor invariants ($q$-$r$) has a characteristic teardrop shape with maximum probability of occurrence along the right discriminant line of the plane \citep{das2020characterization}.
In contrast, the joint PDF of $q_\omega$-$r_\omega$ shows that the highest probability of occurrence is at and around the origin of the plane, which represents straight parallel vortex lines. 
At $Re_\lambda=1$, the PDF resembles that of a Gaussian field reflecting the random forcing of the flow field.
As $Re_\lambda$ increases from $1$ to $25$ the region close to origin becomes progressively more densely populated, indicating an increase in probability of straight vortex lines in the flow. 
At $Re_\lambda=25$, the joint PDF approaches its characteristic shape. The characteristic form is symmetric in $r_\omega$ and resembles a ``bell-like" shape. 
In the next range of Reynolds numbers, i.e. for $Re_\lambda \in (25,225)$, the joint PDF contours undergo finer refinements of this shape. 
At $Re_\lambda = 225$, the joint PDF attains a self-similar shape and is invariant above this Reynolds number. This is demonstrated by superposing line contours in figure \ref{fig:jpdf}(i). It is evident that the joint PDFs of $Re_\lambda =225,385$ and $588$ are nearly identical.
This is similar to the findings of \cite{das2019reynolds} for the $q$-$r$ joint PDF, which also asymptotes to a self-similar shape at the same $Re_\lambda (= 225)$.
It is further evident from figure \ref{fig:jpdf} that from $Re_\lambda = 1$ to $86$ the joint PDF shrinks closer to the origin, while from $Re_\lambda = 86$ to $225$ the PDF expands away from the origin before it achieves a characteristic invariant distribution.
The characteristic PDF at $Re_\lambda \geq 225$ has the highest density near the origin and the densities decrease as we move away from the origin. This indicates a clear preference of turbulence to attain local vortex line shapes that are straight. 
The PDF indicates that focal topologies occupy a significantly larger area in the invariants plane than the non-focal topologies.

\begin{figure}
    \centering
    \includegraphics[width=0.5\textwidth, keepaspectratio]{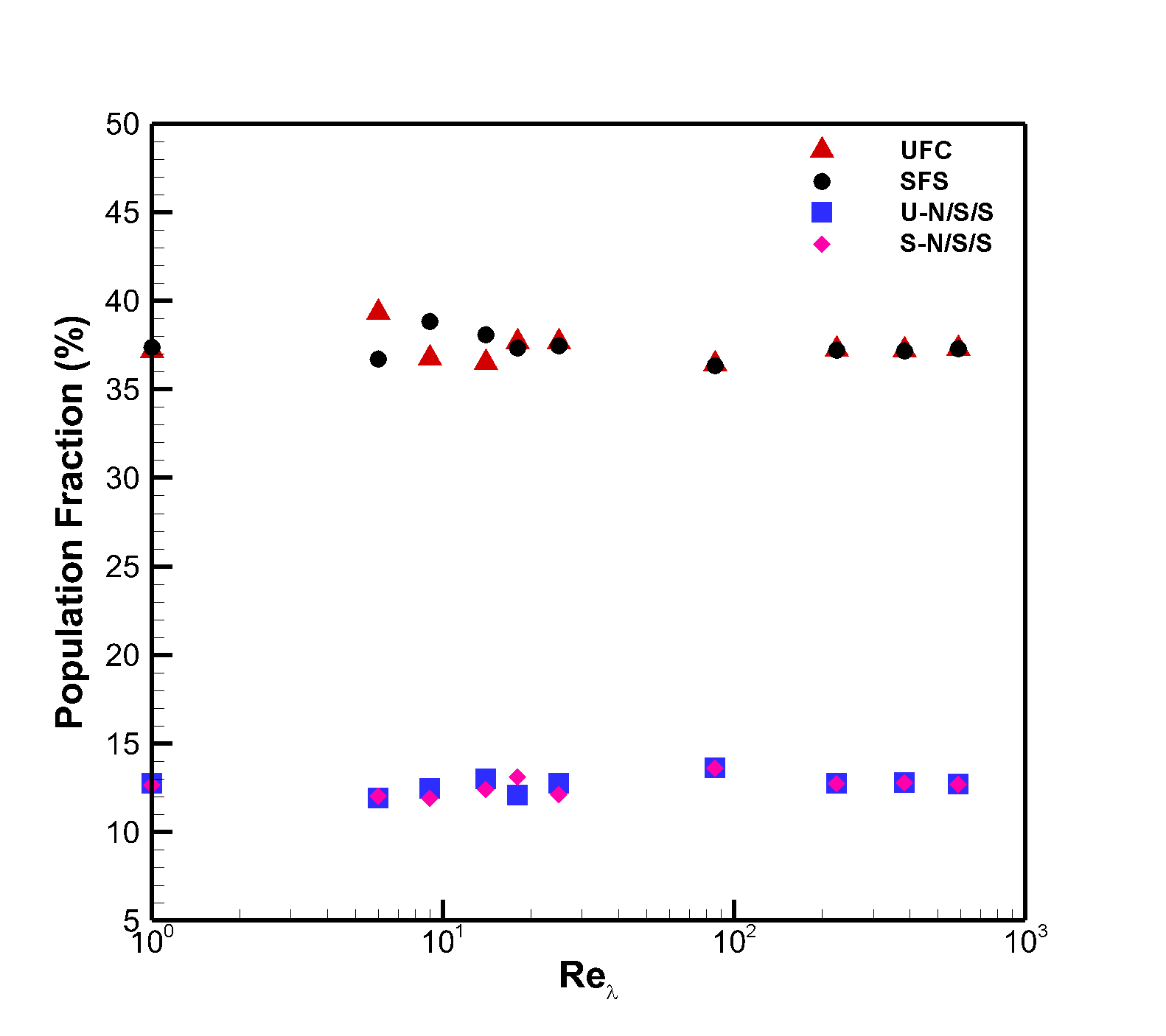}
    \caption{Population fraction of non-degenerate vortex line topologies for Forced isotropic turbulence at different $Re_\lambda$}
    \label{fig:pop_fit}
\end{figure}

The percentage of points in the turbulent flow field belonging to the four different vortex line topologies are plotted as a function of $Re_\lambda$ in figure \ref{fig:pop_fit}. 
The vortex line topology percentages do not show a strong dependence on Reynolds number. There is a noticeable variation in the fractions of SFS and UFC topologies only for $Re_\lambda \leq 25$. However, the sum total of the two focal topologies (SFS and UFC) and that of the two non-focal topologies (SN/S/S and UN/S/S) remain nearly constant at all Reynolds numbers.  
As inferred from the $q_\omega$-$r_\omega$ joint PDF, the focal topologies (SFS and UFC) indeed dominate over the non-focal topologies (SN/S/S and UN/S/S).
The focal vortex lines occupy about $75 \%$ of the flow field, while only $25 \%$ of the field is constituted by non-focal vortex lines.
The symmetry of the probability distribution with respect to $r_\omega$ is further evident in figure \ref{fig:pop_fit}, particularly at high Reynolds numbers ($Re_\lambda \geq 86$), as the population fractions of stable and unstable topologies obtained are exactly equal.

Overall, the local vortex line shape exhibits a characteristic bell shape that is invariant at sufficiently high Reynolds numbers. 
Vortex lines in a turbulent flow field are equally likely to be stable or unstable. However, turbulence exhibits a strong preference for focal vortex lines over non-focal vortex lines.

\subsection{Taylor-Green Vortex Breakdown}


\begin{figure}
    \centering
     \subfloat[]{\includegraphics[width=0.60\textwidth, height=0.4\textheight,keepaspectratio]{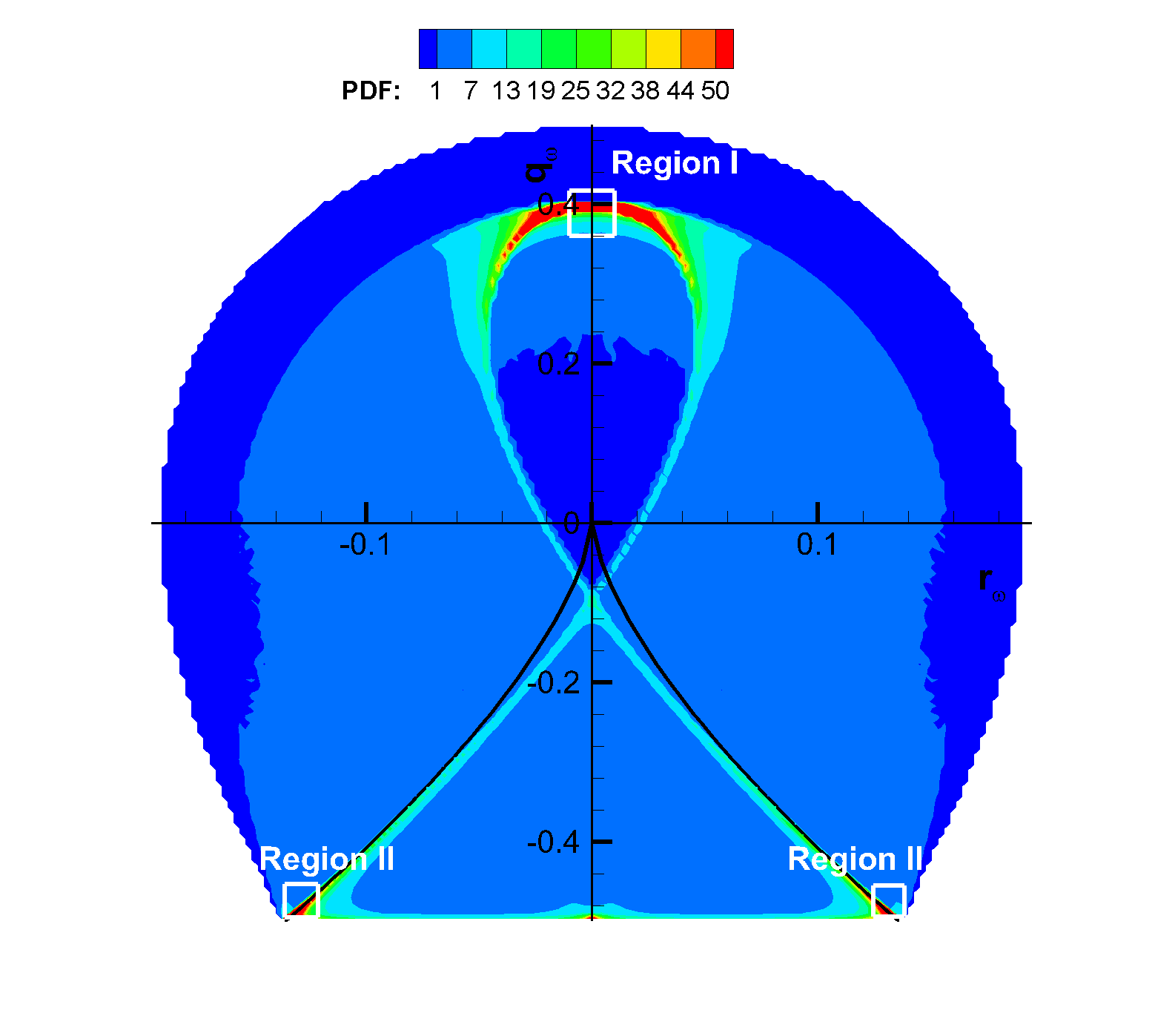}} \\
      \subfloat[]{\includegraphics[width=0.45\textwidth, height=0.4\textheight,keepaspectratio]{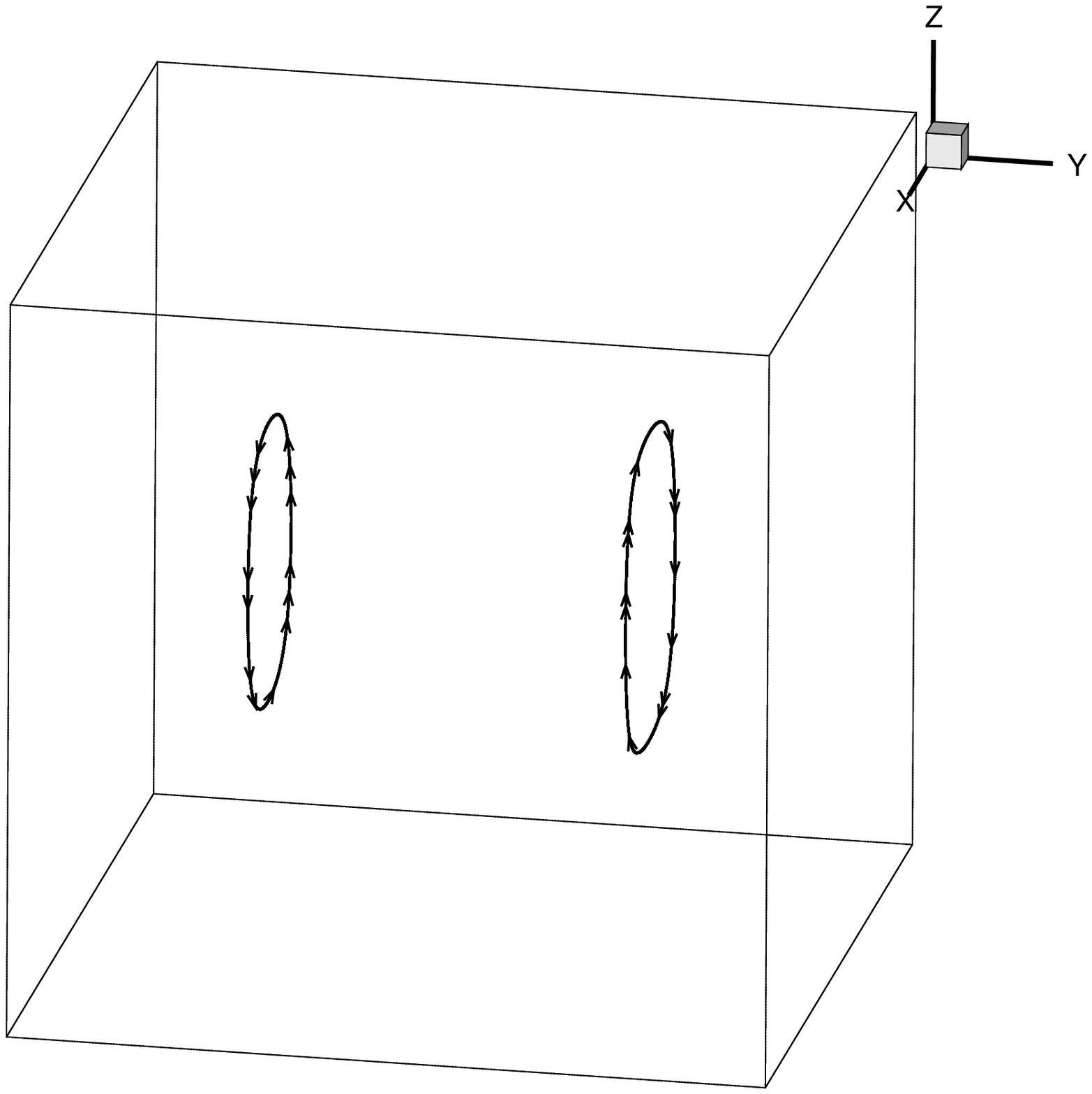}}
       \subfloat[]{\includegraphics[width=0.45\textwidth, height=0.4\textheight,keepaspectratio]{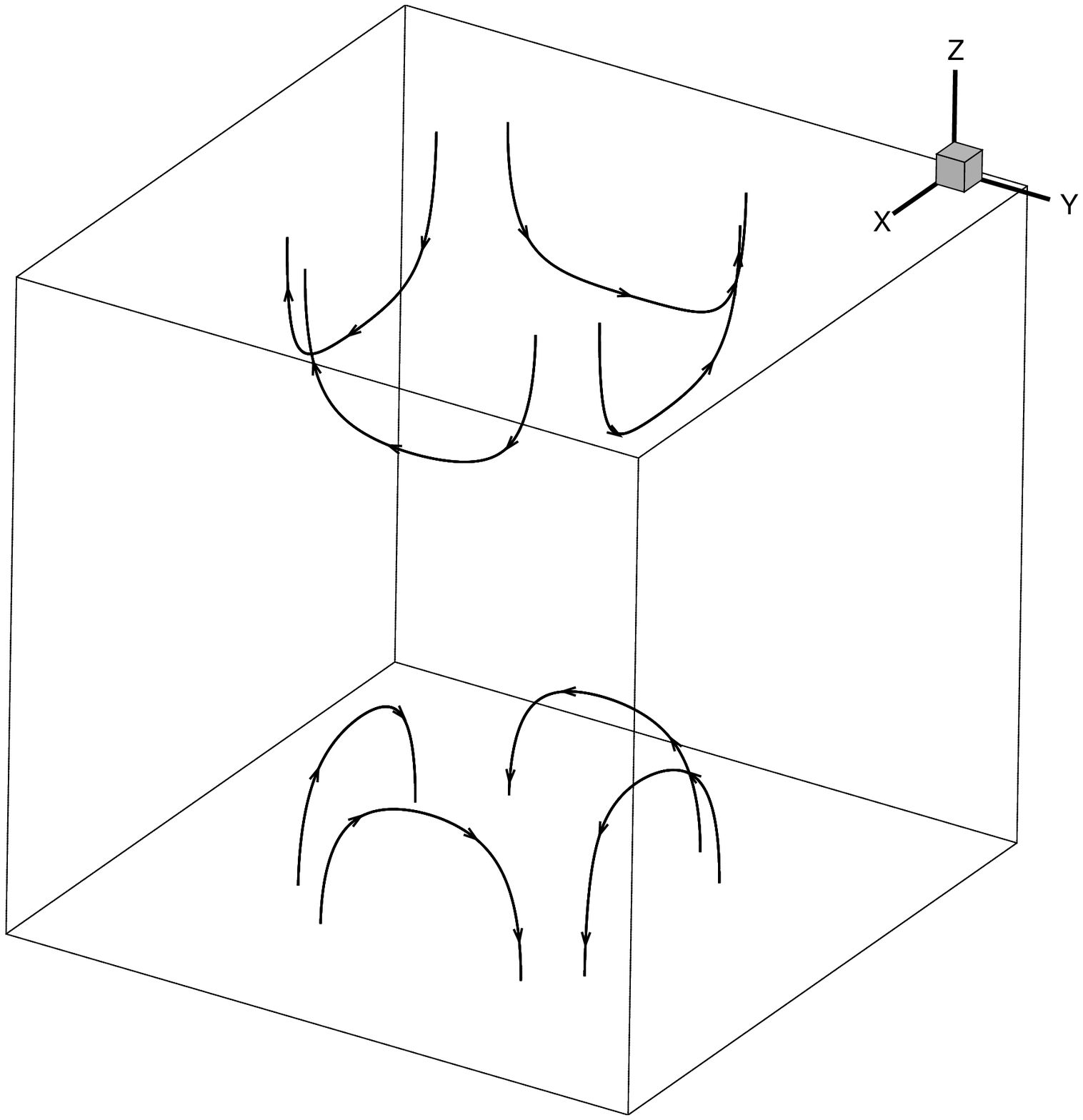}}
    \caption{(a) Filled contour plot of $q_\omega-r_\omega$ joint PDF for initial Taylor-Green field; vortex lines in physical space dominated by (b) region \rom{1} and (c) region \rom{2} of the $q_\omega-r_\omega$ joint PDF}
    \label{fig:vline_TG}
\end{figure}

The joint PDF distribution of $q_\omega-r_\omega$ for the initial Taylor Green field is plotted in figure \ref{fig:vline_TG}(a). The initial PDF is densely populated in two regions: (\rom{1}) at the ordinate axis near $q_\omega=0.4$; and (\rom{2}) at the intersection of the lower boundary and discriminant lines ($q_\omega=-1/2, r_\omega= \pm 1/(3\sqrt{6})$). The local vortex line shapes corresponding to region \rom{2} is axisymmetric vortex expansion or axisymmetric vortex compression depending on the sign of $r_\omega$, whereas,  region \rom{1} corresponds to vortex line shapes that are close to planar elliptic. 
We identify planes in the physical domain wherein the joint PDF of that plane is concentrated in either region \rom{1} or region \rom{2}. 
The actual vortex lines in such planes, derived from the vorticity field, are then plotted. Figure \ref{fig:vline_TG}(b) plots the vortex lines in planes with joint PDF concentrated in region \rom{1}. We observe that the actual vortex lines are planar elliptic similar to the shape predicted by the classification framework in the $q_\omega$-$r_\omega$ space. 
Similarly, the vortex lines in planes with joint PDF concentrated in region \rom{2} are shown in figure \ref{fig:vline_TG}(c). These vortex line shapes resemble that of axisymmetric vortex compression/expansion, consistent with their $q_\omega$-$r_\omega$ values. 

\begin{figure}
    \centering
   \subfloat[]{\includegraphics[width=0.30\textwidth, height=0.4\textheight,keepaspectratio]{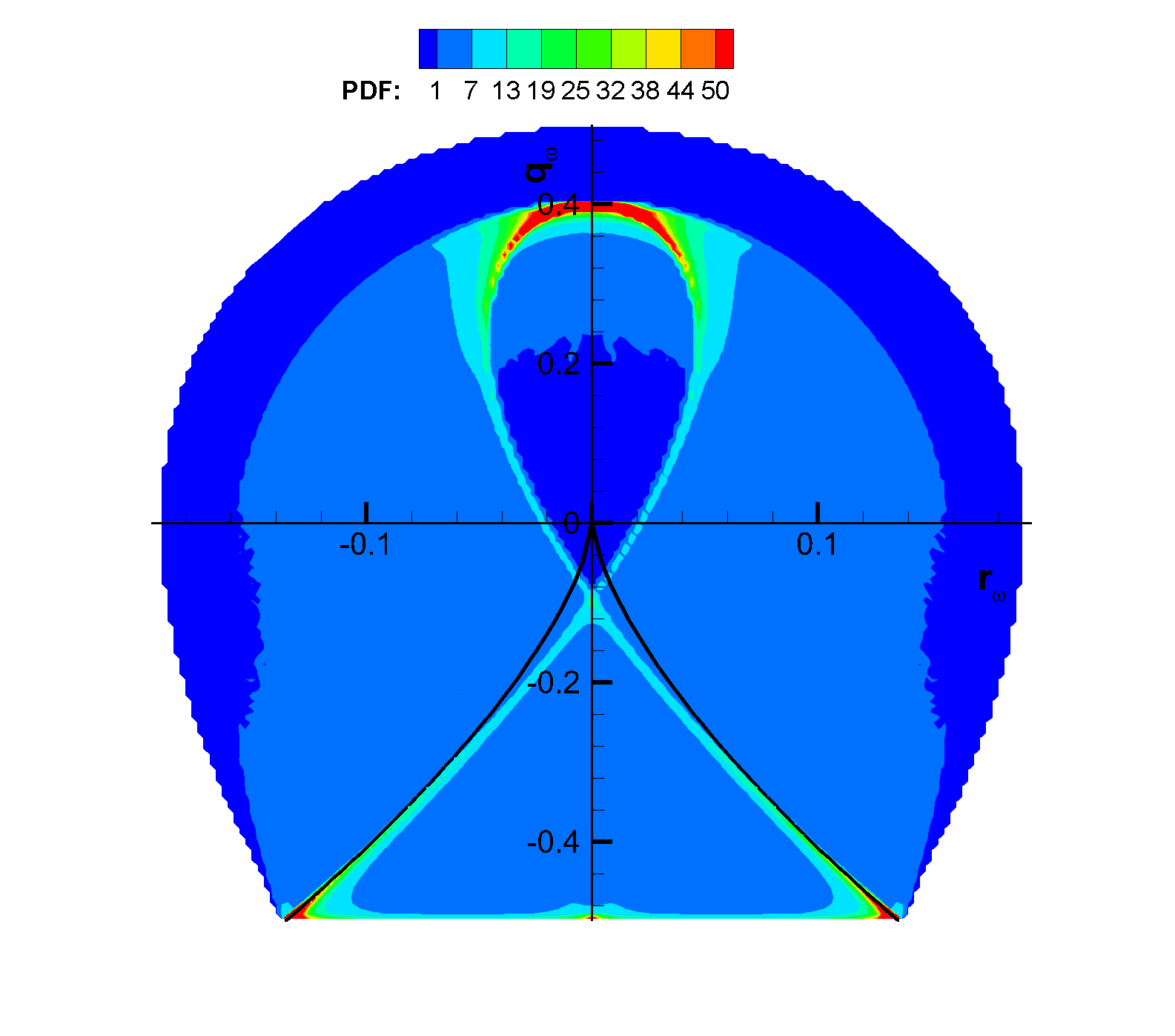}}
   \subfloat[]{\includegraphics[width=0.30\textwidth, height=0.4\textheight,keepaspectratio]{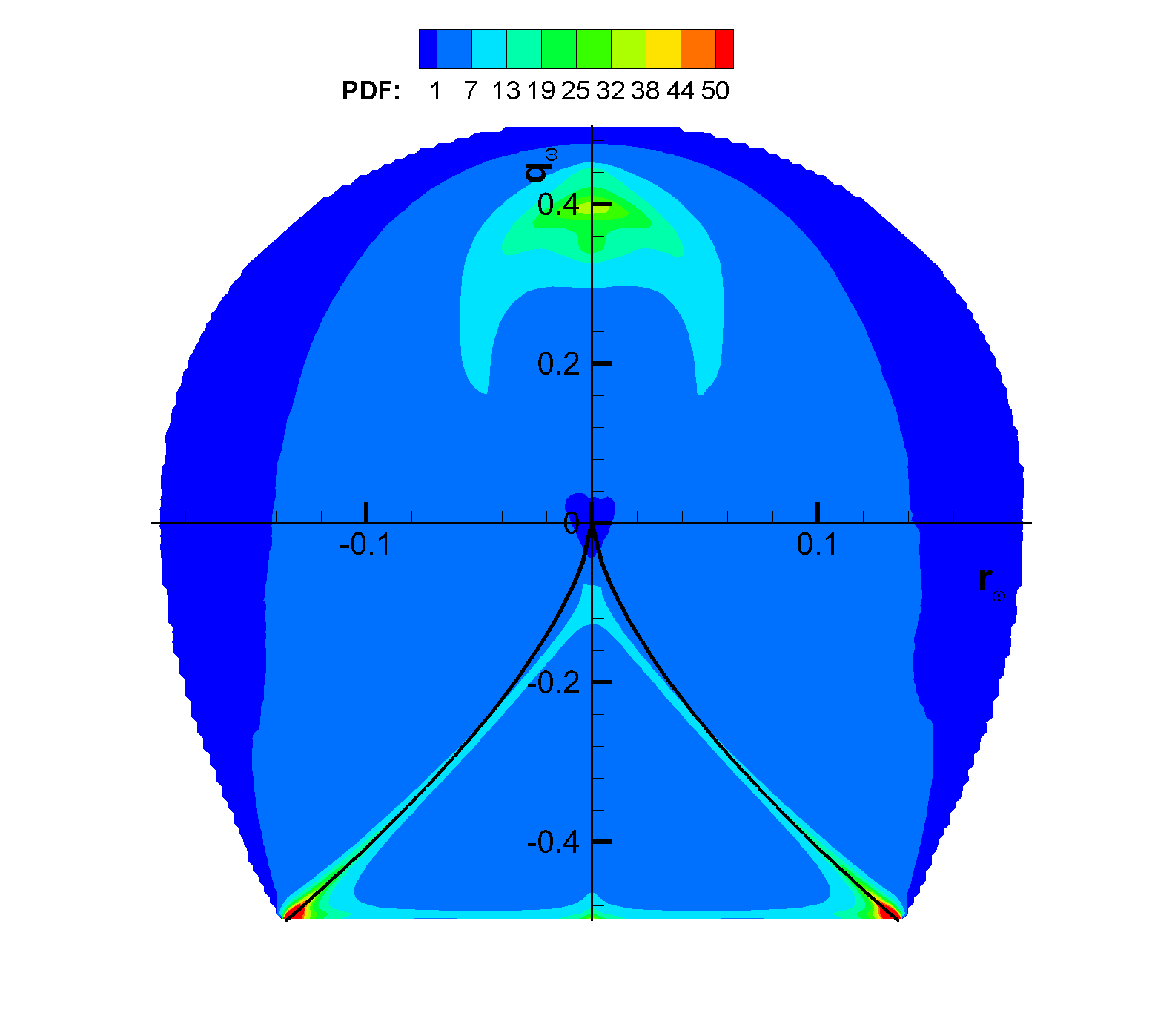}} 
    \subfloat[]{\includegraphics[width=0.30\textwidth, height=0.4\textheight,keepaspectratio]{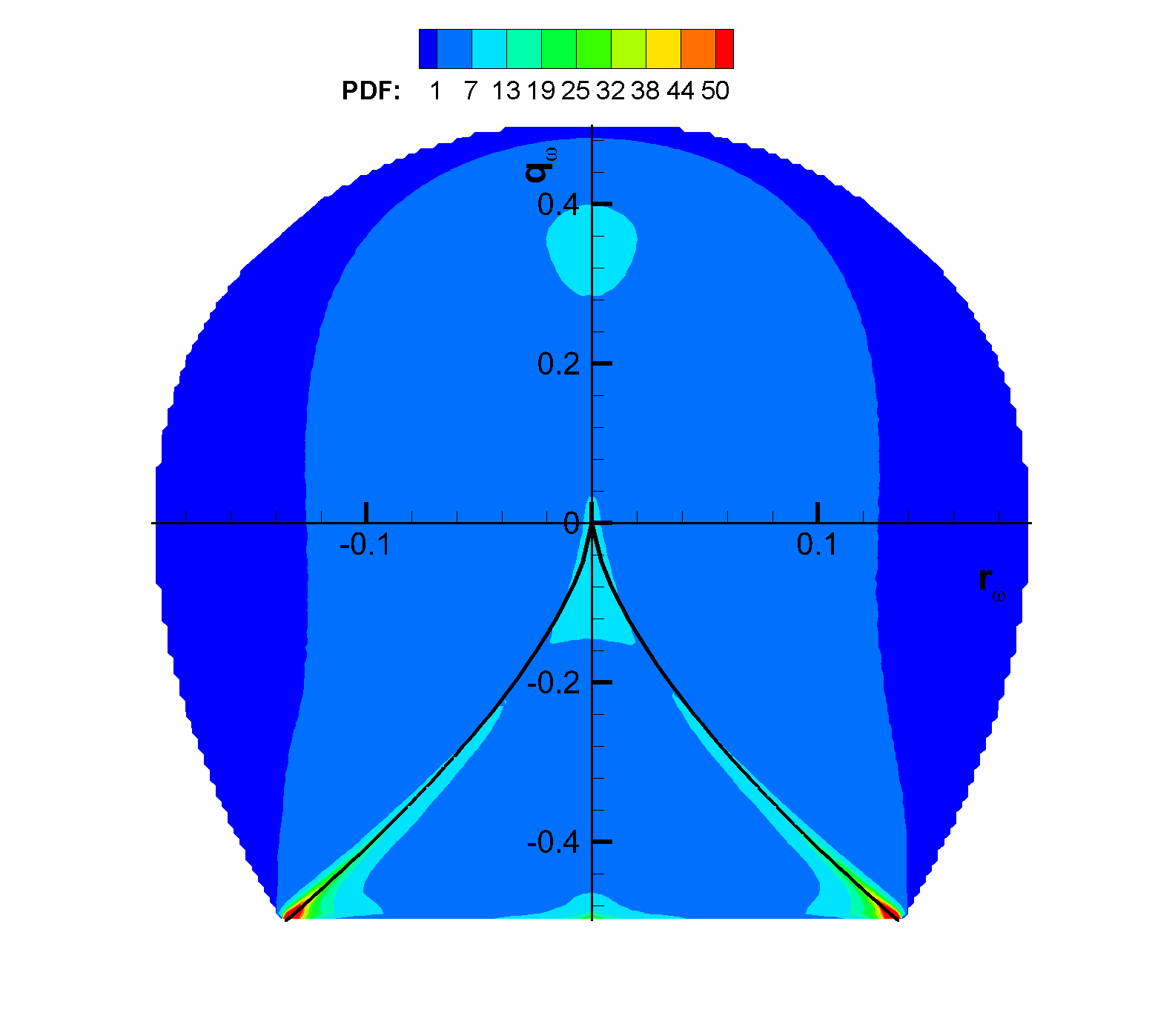}} \\
    \subfloat[]{\includegraphics[width=0.30\textwidth, height=0.4\textheight,keepaspectratio]{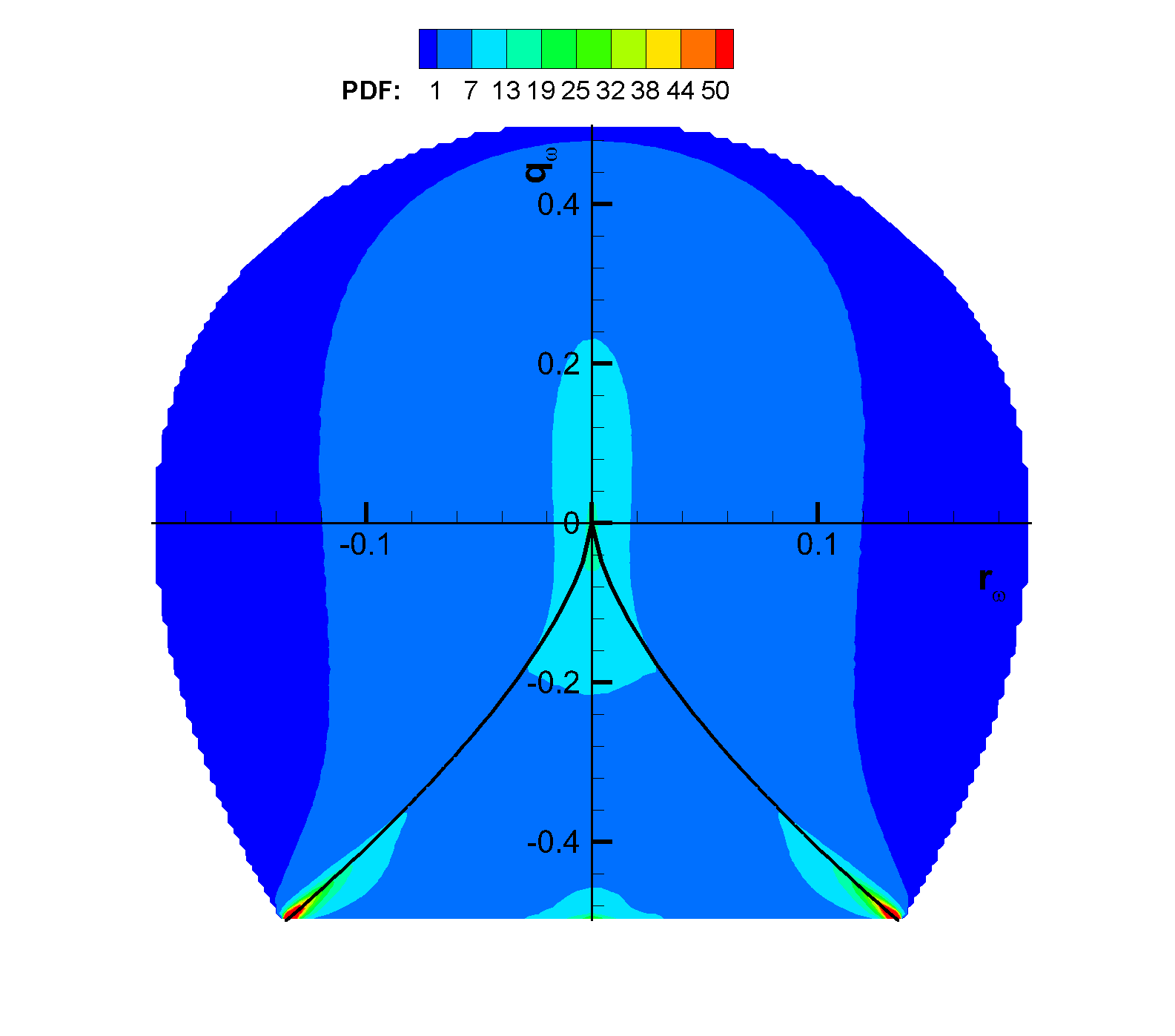}} 
    \subfloat[]{\includegraphics[width=0.30\textwidth, height=0.4\textheight,keepaspectratio]{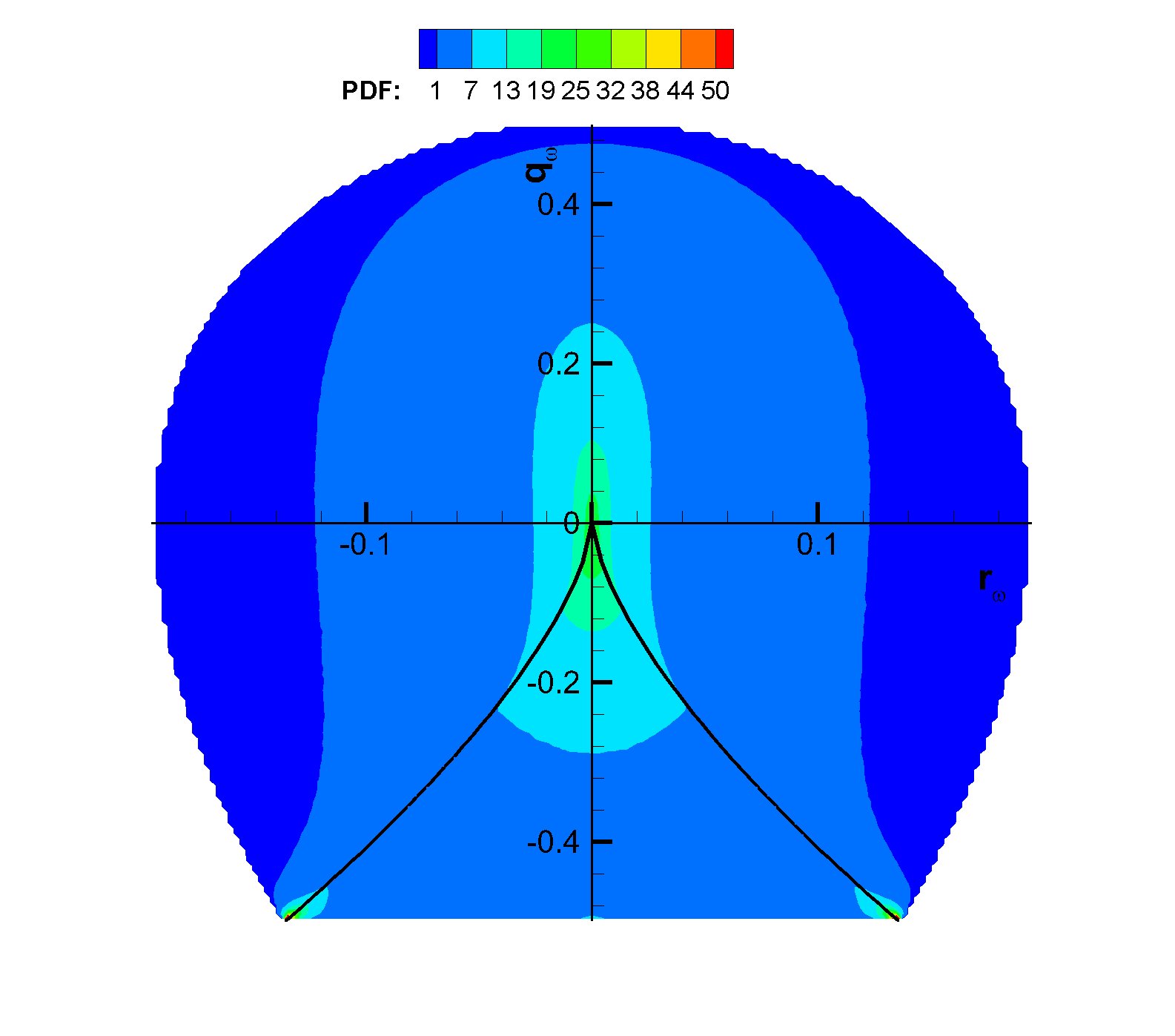}}
    \subfloat[]{\includegraphics[width=0.30\textwidth, height=0.4\textheight,keepaspectratio]{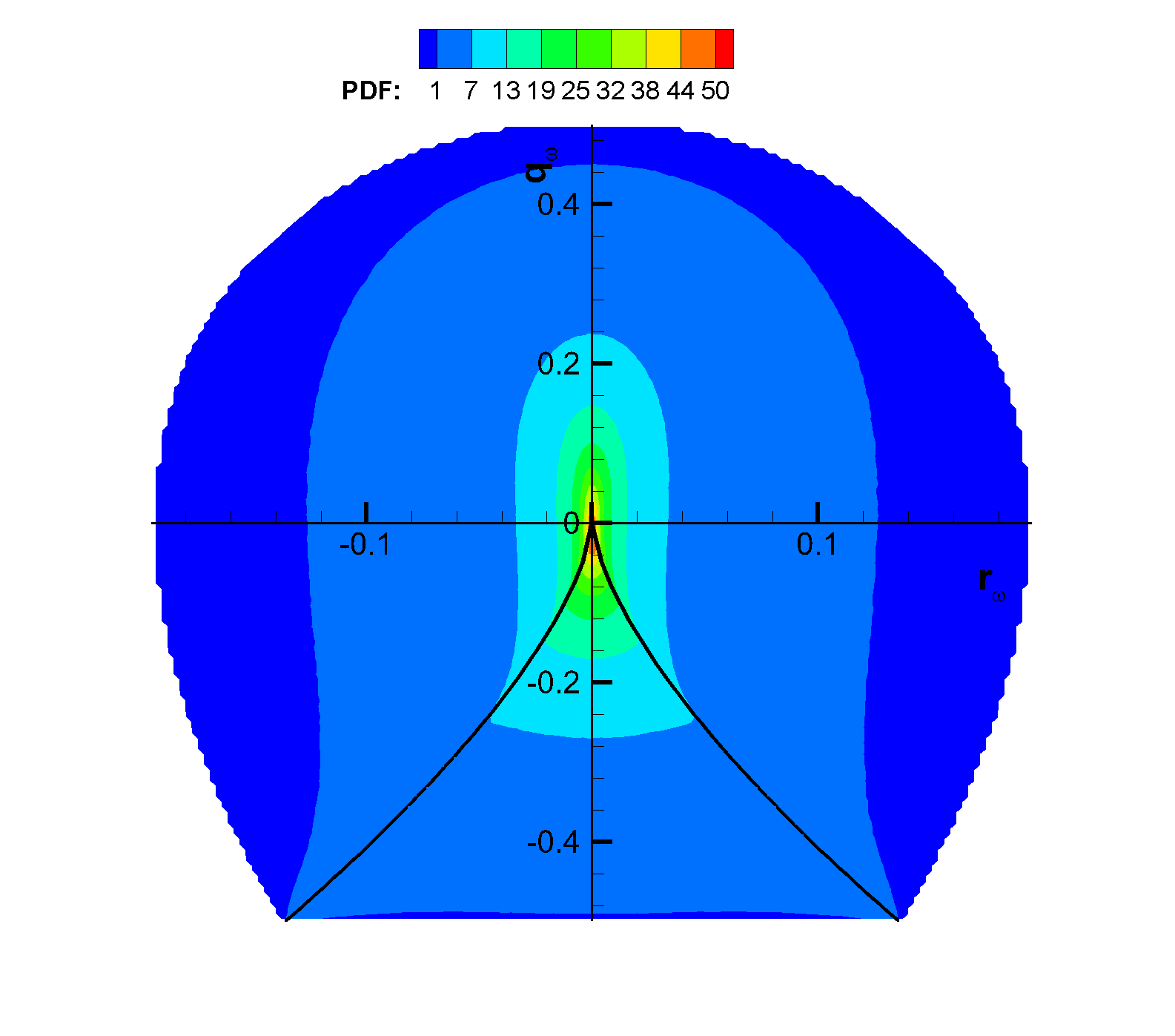}} \\
    \subfloat[]{\includegraphics[width=0.30\textwidth, height=0.4\textheight,keepaspectratio]{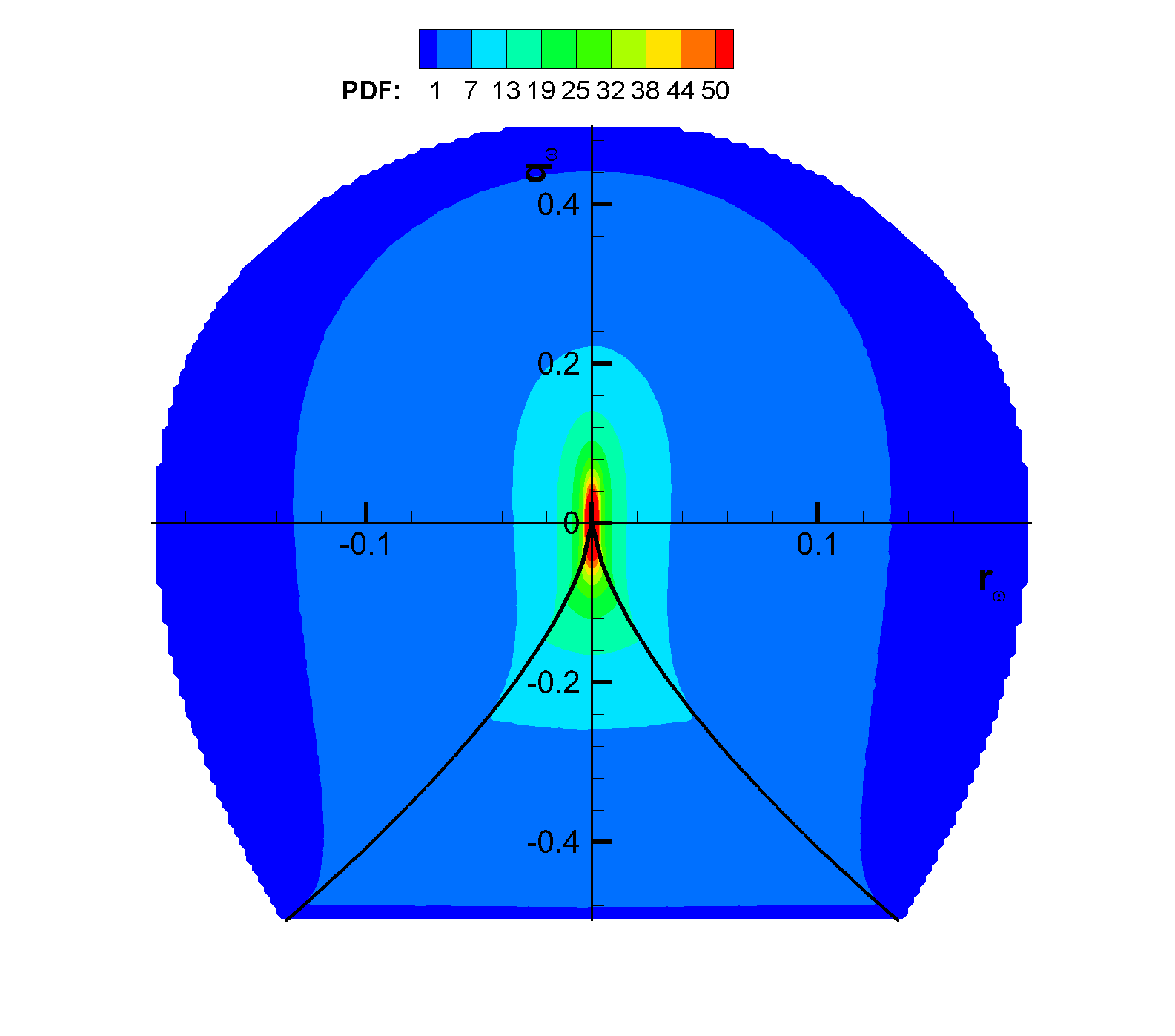}}
    \subfloat[]{\includegraphics[width=0.30\textwidth, height=0.4\textheight,keepaspectratio]{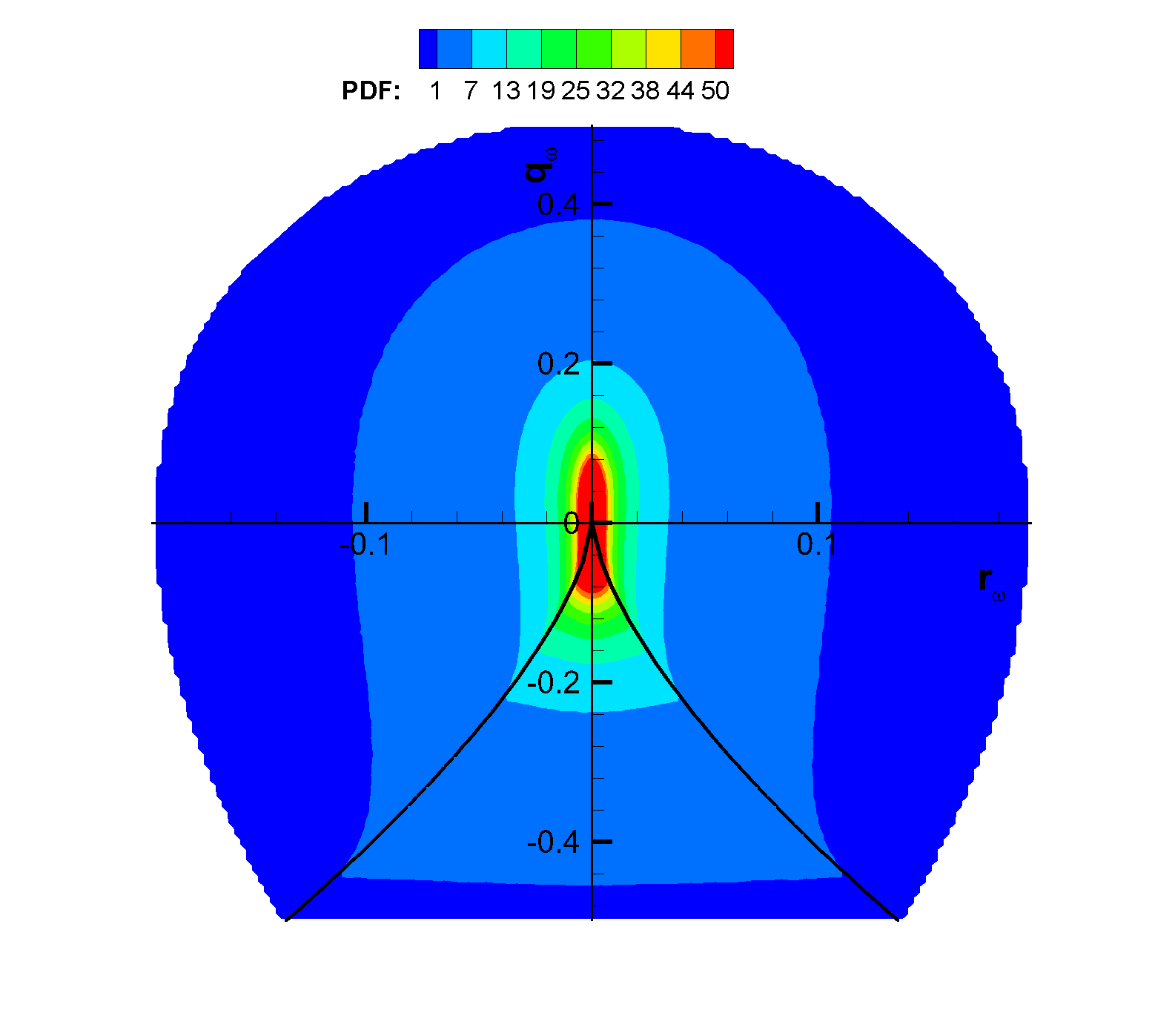}}
    \subfloat[]{\includegraphics[width=0.30\textwidth, height=0.4\textheight,keepaspectratio]{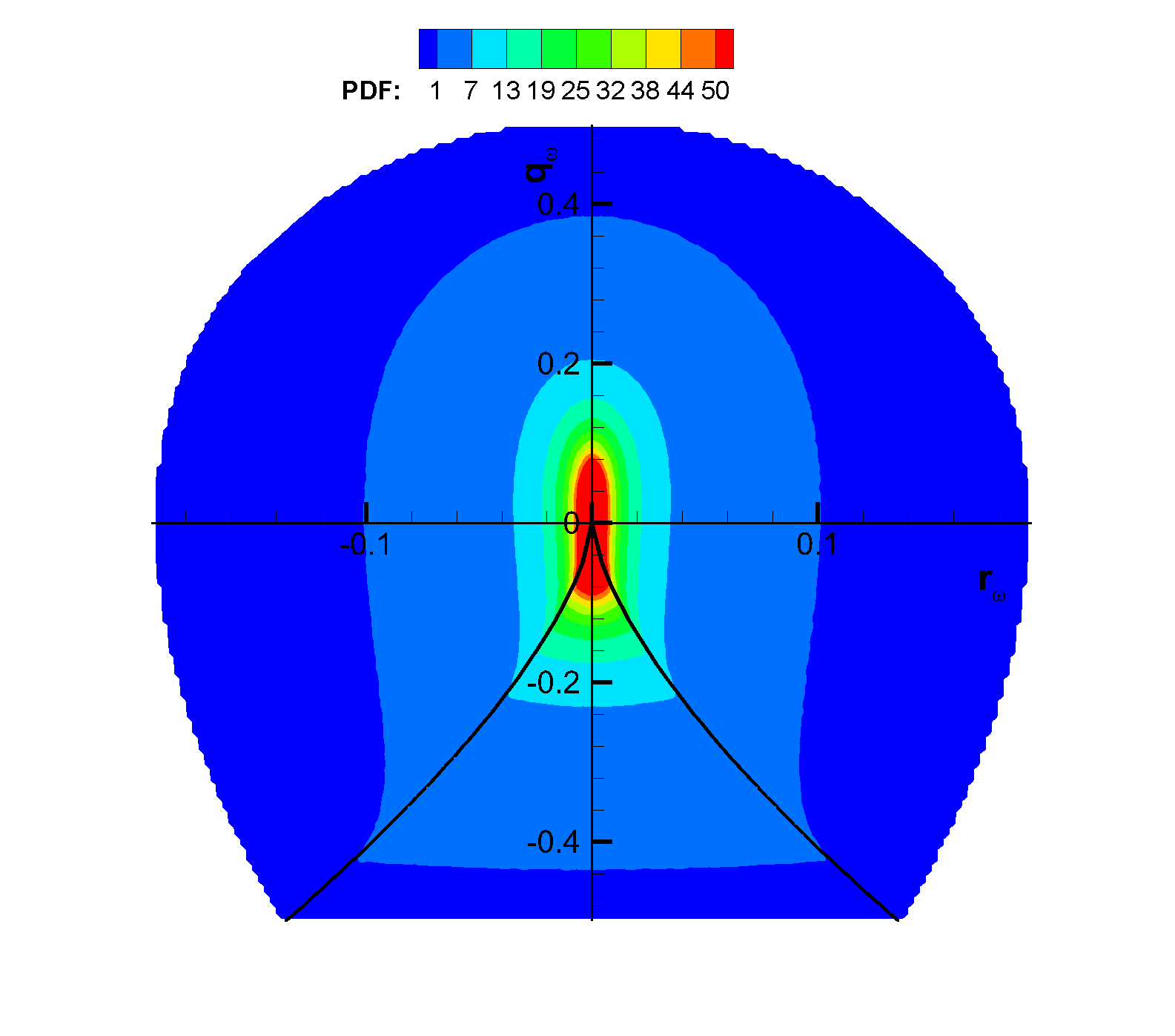}}
    \caption{$q_\omega-r_\omega$ joint PDF filled contour plots for Taylor-Green vortex simulation at (a) $t^*=0$,  (b) $t^*=0.132$, (c) $t^*=0.145$, (d) $t^*=0.160$, (e) $t^*=0.185$, (f) $t^*=0.740$, (g) $t^*=5$, (h) $t^*=9.23$ and (i) $t^*=11.57$ }
    \label{fig:jpdf_tg}
\end{figure}

We now present the evolution of $q_\omega$-$r_\omega$ joint PDF as the flow starts breaking down toward turbulence. 
The initial field (figure \ref{fig:jpdf_tg}a) is mostly constituted by vortex line shapes belonging to the regions \rom{1} and \rom{2} of $q_\omega$-$r_\omega$ plane as discussed above. 
The vortex line shapes belonging to region \rom{1}, predominantly planar elliptic in nature, begin to change first, 
resulting in a reduction of PDF values (figure \ref{fig:jpdf_tg}b-c).
These vortex line shapes are replaced by straight parallel vortex lines, as indicated by the emerging PDF values near the origin (figure \ref{fig:jpdf_tg} c-d).
Following this, the region \rom{2} (axisymmetrically expanding/compressing) vortex lines are replaced by locally straight vortex lines (figure \ref{fig:jpdf_tg}d-f).
At $t^* \approx 0.74$, the characteristic bell-shape of the PDF begins to materialize. Note that this happens quite early on in the timeline of breakdown to turbulence, well before the peak dissipation is achieved at $t^*=9.23$.
The PDF then undergoes further refinements and asymptotes to a self-similar form shortly after the peak of dissipation (figure \ref{fig:jpdf_tg} g-i).



 \begin{figure}
    \centering
    {\includegraphics[width=\textwidth, height=0.38\textheight, keepaspectratio]{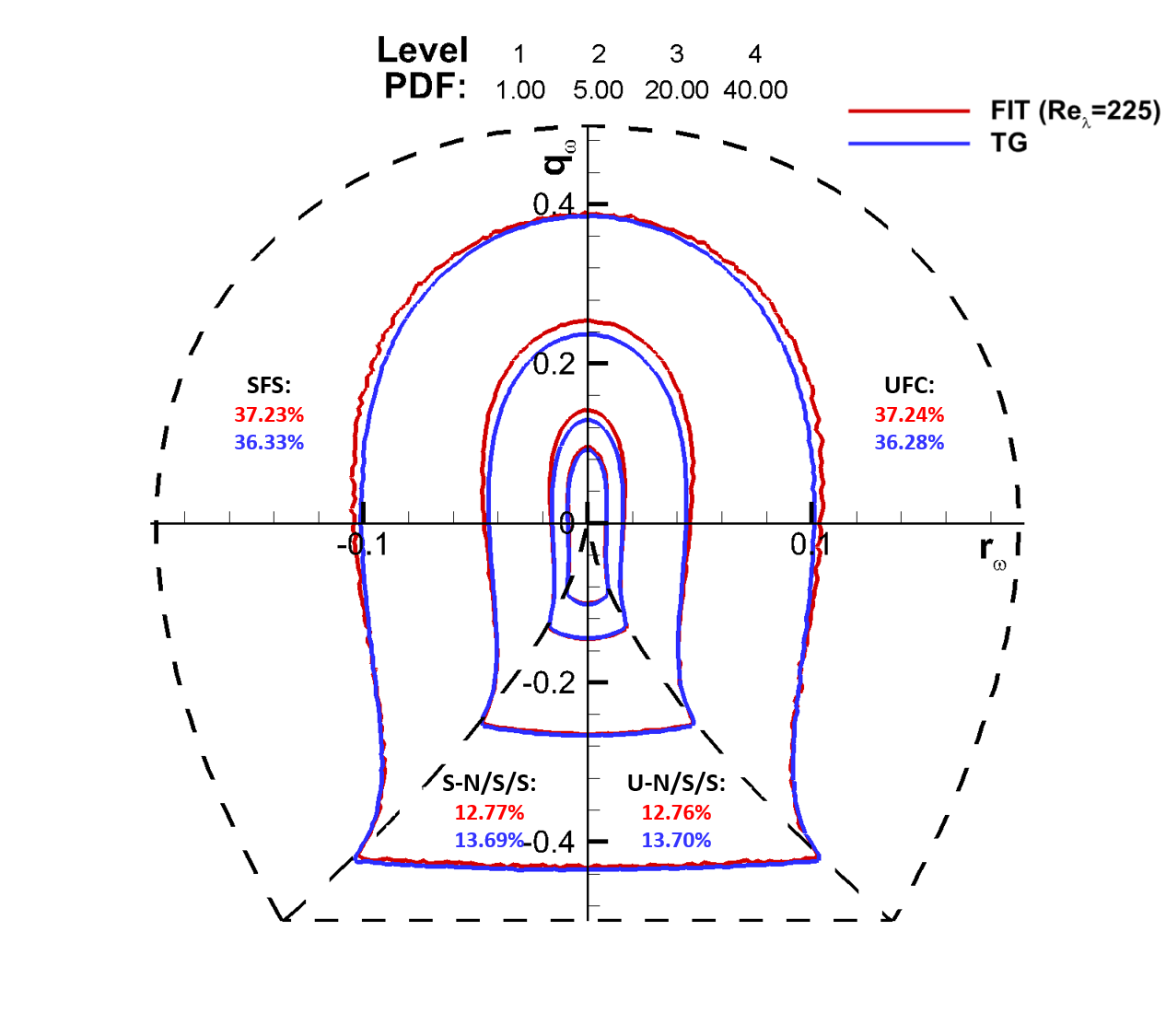}}
    \caption{$q_\omega-r_\omega$ joint PDF line contour plots and population fractions of non-degenerate topologies for Taylor-Green vortex flow (TG) at $t^*=11.57$ (shortly after peak dissipation) and forced isotropic turbulence (FIT) at $Re_\lambda=225$.}
    \label{fig:jpdf_sim}
\end{figure}

We now compare the joint probability distribution of $q_\omega$-$r_\omega$ of the Taylor-Green vortex flow field shortly after peak dissipation with that of the forced isotropic turbulent flow field at $Re_\lambda = 225$. 
As shown in figure \ref{fig:jpdf_sim}, the PDFs for both the flows are nearly identical, including the population percentages for the four vortex line topologies. 
This indicates that the characteristic bell-shape of the $q_\omega$-$r_\omega$ distribution is unique in turbulent flow fields much like the characteristic teardrop-shape distribution observed for the velocity gradient invariants.

\section{Local vortex line shape in vortex re-connection}

We now examine the local vortex line structure during the important process of vortex line reconnection. As mentioned in the Introduction, reconnection plays a key role in many flows of interest. In these flows, reconnection can occur between two vortices of different initial alignments. Here, we examine two canonical initial orientations previously studied in literature and examine the reconnection process from the local vortex line shape point of view. The focus is primarily on the mechanism of bridging \citep{melander1988cut,kida1991collision,boratav1992reconnection} in vortex  reconnection. 

\subsection{Anti-parallel vortex tubes}
\begin{figure}
    \centering
    \subfloat[]{\includegraphics[ trim=0 10 10 0, clip, width=0.33\textwidth, keepaspectratio]{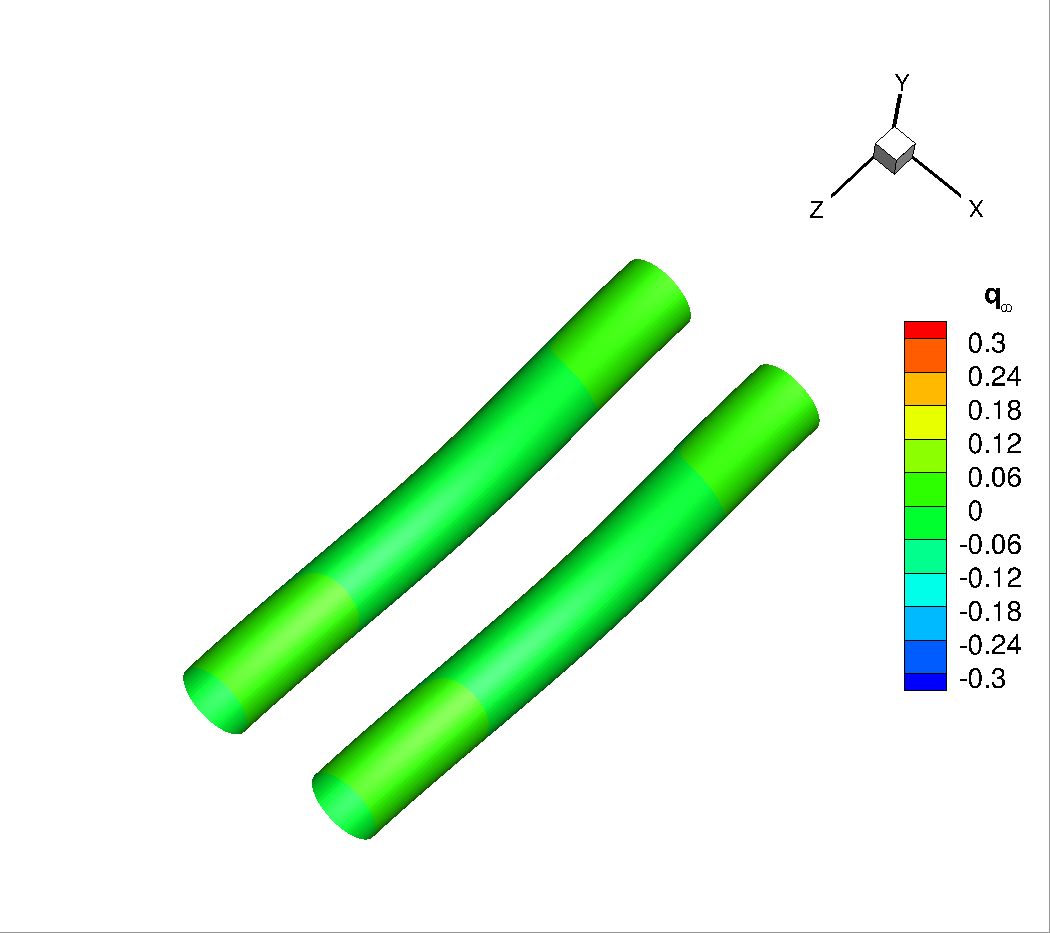}}
    \subfloat[]{\includegraphics[ trim=0 10 10 0, clip, width=0.33\textwidth, keepaspectratio]{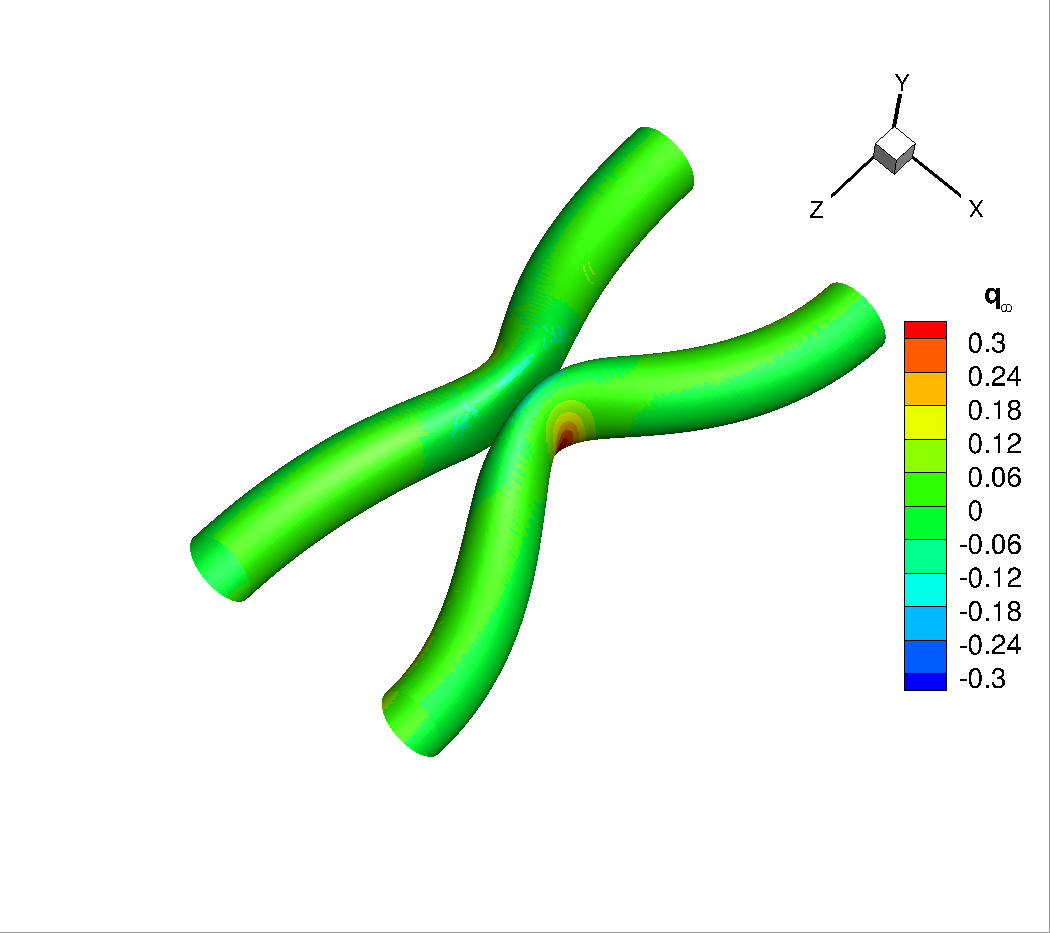}}
    \subfloat[]{\includegraphics[ trim=0 10 10 0, clip, width=0.33\textwidth, keepaspectratio]{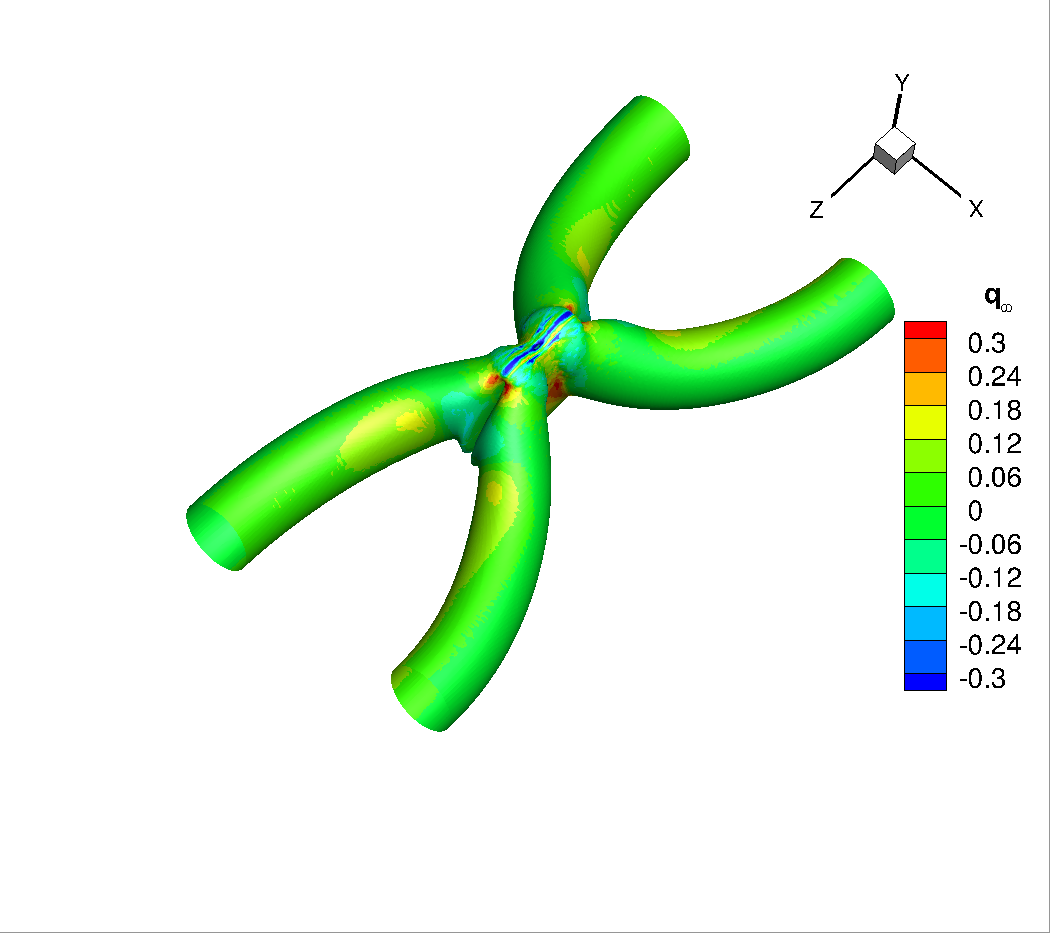}}\\
    \subfloat[]{\includegraphics[ trim=0 10 10 0, clip, width=0.33\textwidth, keepaspectratio]{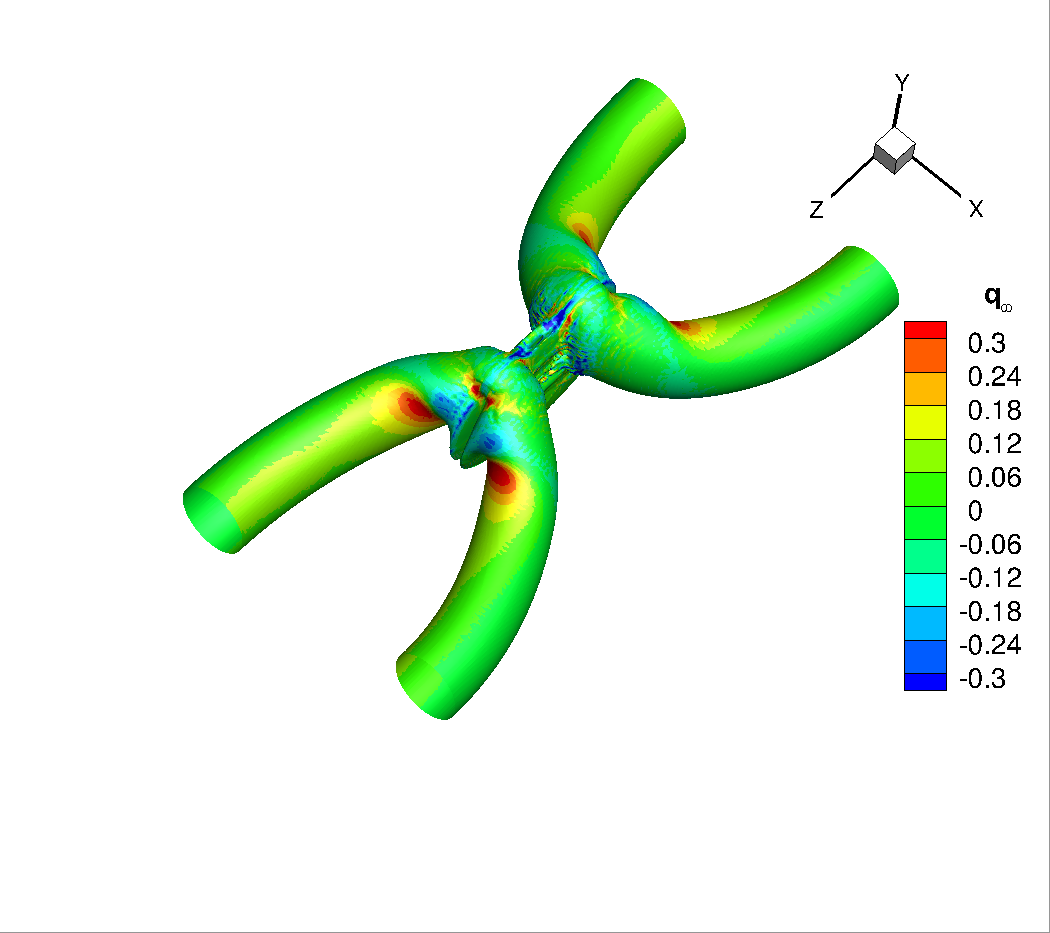}}
    \subfloat[]{\includegraphics[ trim=0 10 10 0, clip, width=0.33\textwidth, keepaspectratio]{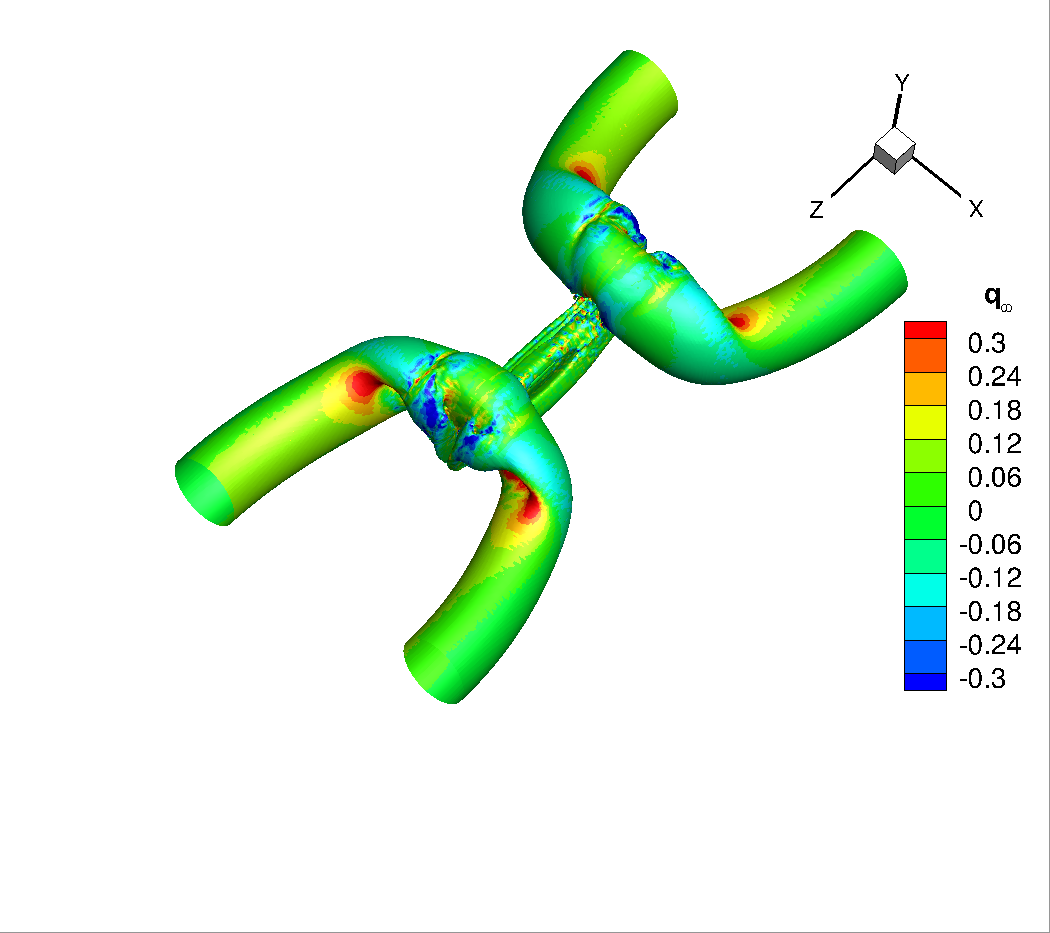}}
    \subfloat[]{\includegraphics[ trim=0 10 10 0, clip, width=0.33\textwidth, keepaspectratio]{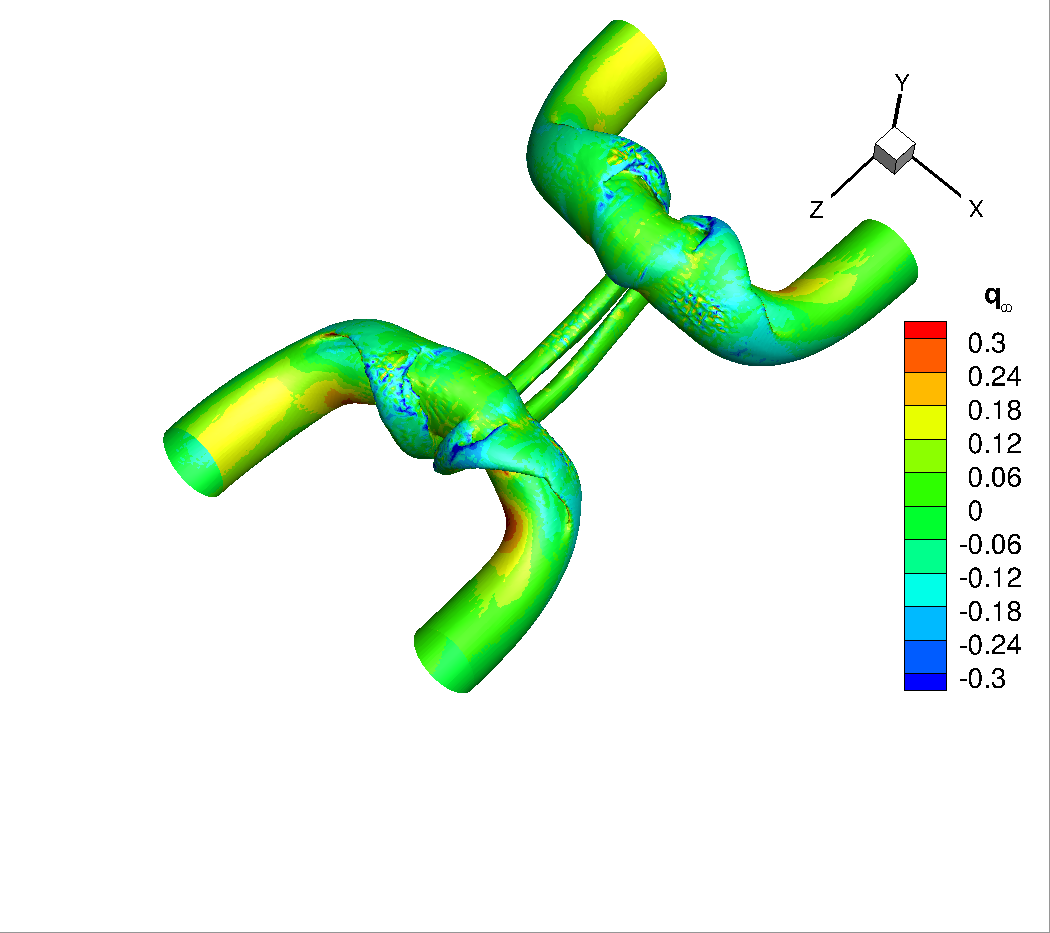}}\\
    \caption{$|\omega|$ isosurfaces at $30\%$ of maximum initial vorticity colored by $q_\omega$ at t = (a) $0$, (b) $3.6$, (c) $4.4$, (d) $4.8$, (e) $5.4$ and (f) $6$.}
    \label{fig:qomg_p}
\end{figure}

\begin{figure}
    \centering
    \subfloat[]{\includegraphics[width=0.42\textwidth]{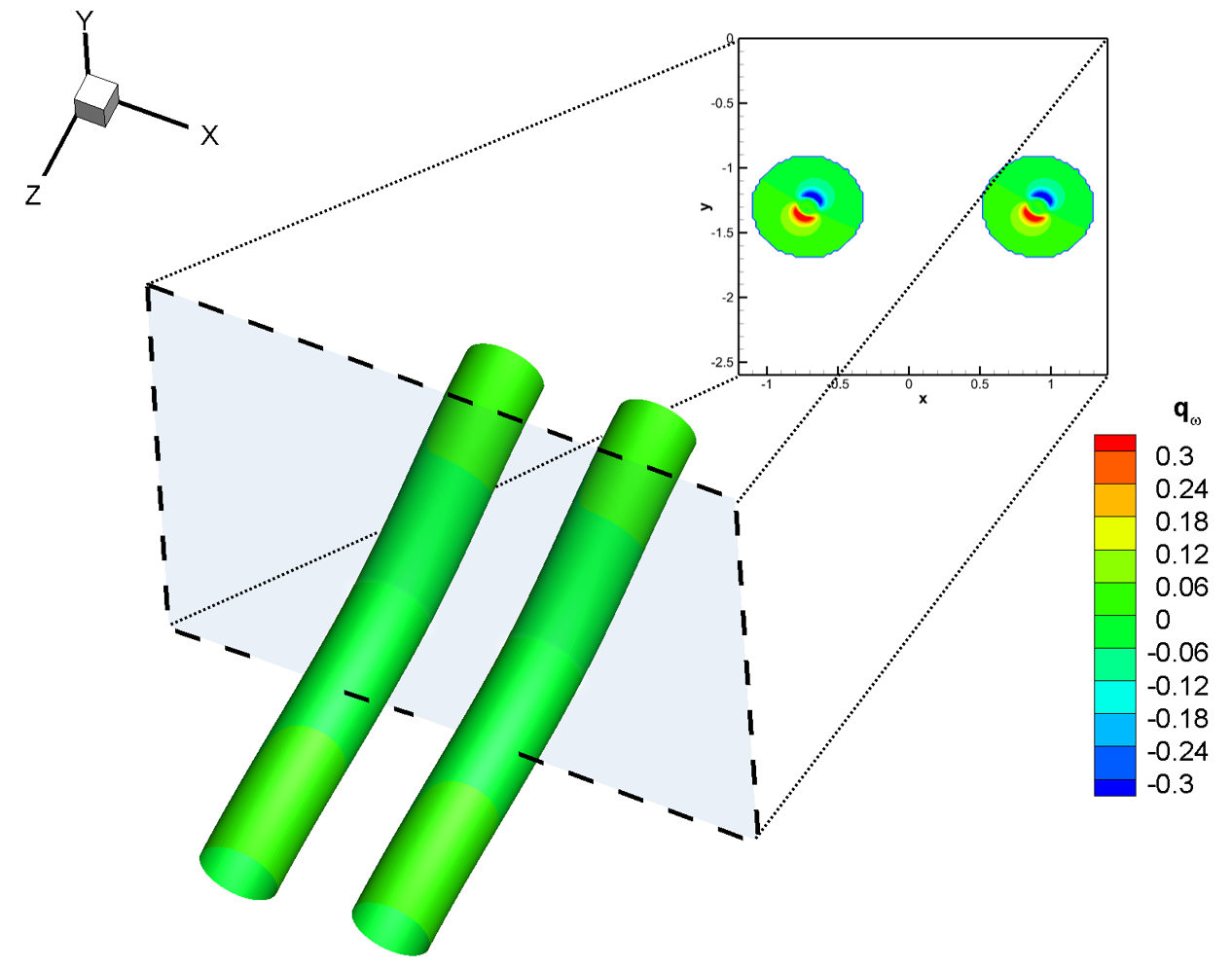}}
    \subfloat[]{\includegraphics[width=0.42\textwidth]{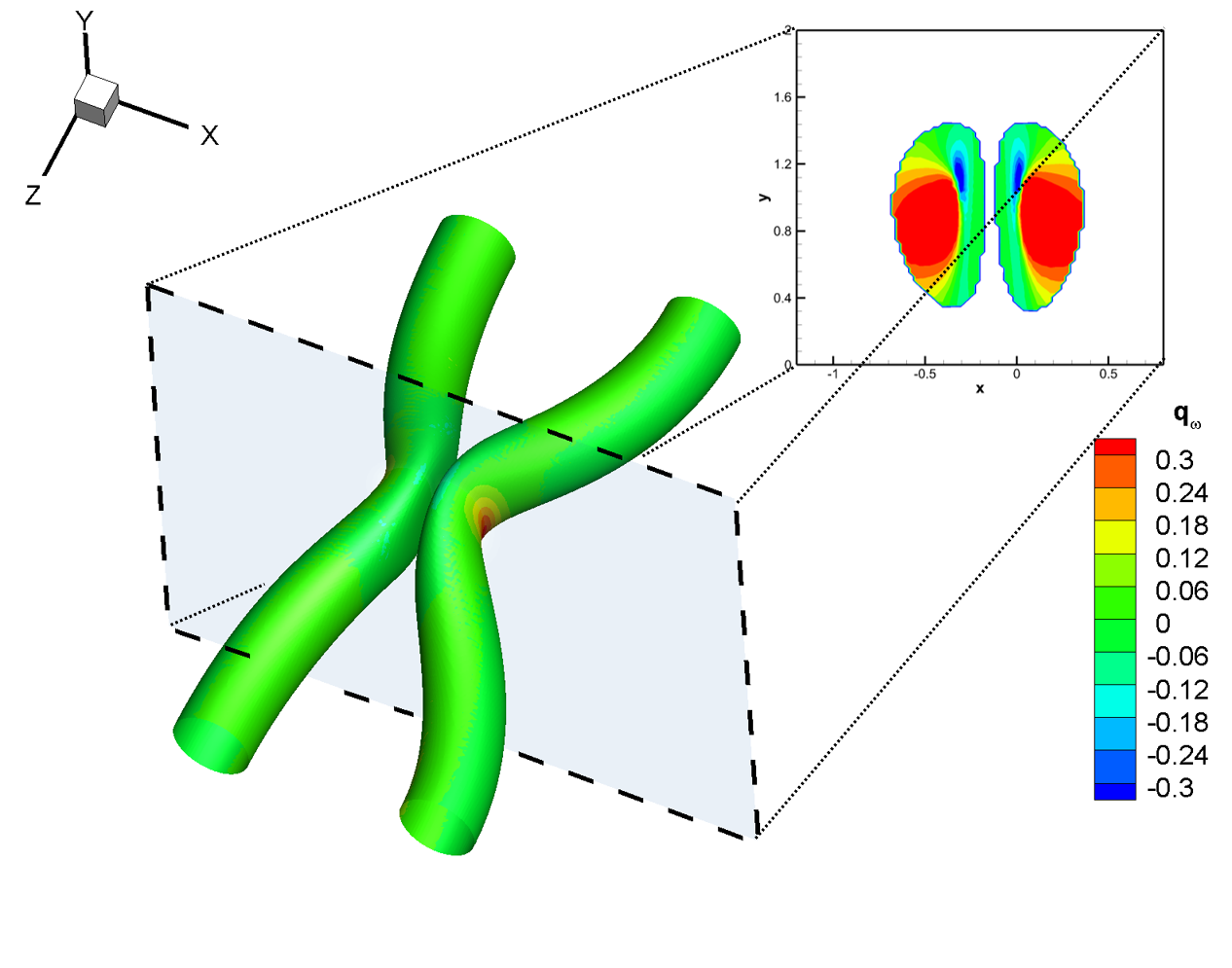}}\\
    \subfloat[]{\includegraphics[width=0.42\textwidth]{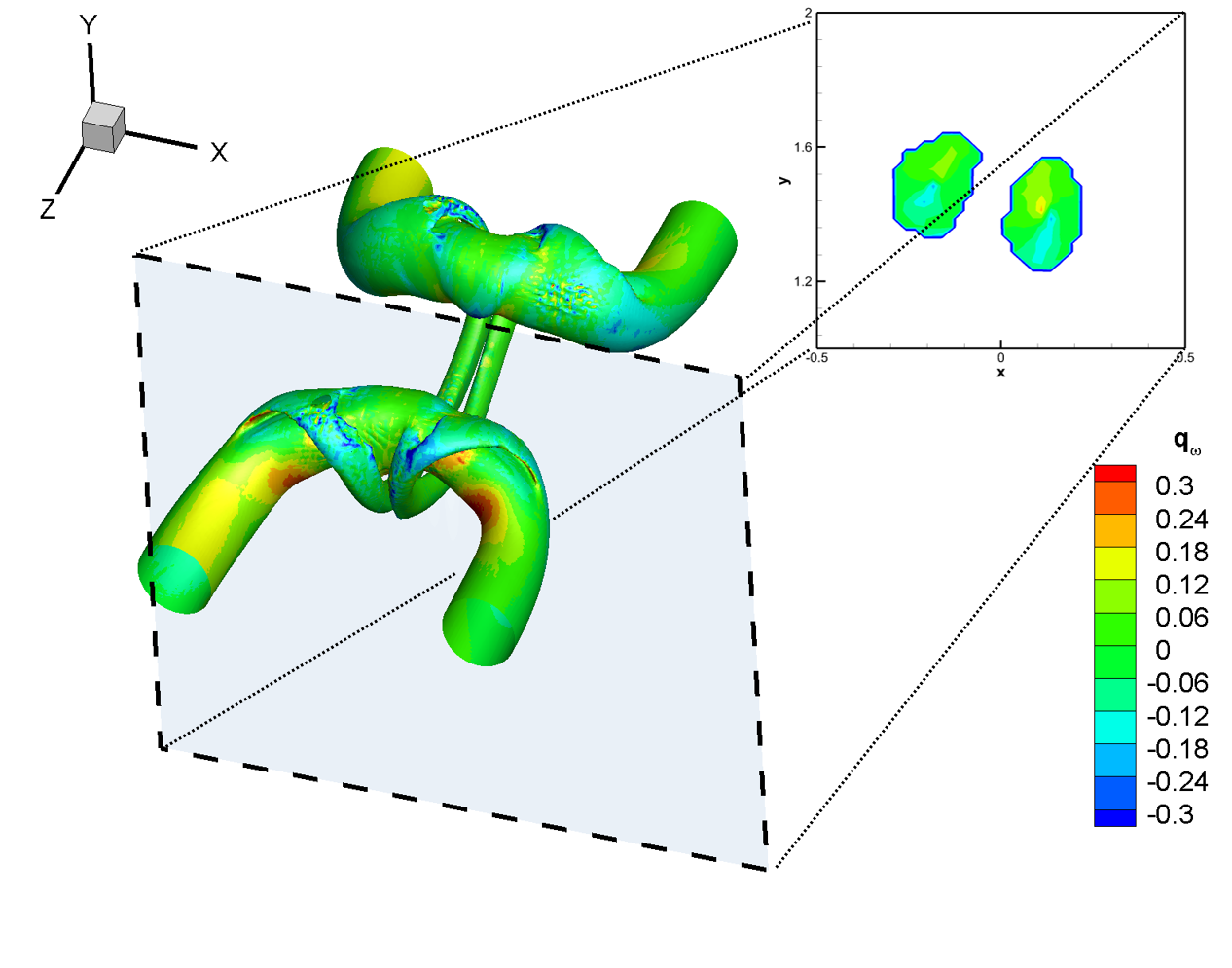}}
    \caption{$q_\omega$ contours in the symmetry plane at $t=$ (a) $0$, (b) $3.6$ and (c) $6$. Contours are only shown in regions wherein $|\omega|>0.3\omega_0$}
    \label{fig:p_qwcontour_symm}
\end{figure}

\begin{figure}
    \centering
    \subfloat[]{\includegraphics[width=0.45\textwidth]{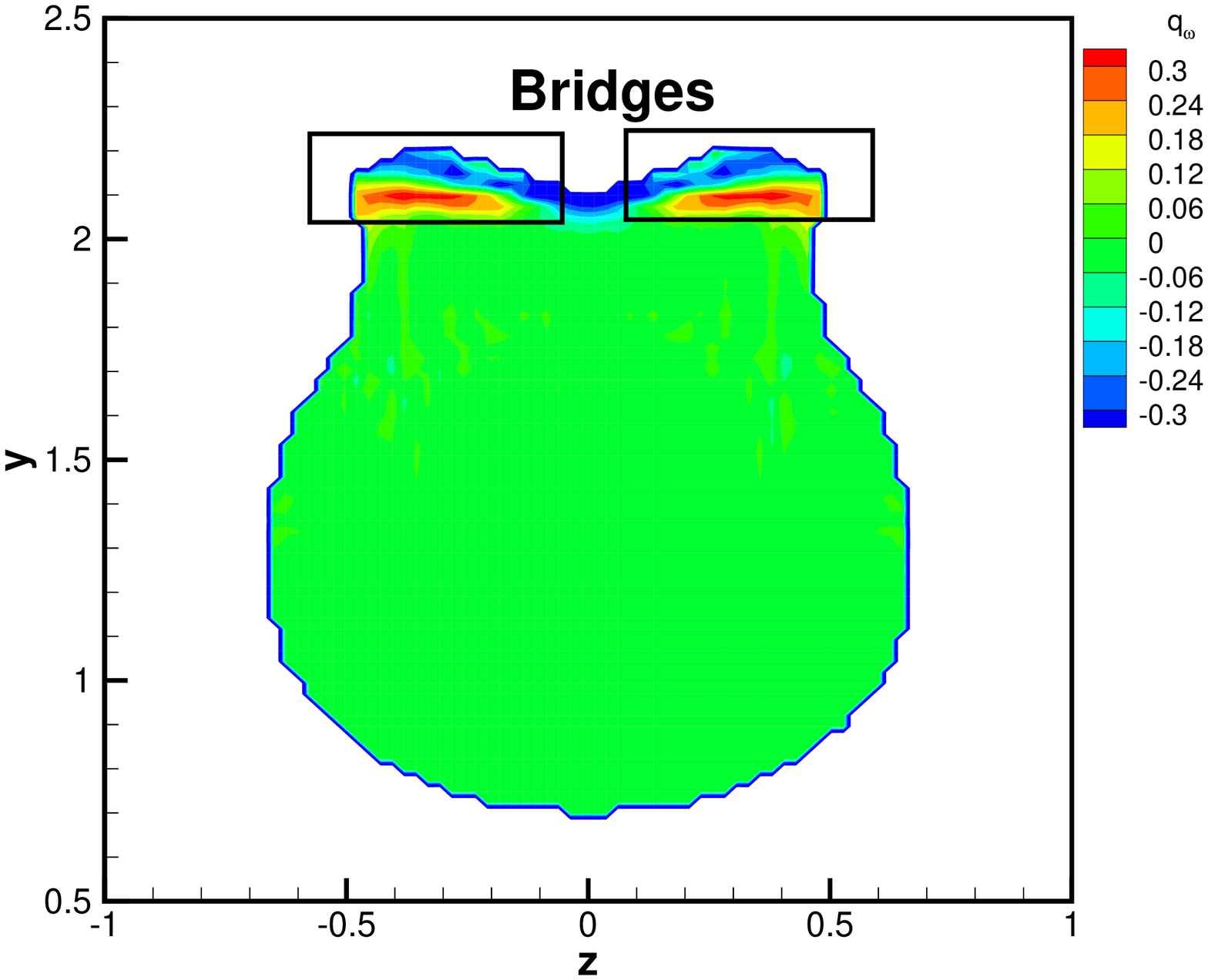}}
    \subfloat[]{\includegraphics[width=0.45\textwidth]{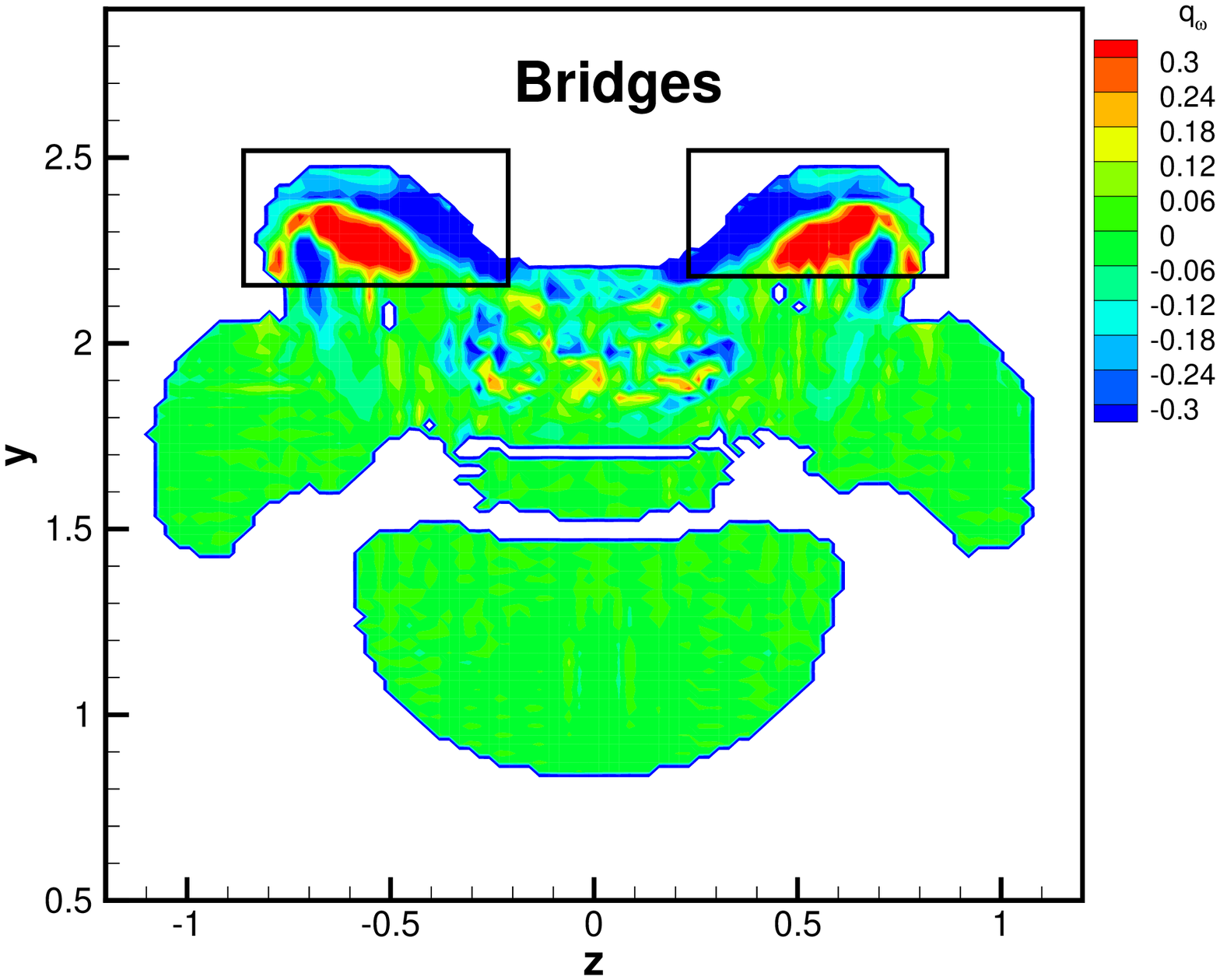}}\\
    \subfloat[]{\includegraphics[width=0.45\textwidth]{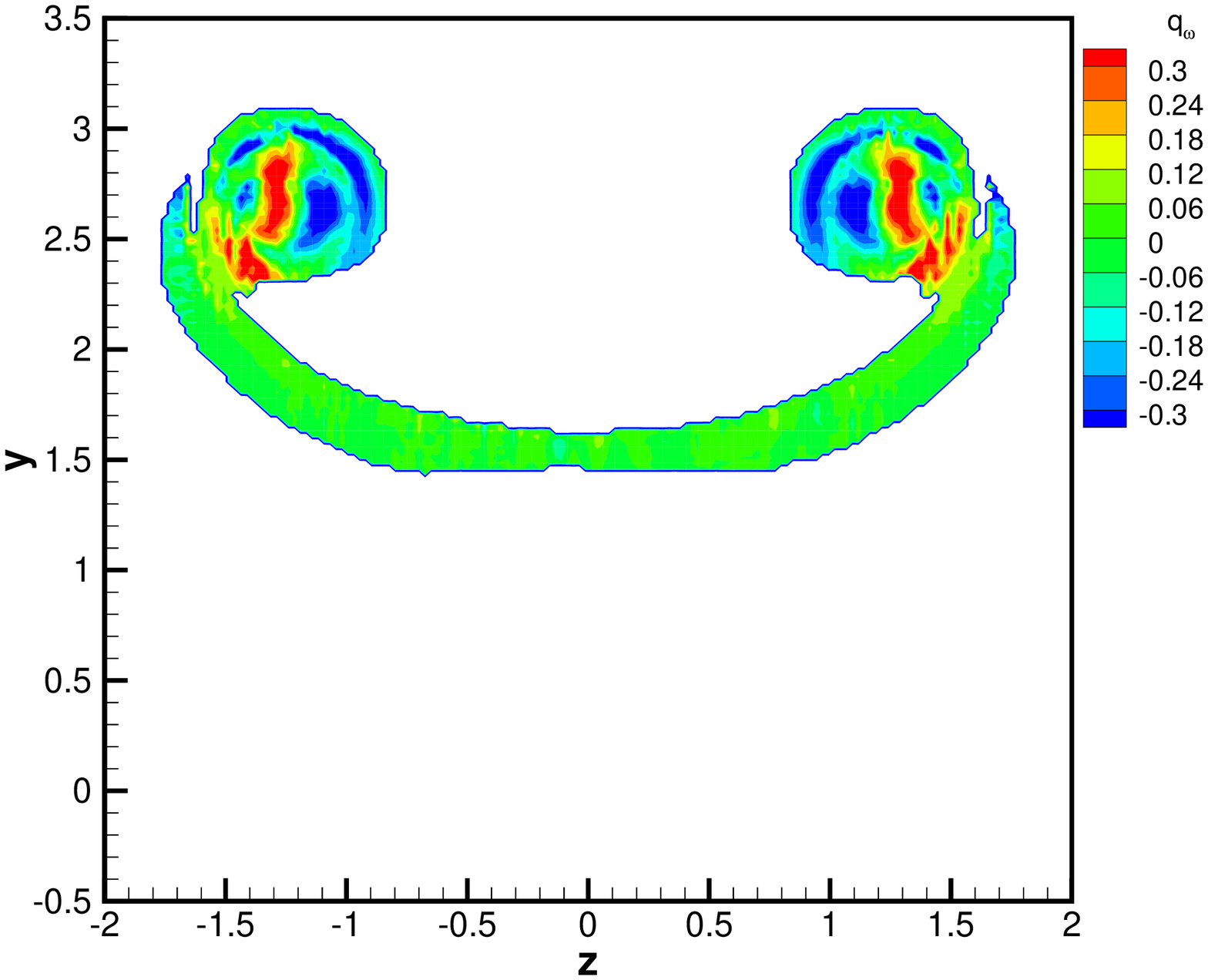}}
    \caption{$q_\omega$ contours in the dividing plane at $t=$ (a) $4.4$, (b) $4.8$ and (c) $6$. Contours are only shown in regions wherein $|\omega| > 0.3 \omega_0$}
    \label{fig:p_qwcontour_div}
\end{figure}

We now examine the evolution of local vortex line shapes during vortex reconnection via bridging in sinusoidally perturbed anti-parallel vortex tubes. The key events leading up to and beyond bridging are traced by analyzing the isosurfaces of vorticity magnitude at different instants of time in figure \ref{fig:qomg_p}. The isosurfaces are colored by $q_\omega$ to track the evolution of local vortex line shapes during this period. We do not show results for $r_\omega$ as it is close to zero everywhere in the tubes at all times. This indicates that the local vortex lines are nearly two dimensional and allows $q_\omega$ to completely characterize the local vortex line shape. 
The different steps in the vortex reconnection process and the corresponding vortex line shapes are discussed below with reference to figures \ref{fig:qomg_p}, \ref{fig:p_qwcontour_symm} and \ref{fig:p_qwcontour_div}: 
\begin{enumerate}
    \item At $t=0$, $q_\omega$ is zero everywhere on the surface, indicating that the local vortex line shape is straight on the tube surface. In figure \ref{fig:p_qwcontour_symm}, we plot contours of $q_\omega$ in the symmetry ($x$-$y$) plane to describe the vortex line shapes in the tube's cross section. Initially, the dominant vortex line shape in the tube's cross-section is straight, except for small regions near the core (red and blue crescent shaped regions in figure \ref{fig:p_qwcontour_symm}(a)). In these particular regions, the curvature effects are dominant.
    Due to the strong curvature effects, the linearity assumption of determining local vortex line shapes from $q_\omega$-$r_\omega$ might not hold. As a result, the local vortex line shape is not necessarily reflective of the large scale vortex line shape. 
    \item The two tubes move toward each other by mutual induction and are pressed against each other at $t=3.6$. At this stage, the cores of the tubes in the interaction region are significantly flatter. The vortex line shape is dominantly straight in the tubes everywhere excluding the contact region. The contact region comprises of positive values of $q_\omega$, indicating locally elliptic vortex lines. The vortex line shapes in the contact region are examined in greater detail by the contours of $q_\omega$ in the cross section of the contact zone shown in figure \ref{fig:p_qwcontour_symm}(b). 
    The local vortex line shapes in the contact zone are dominantly planar elliptic as evident from the positive values of $q_\omega$.
    \item Cross-linking between the tubes results in annihilation of vorticity in the symmetry plane. Correspondingly, orthogonal vorticity emerges in the dividing plane resulting in  vortex reconnection.  By $t=4.4$, reconnection is initiated and at the ends of contact zone a hump connects the two tubes. The hump is called a bridge \citep{melander1988cut} and the process is termed vortex reconnection via bridging. Vortex line shapes inside the bridges are analyzed by examining the contours of $q_\omega$ in the dividing plane in figure \ref{fig:p_qwcontour_div}. The vortex-line shapes inside the bridges is dominantly planar elliptic in the inner bridge portions, while planar hyperbolic vortex lines are prevalent in the outer bridge regions. Henceforth, this specific occurrence of paired vortex-line shapes in the bridges will be referred to as ``elliptic-hyperbolic pairing".
    Aside from the bridges, the vortex line shapes everywhere else in the vortex tubes are mostly straight. 
    \item The orthogonal transfer of vorticity from the tubes in the interaction region makes them weaker, and correspondingly the bridges become stronger. This is evident from figure \ref{fig:qomg_p}(d), wherein the bridges have thickened and the interaction region has experienced thinning. At this stage the vortex line shapes in the curved region of the tubes are more elliptic. Moreover, as shown in figure \ref{fig:p_qwcontour_div}(b), ``elliptic-hyperbolic pairing" continues to be the dominant vortex line shapes inside the bridges.
    \item As the bridges strengthen, self-induction causes bridges to pull apart from the interaction region resulting in stretching the remnant of tubes (threads) in the contact zone. The separation between the bridges has increased in figure \ref{fig:qomg_p}(e). 
    The distribution of vortex line shapes in the tubes is similar to figure \ref{fig:qomg_p}(d). 
    \item At the final time step under consideration (figure \ref{fig:qomg_p}f), the bridges have morphed to be part of the two vortex half rings and the hump is indiscernible. The two formed vortex half rings have elliptic vortex line shapes near the curved portions.  The ``elliptic-hyperbolic pairing" is still the dominant vortex line topology inside the morphed bridges (figure \ref{fig:p_qwcontour_div}c). Meanwhile, self induction between the curved threads results in them moving away from each other. The corresponding vortex line shapes inside the threads (figure \ref{fig:p_qwcontour_symm}c) are also predominantly straight.
\end{enumerate}

\begin{figure}
    \centering
    \subfloat[]{\includegraphics[width=0.3\textwidth]{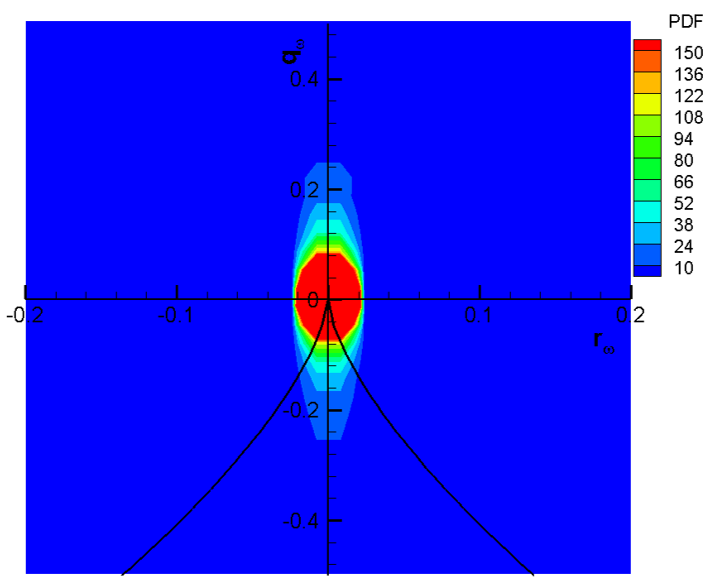}}
    \subfloat[]{\includegraphics[width=0.3\textwidth]{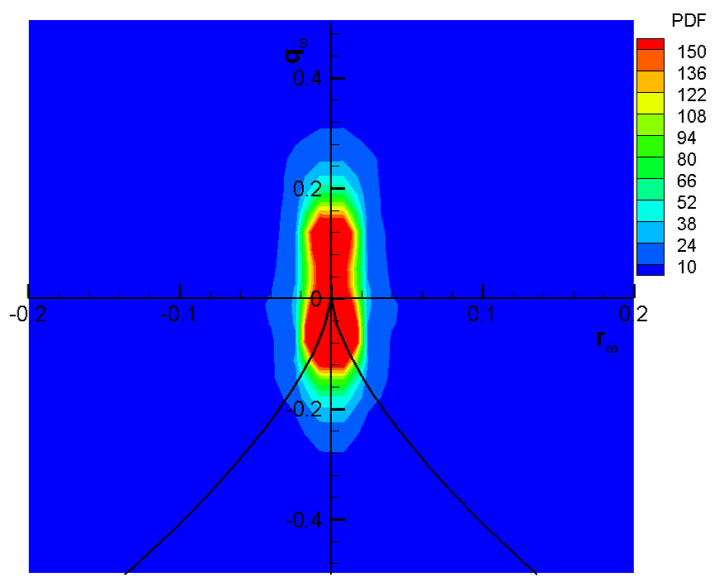}}
    \subfloat[]{\includegraphics[width=0.3\textwidth]{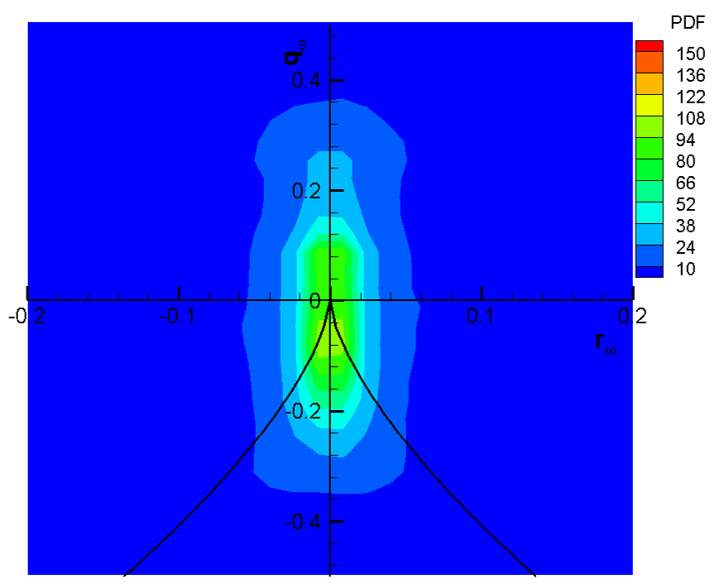}}
    \caption{$q_\omega-r_\omega$ joint PDF filled contours $t=$ (a) $0$, (b) $4.4$ and (c) $6$. Only points with $|\omega|>0.3\omega_0$ are considered.}
    \label{fig:qwpdf_p}
\end{figure}
 We now analyze the evolution of vortex line shapes in the vortex tubes altogether by examining the joint probability distribution of $q_\omega-r_\omega$. We only consider points with vorticity magnitude greater than $30\%$ of the maximum initial vorticity in plotting such joint distributions (figure \ref{fig:qwpdf_p}). Initially, the vortex line shape is locally straight (i.e. $q_\omega \approx r_\omega \approx 0$) almost everywhere. At $t=4.4$, i.e. at the inception of bridging, the joint PDF has expanded along the $q_\omega$ axis while still remaining constrained in the $r_\omega$ axis. 
 This implies that the local vortex line shapes are highly likely to be planar. In addition, the vortex line shapes are no longer restricted to only straight lines, rather they are strongly likely to be elliptic or hyperbolic in nature. 
 Finally, by $t=6$, the likelihood of straight vortex lines has decreased further as more elliptic and hyperbolic vortex lines appear.

\subsection{Orthogonally interacting tubes}

\begin{figure}
    \centering
    \subfloat[]{\includegraphics[ trim=0 10 10 0, clip, width=0.33\textwidth, keepaspectratio]{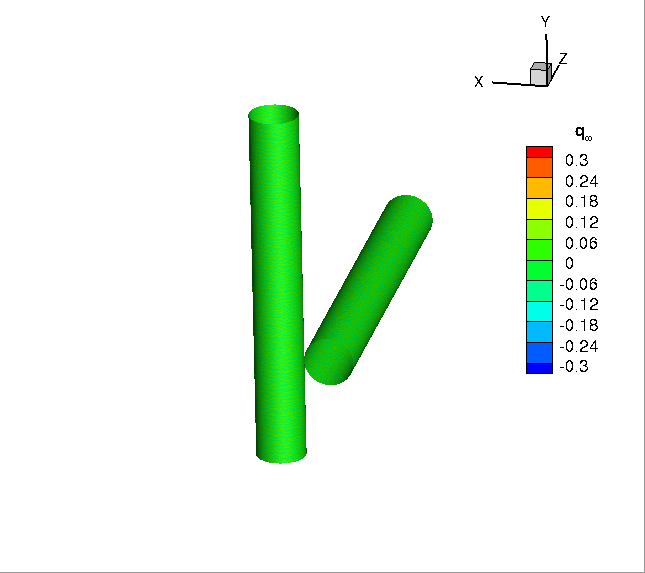}}
    \subfloat[]{\includegraphics[ trim=0 10 10 0, clip, width=0.33\textwidth, keepaspectratio]{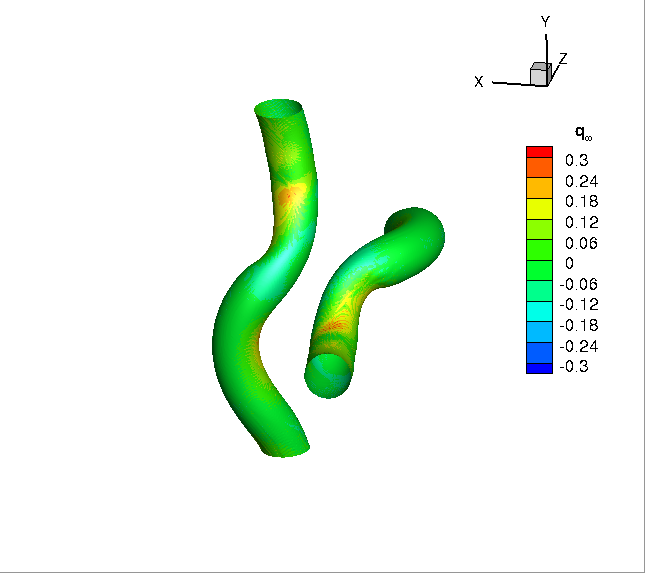}}
    \subfloat[]{\includegraphics[ trim=0 10 10 0, clip, width=0.33\textwidth, keepaspectratio]{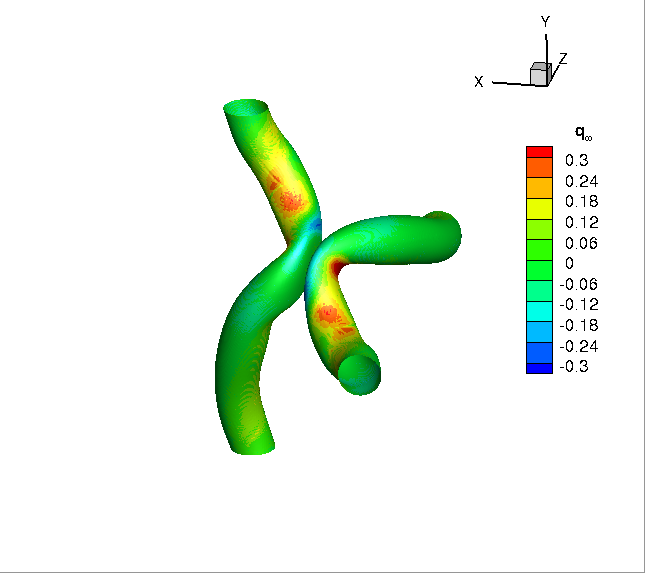}}\\
    \subfloat[]{\includegraphics[ trim=0 10 10 0, clip, width=0.33\textwidth, keepaspectratio]{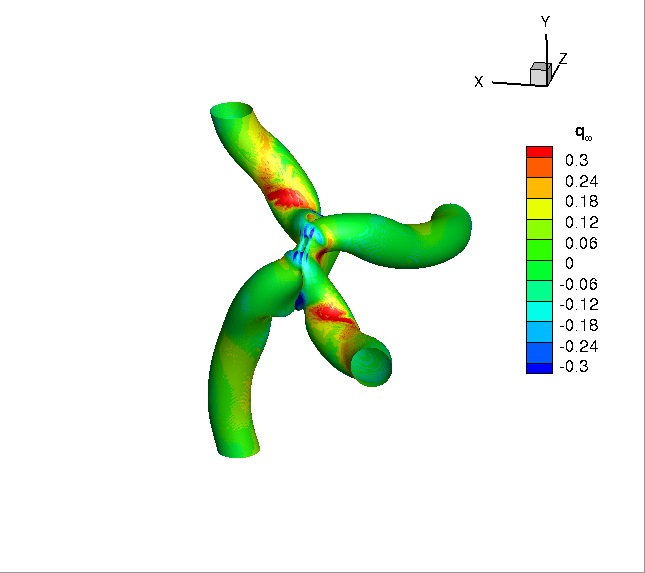}}
    \subfloat[]{\includegraphics[ trim=0 10 10 0, clip, width=0.33\textwidth, keepaspectratio]{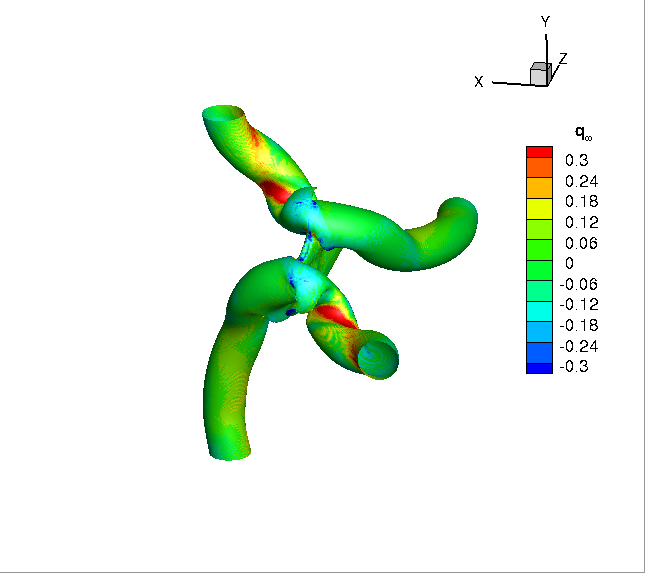}}
    \subfloat[]{\includegraphics[ trim=0 10 10 0, clip, width=0.33\textwidth, keepaspectratio]{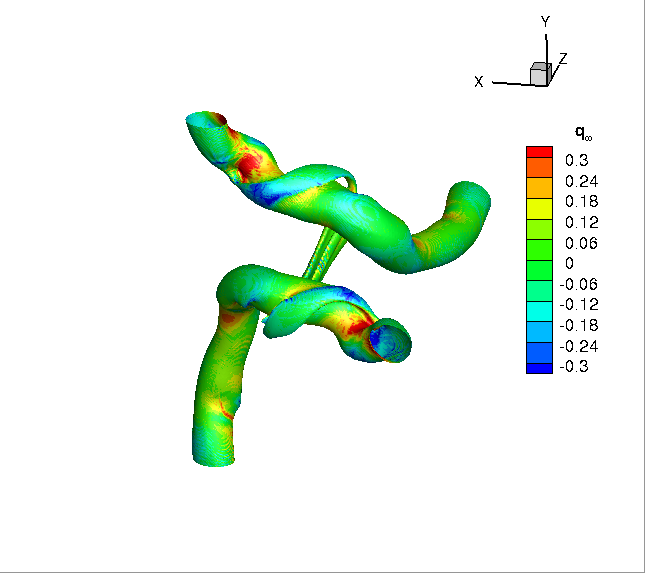}}\\
    \caption{$|\omega|$ isosurfaces at $40\%$ of maximum initial vorticity colored by $q_\omega$ at t = (a) $0$, (b) $2.64$, (c) $4.32$, (d) $4.92$, (e) $5.28$ and (f) $6$.}
    \label{fig:qomg_o}
\end{figure}

\begin{figure}
    \centering
    \subfloat[]{\includegraphics[width=0.3\textwidth]{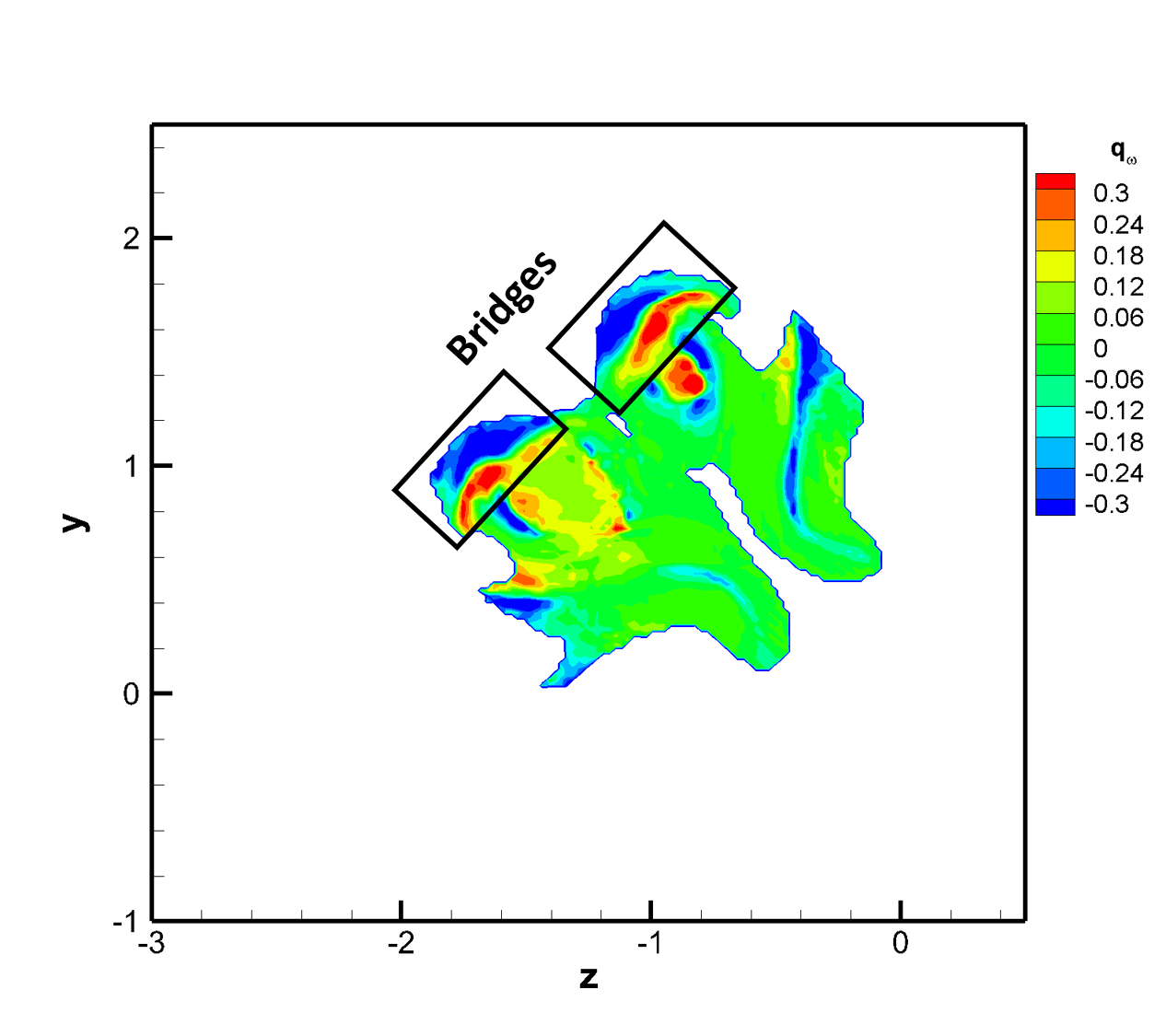}}
    \subfloat[]{\includegraphics[width=0.3\textwidth]{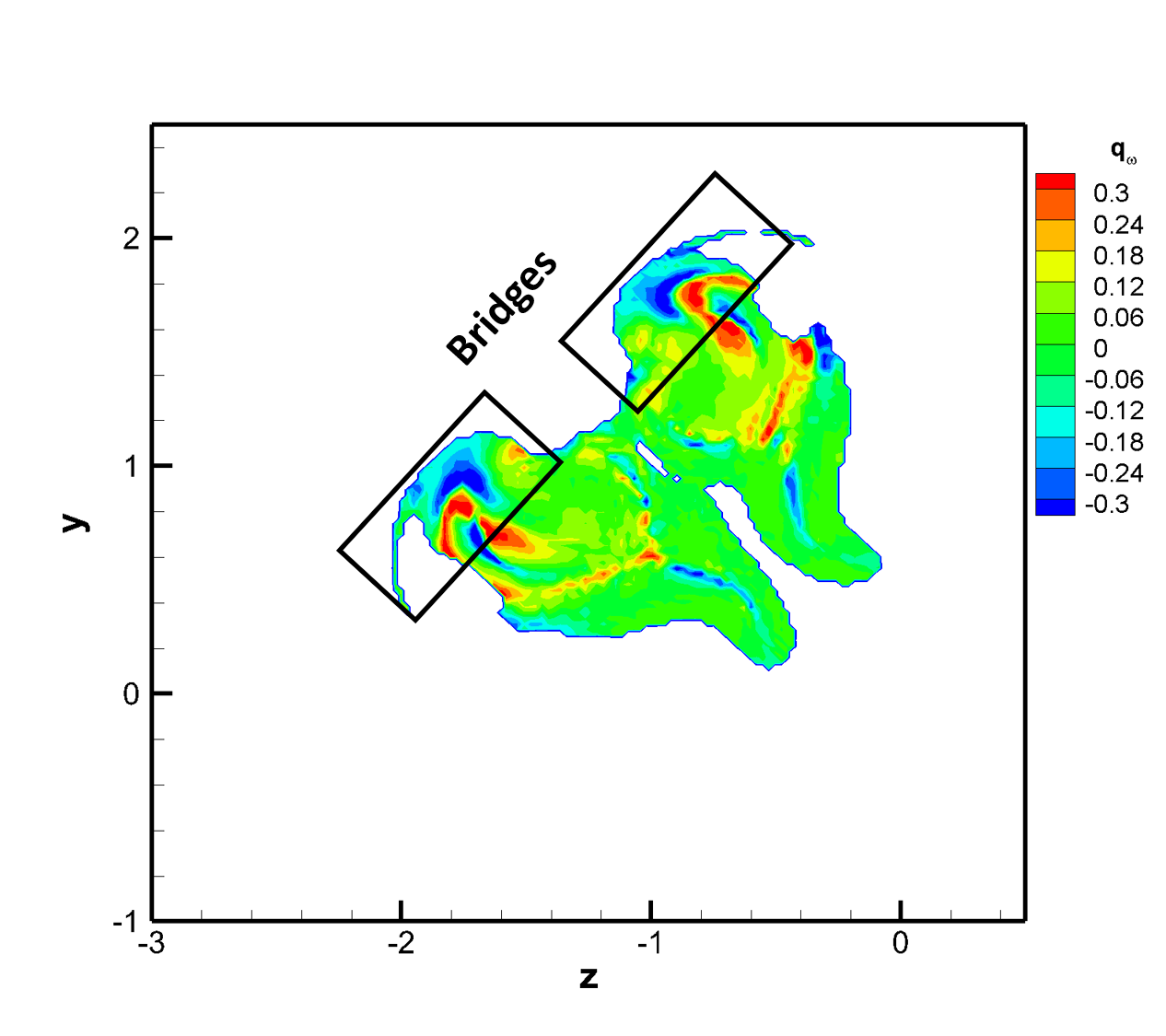}}
    \subfloat[]{\includegraphics[width=0.3\textwidth]{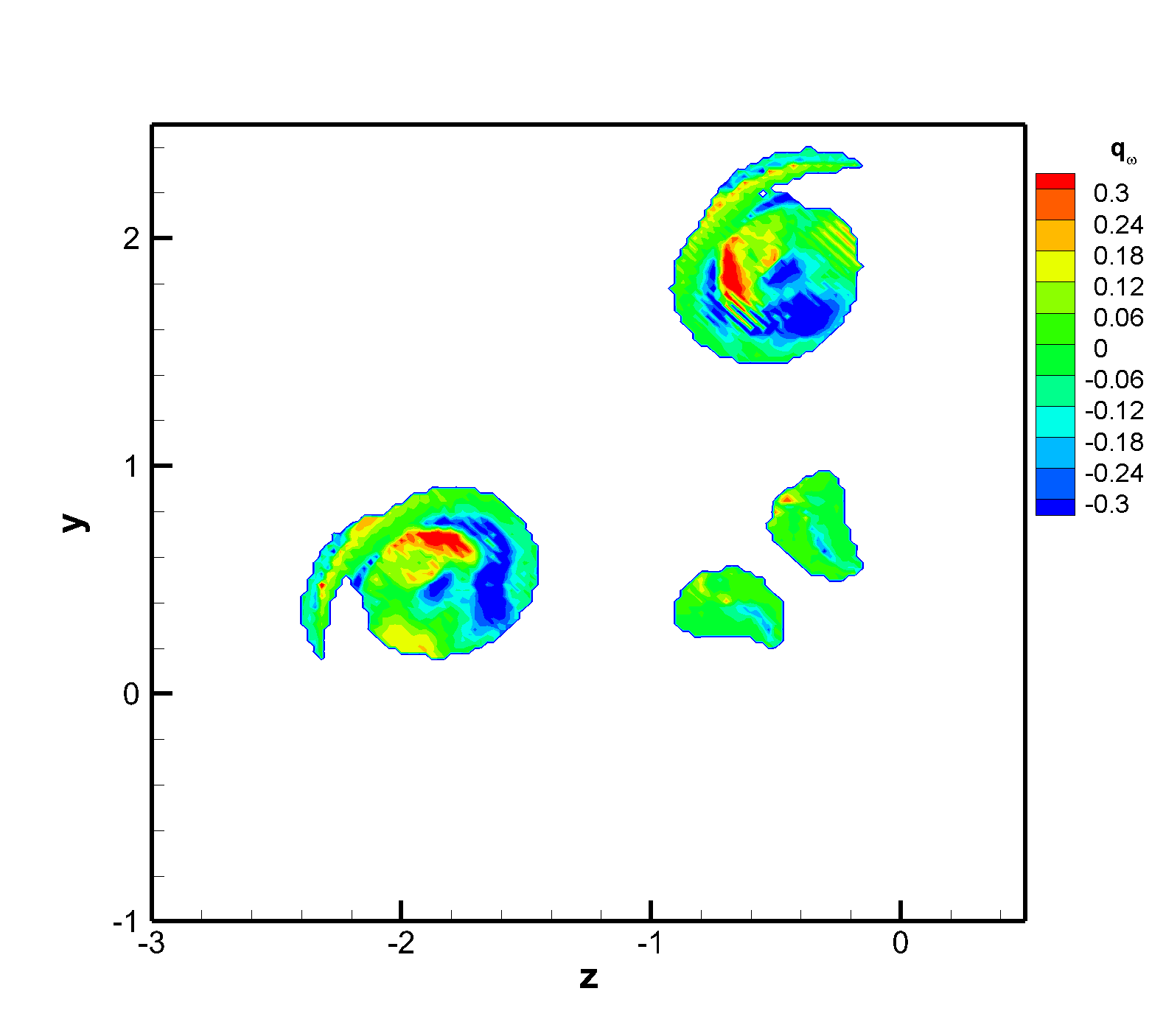}}
    \caption{$q_\omega$ contours in the dividing plane at $t=$ (a) $5.16$, (b) $5.28$ and (c) $6$. Contours are only shown in regions wherein $|\omega|>0.4\omega_0$}
    \label{fig:o_qwcontour_div}
\end{figure}

\begin{figure}
    \centering
    \subfloat[]{\includegraphics[width=0.3\textwidth]{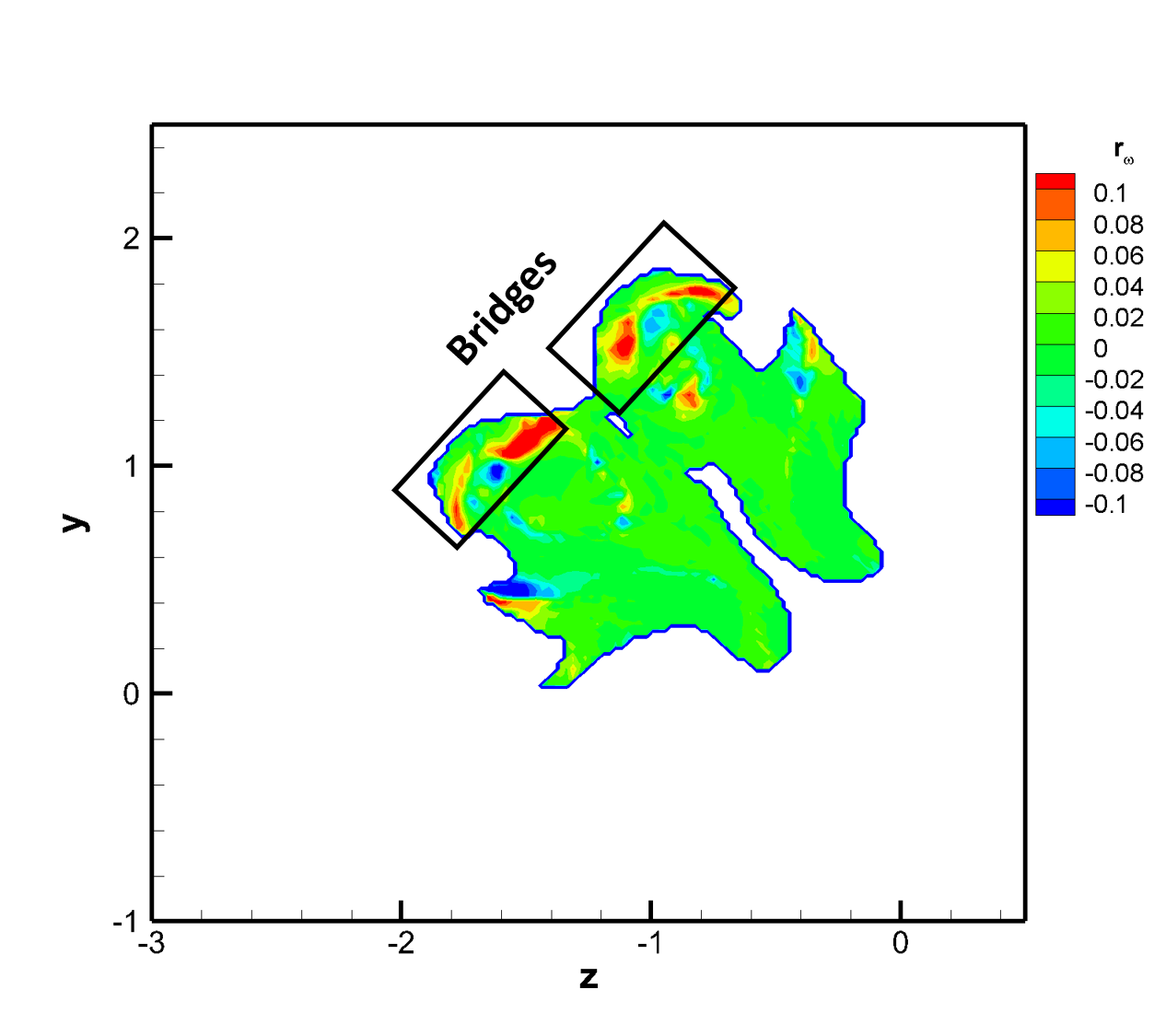}}
    \subfloat[]{\includegraphics[width=0.3\textwidth]{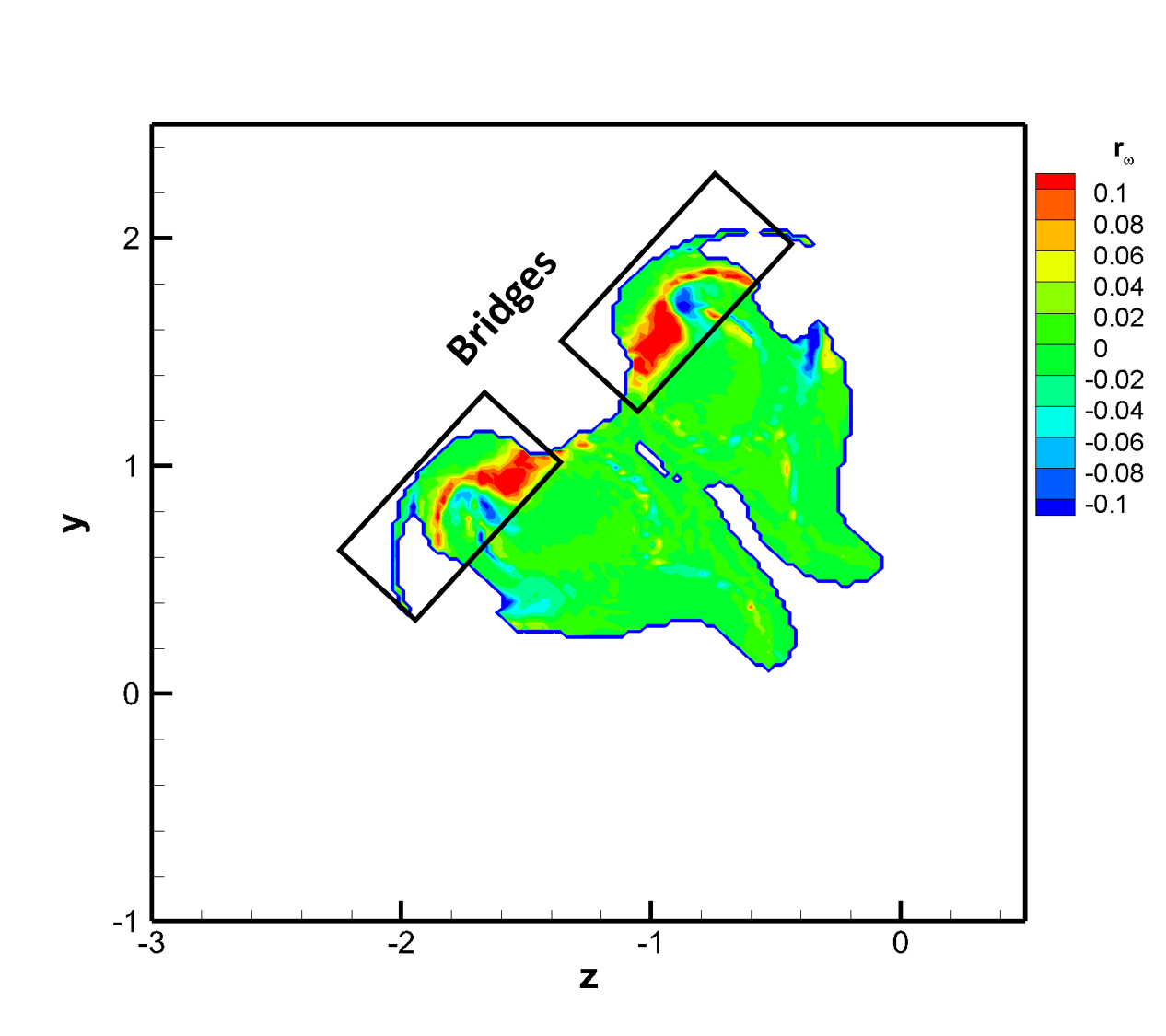}}
    \subfloat[]{\includegraphics[width=0.3\textwidth]{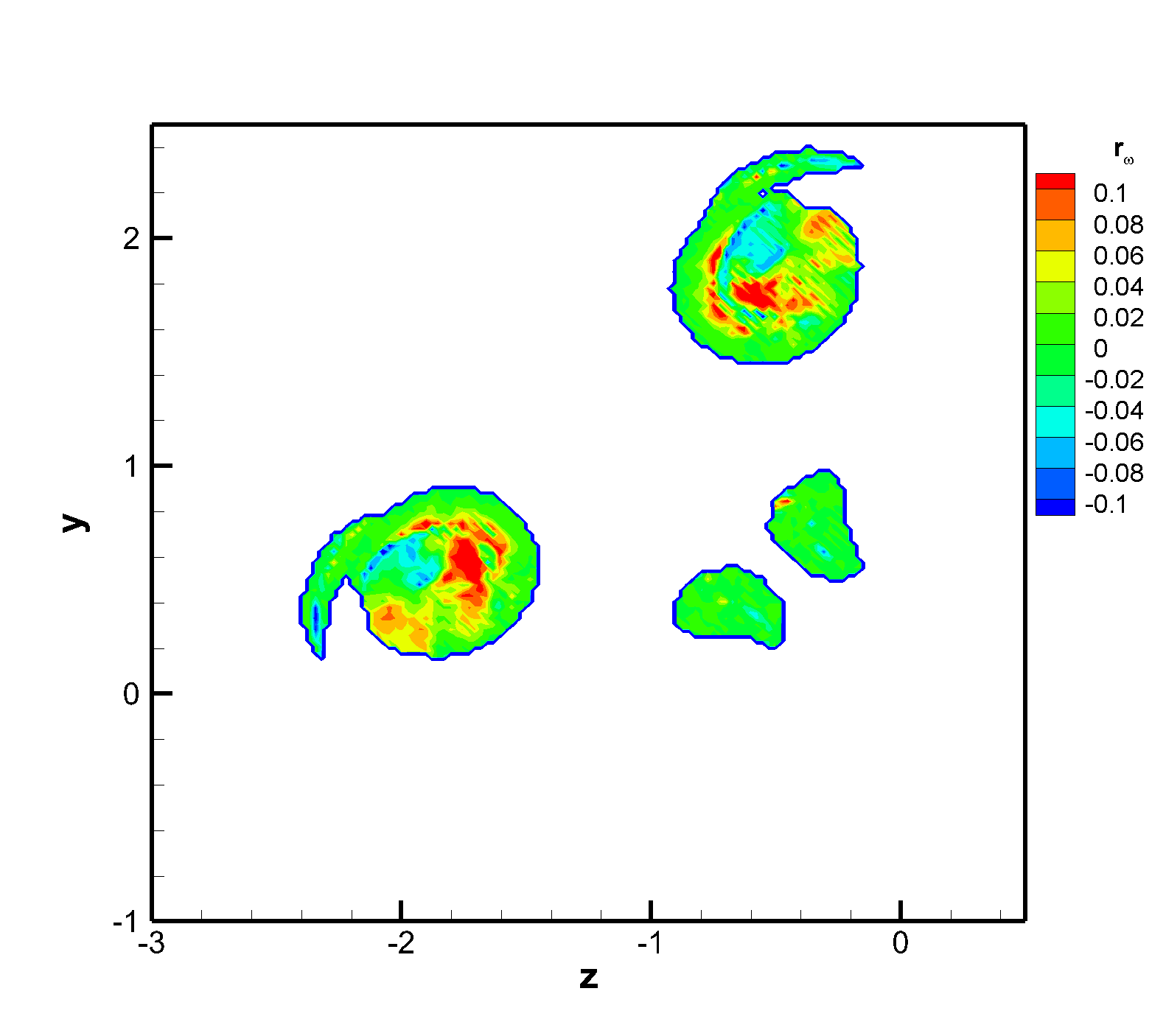}}
    \caption{$r_\omega$ contours in the dividing plane at $t=$ (a) $5.16$, (b) $5.28$ and (c) $6$. Contours are only shown in regions wherein $|\omega|>0.4\omega_0$}
    \label{fig:o_rwcontour_div}
\end{figure}

In this section we consider the reconnection of orthogonally offset vortex tubes. 
The evolution of isosurfaces of vorticity magnitude colored by $q_\omega$ at different stages leading up to and beyond reconnection of the tubes are shown in figure \ref{fig:qomg_o}.
Unlike the previous case of anti-parallel vortex tubes, there exist certain regions at the vortex tube surface wherein $r_\omega$ is non-zero, suggesting that the local vortex line shapes in such regions are not necessarily planar and likely to be three dimensional. However, such regions are few and far between, and locally the vortex lines are predominantly two dimensional at the surface of the tubes. The different stages of reconnection are described as follows with reference to figures \ref{fig:qomg_o}, \ref{fig:o_qwcontour_div} and \ref{fig:o_rwcontour_div}:
\begin{enumerate}
    \item Initially, the local vortex lines are straight everywhere in the tubes. The vortex tubes move and deform under the influence of each other's velocity field. At $t=2.64$, the local vortex line shape continues to be straight almost everywhere except for the highly curved regions in the vortex tube wherein it is elliptic as indicated by positive $q_\omega$ values.
    \item  Vortex Z under the influence of the velocity field of Vortex Y starts moving along the direction of it's binormal. This leads to the configuration as shown in figure \ref{fig:qomg_o}(c) wherein the mid regions of the vortex tubes are parallel and vorticity in the tubes are directed in opposite directions. 
    The cores of the vortex tubes at the mid regions are flat and pressed against each other creating an ideal setting for reconnection via bridging. At this stage, the vortex lines shapes in the tubes are dominantly straight except for the elliptic vortex lines in the curved regions of the tubes. As in the previous case of reconnection in anti-parallel vortex tubes, vortex line shapes in the contact region are also elliptic.
    \item By $t=4.92$, bridging is initiated and the ends of the contact regions are connected by bridges. The local vortex line shapes in the upper bridge surface is dominantly hyperbolic as indicated by negative values of $q_\omega$ there. We also plot contours of $q_\omega$ (figure \ref{fig:o_qwcontour_div}) and $r_\omega$ (figure \ref{fig:o_rwcontour_div}) in the dividing plane slicing through the bridges. 
    At $t=5.16$, just after the onset of bridging, $q_\omega$ is dominantly negative in the outer region of bridges while it is positive in the inner regions. At this stage, the contours of $r_\omega$ shown in figure \ref{fig:o_rwcontour_div}(a) demonstrate that 
    even though $r_\omega$ is not exactly zero everywhere inside the bridges, it is very close to zero in the non-zero $q_\omega$ regions of the bridges.
    Therefore, in such regions the vortex line shapes are nearly planar and the dominant vortex line shape in the bridges is clearly constituted by ``elliptic-hyperbolic pairing".
    \item 
    The sequence of events beyond bridging is similar to the anti-parallel case. Further annihilation of vorticity in the symmetry plane accompanied by generation of orthogonal vorticity in the dividing plane makes the bridges stronger, while, simultaneously weakening the mid section of the tube (figure \ref{fig:qomg_o}(e)). The vortex line shapes in the tubes in the bridges continue to show ``elliptic-hyperbolic pairing".  
    \item By $t=6$, the bridges have integrated with the tubes and the hump is indiscernible. Self induction has pushed the bridges away from each other consequently stretching the mid sections of the tubes into slender threads. At this stage, the vortex line shapes at the surface of the reconnected tubes are mostly straight lines barring the highly curved regions of the tubes wherein elliptic vortex lines occur. Overall, the vortex line shapes in the reconnected region are still dominated by ``elliptic-hyperbolic pairing" (figure \ref{fig:o_qwcontour_div}c).
\end{enumerate}

The sequence of events leading up to and beyond vortex reconnection via bridging in this case are similar to the anti-parallel case.
Additionally, in both the cases a specific configuration of elliptic and hyperbolic vortex lines are prominent in the bridges.
This leads us to conclude that the emergence of this ``elliptic-hyperbolic pairing" in the bridges is independent of the initial orientation of the vortex tubes.
  

\section{Summary and Conclusions}

The paper seeks to characterize infinitesimal vortex line topology by adapting the local streamline topology classification method of \cite{chong1990general}.  The differences between velocity and vorticity fields are identified and the critical point analysis is suitably modified. Topology classification in terms of second and third invariants of the vorticity-gradient tensor is reiterated and the invariant evolution equations are derived.    Vortex line geometry classification, which is related to topology, is introduced using suitably normalized vorticity-gradient invariants \citep{das2019reynolds}. 
An extensive investigation of vortex line geometry distribution in forced isotropic turbulence is conducted over a wide range of Reynolds numbers. Specifically, the joint pdf of the second and third normalized vorticity-gradient invariants ($q_\omega$-$r_\omega$) is examined. At very low Reynolds numbers (order unity), the joint pdf form is similar to that of a Gaussian field.  With increasing Reynolds number, the pdf form changes and attains self-similarity beyond $Re_\lambda > 200$. It was shown in \cite{das2019reynolds} that the joint pdf of normalized velocity-gradients invariants also attains self-similarity beyond $Re_\lambda > 200$. The high Reynolds number vorticity-invariants’ pdf is of symmetric bell-shape with the highest probability density at locally parallel vortex lines.  
The topology and geometry distribution during Taylor-Green vortex breakdown toward turbulence is next examined. Initially, the flow field is constituted of only two specific vortex line geometric shapes. With time the flow deforms and convolutes the vortex lines, creating smaller scales of motion. Consequently, the vortex line elements of different topologies and geometric shapes are generated. The joint pdf of  $q_\omega$-$r_\omega$  gradually takes the characteristic bell-shape seen in forced isotropic turbulence. This finding suggests that that the bell-shape distribution is likely a universal characteristic of turbulence across different types of turbulent flows analogous to the tear-drop shape of the velocity-gradient invariants’ joint pdf.
The study next examines the vortex reconnection phenomenon, specifically the bridging process that initiates the merger. Different initial vortex-tube configurations are considered. It is demonstrated the structure of vortex filaments inside the bridges in both cases considered are distributed in a similar fashion. The bridge is constituted of two-dimensional elliptic vortex lines on one side and hyperbolic vortex lines on the other. 

\newpage
\vspace{0.4cm}
\noindent \textbf{Acknowledgements}
\vspace{0.1cm}

The authors would like to thank Prof. Diego Donzis of Texas A\&M University for providing the DNS data of forced isotropic turbulence used in this study.
Portions of this research were conducted with the advanced computing resources provided by Texas A\&M High Performance Research Computing.

\vspace{0.4cm}
\noindent
Declaration of Interests. The authors report no conflict of interest.


\appendix

\section{}\label{app:A}
In a frame of reference rotating with angular velocity $\Omega$, the local vorticity ($\vec{\omega}_R$) field is related to the inertial vorticity field by
\begin{equation}
    \label{eq_app:vort_rel}
    \vec{{\omega}}_R=\vec{\omega}-2\vec{\Omega}
\end{equation}
We select a coordinate frame rotating with angular velocity $\vec{\Omega}=\vec{\omega}(\vec{x}_0)/2$. In such a frame the local vorticity field ($\vec{\omega}_R$) is given by the following equation. 
\begin{equation}
    \label{eq_app:vort_rel1}
    \vec{{\omega}}_R=\vec{\omega}-\vec{\omega}(\vec{x}_0)
\end{equation}
The local vorticity in such a frame is the same as the ``relative vorticity field'' $\vec{\tilde{\omega}}(\vec{x};\vec{x}_0)$ as defined in \eqref{eq:def_vortrel}. We now derive equations for vortex lines in the rotating frame of reference. Vortex lines as observed from a rotating frame of reference are curves tangent to the local vorticity in the rotating frame. We denote by $x_i'$ the basis of rotating frame whereas $x_i$ represents the inertial basis. Similarly, $(\omega_R')_i$ denotes the components of local vorticity vector along the rotating basis and $(\omega_R)_i$ denotes the components of vorticity in the rotating frame expressed along the inertial basis. At time $t$, the coordinate basis $x_i$ can be transformed to $x_i'$ by a proper rotation. Let $\boldsymbol{Q}$ be an orthogonal coordinate transformation tensor such that
\begin{equation}
    \label{eq_app:Qdef}
    Q_{ij}=\frac{\partial x_i'}{\partial x_j}
\end{equation} 
$\boldsymbol{Q}$ obeys the standard transformation rules transforming vectors between the two bases $x_i$ and $x_i'$ as follows
\begin{equation}
    \label{eq_app:vec_trans}
    (\omega_R')_i=Q_{im}(\omega_R)_m
\end{equation}
 In a rotating frame, the differential equation governing vortex lines is as follows:
\begin{equation}
\label{eq_app:vline_rot}
    \frac{dx_i'}{ds}=(\omega_R')_i
\end{equation}
We multiply \eqref{eq_app:vline_rot} by $Q_{im}$ to cast it along the inertial basis.
\begin{equation}
\label{eq_app:vline_rot_i}
    \frac{dx_m}{ds}=(\omega_R)_m
\end{equation}
where we have used \eqref{eq_app:Qdef} and the transformation identity for vectors \eqref{eq_app:vec_trans}. Solution trajectories obtained by integrating \eqref{eq_app:vline_rot_i} for a frozen vorticity field are vortex lines as observed in a rotating frame of reference. 

Since $\vec{\tilde{\omega}}(\vec{x};\vec{x_0})$ and $\vec{\omega}_R$ are same by definition, equation \eqref{eq_app:vline_rot_i} and \eqref{eq:rvline_diff} are identical. Thus, the so called ``relative vortex lines'' are indeed the vortex lines observed from a frame rotating with angular velocity $\vec{\omega}(\vec{x}_0)/2$.

\section{Evolution equations}\label{app:B}
In this section the evolution equations for the components of vorticity gradient tensor ($\Phi_{ij}$) and it's invariants $Q_\omega$ \& $R_\omega$ are developed. 
The governing equation for vorticity ($\omega_i$) is given by,
\begin{equation}
    \label{eq:vorticity}
    \frac{D \omega_i}{D t}=S_{ik}\omega_{k} +\nu\frac{\partial ^2 \omega_i}{\partial x_k\partial x_k}
\end{equation}
where, $\bm{S}$ is the strain-rate tensor (symmetric part of velocity gradient tensor, $\bm{A}$).
The evolution equation for $\Phi_{ij}$ is obtained by differentiating equation (\ref{eq:vorticity}) with respect to the spatial coordinates $x_j$.
\begin{equation}
    \label{eq:wgt}
    \frac{D \Phi_{ij}}{Dt} =-\Phi_{ik}A_{kj}+\frac{\partial S_{ik}}{\partial x_j}\omega_k+S_{ik}\Phi_{kj}+\nu\frac{\partial \Phi_{ij}}{\partial x_k \partial x_k}
\end{equation}
The 1st term on the right hand side of equation (\ref{eq:wgt}) is \textit{non-linear production} of vorticity gradient, the 2nd and 3rd term represent the effect of vortex stretching on vorticity gradients and the final term is viscous diffusion.

To obtain the equation of the second invariant ($Q_\omega$) of $\bm{\Phi}$, first the equation for inner product of $\Phi_{ij}$ is derived 
\begin{equation}
    \label{eq:innerwgt}
   \begin{split}
           \frac{D}{Dt}(\Phi_{ij}\Phi_{jn})  & =-\left[\Phi_{ik}A_{kj}\Phi_{jn}+\Phi_{ij}A_{jk}\Phi_{kn}\right]+\left[\Phi_{ij}\frac{\partial S_{jk}}{\partial x_n}+\frac{\partial S_{ik}}{\partial x_j}\Phi_{jn}\right]\omega_k \\ 
            & +\left[\Phi_{ij}S_{jk}\Phi_{kn}+S_{ik}\Phi_{kj}\Phi_{jn}\right] -2\nu\frac{\partial \Phi_{ij}}{\partial x_k}\frac{\partial \Phi_{jn}}{\partial x_k} \\
            & +\nu\frac{\partial^2}{\partial x_k \partial x_k}(\Phi_{ij}\Phi_{jn})
    \end{split}
\end{equation}
The equation for $Q_\omega$ can be derived by taking the trace of (\ref{eq:innerwgt})
\begin{equation}
    \label{eq:Qw}
    \frac{D Q_\omega}{Dt}=\Phi_{ij}\left[A_{jk}\Phi_{ki}-\frac{\partial }{\partial x_i}(S_{jk}\omega_{k})\right]+\nu\left[\frac{\partial \Phi_{ij}}{\partial x_k}\frac{\partial \Phi_{ij}}{\partial x_k}+\frac{\partial^2 Q_\omega}{\partial x_k\partial x_k}\right]
\end{equation}
To obtain the equation of the third invariant ($R_\omega$) of $\bm{\Phi}$, first the equation for triple product of $\Phi_{ij}$ is derived
\begin{equation}
    \label{eq:tpwgt}
       \begin{split}
           \frac{D}{Dt}(\Phi_{ij}\Phi_{jn}\Phi_{nl})  & =-\left[\Phi_{ik}A_{kj}\Phi_{jn}\Phi_{nl}+\Phi_{ij}A_{jk}\Phi_{kn}\Phi_{nl}+\Phi_{ij}\Phi_{jn}\Phi_{nk}A_{kl}\right]\\
           & +\left[\Phi_{ij}\frac{\partial S_{jk}}{\partial x_n}\Phi_{nl}+\frac{\partial S_{ik}}{\partial x_j}\Phi_{jn}\Phi_{nl}+\Phi_{ij}\Phi_{jn}\frac{\partial S_{nk}}{\partial x_l}\right]\omega_k \\ 
            & +\left[\Phi_{ij}S_{jk}\Phi_{kn}\Phi_{nl}+S_{ik}\Phi_{kj}\Phi_{jn}\Phi_{nl}\right] \\ 
            & -2\nu\left[\frac{\partial}{\partial x_k}( \Phi_{ij}\Phi_{jn})\frac{\partial \Phi_{nl}}{\partial x_k}+\frac{\partial \Phi_{ij}}{\partial x_k}\frac{\partial \Phi_{jn}}{\partial x_k}\Phi_{nl}\right]  \\
            & +\nu\frac{\partial^2}{\partial x_k \partial x_k}(\Phi_{ij}\Phi_{jn}\Phi_{nl})
    \end{split}
\end{equation}
The equation for $R_\omega$ can be derived by taking the trace of (\ref{eq:tpwgt})
\begin{equation}
    \label{eq:Rw0}
    \begin{split}
    \frac{D R_\omega}{Dt} & =\Phi_{ij}\left[A_{jk}\Phi_{kn}-\frac{\partial}{\partial x_n}(S_{jk}\omega_k)\right]\Phi_{ni}\\
    & +\frac{2\nu}{3}\left[\frac{\partial}{\partial x_k}(\Phi_{ij}\Phi_{jn})\frac{\partial \Phi_{ni}}{\partial x_k}+\frac{\partial \Phi_{ij}}{\partial x_k}\frac{\partial \Phi_{jn}}{\partial x_k}\Phi_{ni}\right] +\nu\frac{\partial^2 R_\omega}{\partial x_k\partial x_k}
    \end{split}
\end{equation}
Using Cayley-Hamilton theorem,
\begin{equation}
    \label{eq:ch}
    \Phi_{ij}\Phi_{jk}\Phi_{kn}+Q_\omega\Phi_{ij}\Phi_{jn}+R_\omega\delta_{in}=0
\end{equation}
equation (\ref{eq:Rw0}) can be further simplified to attain the evolution equation of $R_\omega$
\begin{equation}
    \label{eq:Rw}
    \begin{split}
    \frac{D R_\omega}{Dt}  =Q_\omega\Phi_{ij}W_{ij} -\Phi_{ij}\frac{\partial S_{jk}}{\partial x_n}\omega_k\Phi_{ni}
     +\frac{2\nu}{3}\left[\frac{\partial}{\partial x_k}(\Phi_{ij}\Phi_{jn})\frac{\partial \Phi_{ni}}{\partial x_k}+\frac{\partial \Phi_{ij}}{\partial x_k}\frac{\partial \Phi_{jn}}{\partial x_k}\Phi_{ni}\right]
     +\nu\frac{\partial^2 R_\omega}{\partial x_k\partial x_k}
    \end{split}
\end{equation}

\bibliographystyle{jfm}
\bibliography{main1}

\end{document}